\documentclass[a4paper,fleqn]{cas-dc}

\usepackage[numbers]{natbib}

\usepackage{algorithm2e}
\usepackage{algcompatible}

\usepackage{environ}
\usepackage{graphicx}

\allowdisplaybreaks[1]
\sloppypar

\newcommand{\apriori}{\textit{a priori}}
\newcommand{\eg}{\textit{e.g.}}
\newcommand{\etal}{\textit{et al.}}
\newcommand{\etc}{\textit{etc.}}
\newcommand{\ie}{\textit{i.e.}}

\renewcommand{\@}{\partial}
\newcommand{\argmin}{\mathop{\mathrm{argmin}}}
\newcommand{\bydef}{\;\triangleq\;}

\renewcommand{\d}{\mathrm{d}}

\newcommand{\Df}[2]{\dfrac{\d #1}{\d #2}}
\newcommand{\Ddf}[2]{\dfrac{\d^2{#1}}{\d{#2}^2}}
\newcommand{\df}[2]{\dfrac{\partial #1}{\partial #2}}
\newcommand{\ddf}[2]{\dfrac{\partial^2 #1}{\partial #2^2}}
\newcommand{\diag}{\mathrm{diag}}       
\newcommand{\dirac}{{\delta}}  

\newcommand{\e}{\mathrm{e}}
\newcommand{\Heav}{\mathrm{H}}   
\newcommand{\inner}[2]{\left\langle #1 \,\Big|\, #2 \right\rangle} 
\newcommand{\intinf}{\int\limits_{-\infty}^{\infty}}
\newcommand{\intoinf}{\int\limits_{0}^{\infty}}
\newcommand{\kron}[2]{{\delta_{#1,#2}}} 
\newcommand{\Lamb}{\mathrm{W}_0} 
\newcommand{\Ltwo}{L^2}          
\newcommand{\Mx}[1]{\begin{pmatrix}#1\end{pmatrix}}
\newcommand{\Min}[2]{\left(#1\wedge #2\right)}
\newcommand{\mx}[1]{\mathbf{#1}}
\providecommand{\norm}[1]{\left\lVert#1\right\rVert}
\renewcommand{\O}[1]{\mathcal{O}\left(#1\right)}

\newcommand{\Real}{\mathbb{R}}

\newcommand{\sech}{\mathrm{sech}}
\newcommand{\T}{^{\!\top}}
\newcommand{\Tr}{^{\!\top}}
\newcommand{\X}[1]{\cdot10^{#1}} 

\newcommand{\alref}[1]{\ref{alg:#1}}
\newcommand{\alg}[1]{Algorithm~\ref{alg:#1}}
\newcommand{\algs}[1]{Algorithms~\ref{alg:#1}}
\newcommand{\alglabel}[1]{\label{alg:#1}}
\newcommand{\Algstep}[1]{\State $\bullet$ \emph{#1}}
\newcommand{\Algwidget}[2]{\Statex \textbf{#1} \emph{#2}}

\newcommand{\Appn}[1]{Appendix~\secref{#1}}

\newcommand{\eq}[1]{(\ref{eq:#1})}
\newcommand{\eqlabel}[1]{\label{eq:#1}}
\def\eqreftwo(#1,#2){(\ref{eq:#1},\ref{eq:#2})}
\newcommand{\eqtwo}[1]{\eqreftwo(#1)}
\def\eqrefthree(#1,#2,#3){(\ref{eq:#1},\ref{eq:#2},\ref{eq:#3})}
\newcommand{\eqthree}[1]{\eqrefthree(#1)}
\def\eqreffour(#1,#2,#3,#4){(\ref{eq:#1},\ref{eq:#2},\ref{eq:#3},\ref{eq:#4})}

\def\eqreffive(#1,#2,#3,#4,#5){(\ref{eq:#1},\ref{eq:#2},\ref{eq:#3},\ref{eq:#4},\ref{eq:#5})}

\NewEnviron{eqsplit}{\begin{equation}\begin{split} \BODY \end{split}\end{equation}}

\newcommand{\Fig}[1]{Fig.~\ref{fig:#1}}
\newcommand{\fig}[1]{fig.~\ref{fig:#1}}
\newcommand{\figlabel}[1]{\label{fig:#1}}

\newcommand{\dblfigure}[3]{\begin{figure*}\centerline{#1}\caption[]{#2}\figlabel{#3}\end{figure*}}
\newcommand{\sglfigure}[3]{\begin{figure}\centerline{#1}\caption[]{#2}\figlabel{#3}\end{figure}}

\newcommand{\Secn}[1]{Section~\secref{#1}}

\newcommand{\seclabel}[1]{\label{sec:#1}}
\newcommand{\secref}[1]{\ref{sec:#1}}

\newcommand{\steps}[1]{\textrm{\textit{\textbf{\underline{#1}}}}}

\newcommand{\leri}[1]{\left(#1\right)}
\newcommand{\rile}[1]{\right. \\ & \left.#1}

\ifx\notcolor\undefined\newcommand{\notcolor}{blue}\else\fi
\newcommand{\+}[2]{\def#1{{#2}}}
\newcommand{\1}[2]{\def#1##1{{#2}}}
\newcommand{\2}[2]{\def#1##1##2{{#2}}}
\newcommand{\3}[2]{\newcommand{#1}[3]{{#2}}}

\+{\0}{\mx{0}}      
\+{\A}{A}           
\+{\Ar}{\mx{A}}     
\+{\Al}{\tilde{A}}  
\+{\a}{a}           
\+{\ai}{A}          
\+{\alp}{\alpha}    
\+{\ampu}{U\st}     
\+{\ampuc}{U_{\mathrm{s}}^*}  
\+{\ampucu}{\overline{\ampuc}} 
\+{\ampucl}{\underline{\ampuc}} 
\+{\amput}{U\st\test}  
\+{\ampv}{V\st}     
\+{\ampi}{I\st}     
\+{\ampic}{I_{\mathrm{s}}^*}  
\+{\ampicl}{\underline{\ampic}}  
\+{\B}{B}           
\+{\bet}{\beta}     
\+{\bet}{\beta}     
\2{\br}{{#1}_{#2}}          
\2{\bl}{\tilde{#1}_{#2}\,} 
\3{\Bl}{\tilde{#1}_{#2}^{#3}\,} 

\+{\brc}{\mathrm{Ca}} 
\+{\brd}{d}         
\+{\brf}{f}         
\+{\brh}{h}         
\+{\brj}{j}         
\+{\brm}{m}         
\+{\brx}{x_1}       
\+{\brV}{V}         
\+{\brVc}{\hat\brV} 
\+{\brik}{I_{K_1}}  
\+{\briks}{\overline{\brik}}  
\+{\brin}{I_{Na}}   
\+{\bris}{I_s}      
\+{\brix}{I_{x1}}   
\+{\brald}{\alpha_d}
\+{\brbed}{\beta_d} 
\+{\bralf}{\alpha_f}
\+{\brbef}{\beta_f} 
\+{\bralh}{\alpha_h}
\+{\brbeh}{\beta_h} 
\+{\bralj}{\alpha_j}
\+{\brbej}{\beta_j} 
\+{\bralm}{\alpha_m}
\+{\brbem}{\beta_m} 
\+{\bralx}{\alpha_{x1}}
\+{\brbex}{\beta_{x1}} 
\+{\brgix}{g_{ix}}  
\+{\brgna}{g_{Na}}  
\+{\brgnac}{g_{Na_c}}  
\+{\brgs}{g_{s}}    
\+{\brena}{E_{Na}}  
\+{\bres}{E_{s}}    
\+{\bralp}{\alpha}  
\+{\Compat}{\mx{\Phi}} 
\+{\C}{C}           
\+{\c}{c}           
\+{\ct}{\c\test}    
\+{\cwave}{\c\wave} 
\+{\cP}{c_P}        
\+{\chir}{\chi}     
\+{\chil}{\tilde{\chi}}     
\+{\compat}{\Phi}   
\+{\bD}{\mx{D}}     
\+{\D}{D}           
\+{\Den}{\mathcal{D}}         
\+{\dim}{d}         
\+{\dt}{\Delta_t}   
\+{\dx}{\Delta_x}   
\+{\E}{E}           
\+{\Ec}{\hat{E}}    
\+{\Ep}{E_1}        
\+{\Er}{E_{\mathrm{r}}} 
\2{\ETT}{\mathcal{\E}^{#1}_{#2}} 
\2{\ETTnum}{\mathcal{\numer{\E}}^{#1}_{#2}}
\+{\best}{\mx{e}}    
\+{\est}{e}    
\+{\Est}{E}    
\+{\exty}{\tau}     
\+{\Fem}{\Phi}      
\+{\ff}{\mx{f}}     
\+{\f}{f}           
\+{\feA}{\mx{\fea}} 
\+{\fea}{A}         
\+{\feB}{\mx{\feb}} 
\+{\feb}{B}         
\+{\feC}{\mx{\fec}} 
\+{\fec}{C}         
\+{\feF}{\mx{\fef}} 
\+{\fef}{F}         
\+{\fE}{f_{\E}}     
\+{\fh}{f_{\h}}     
\+{\fpin}{\mu}      
\+{\fone}{\f_1}     
\+{\ftwo}{\f_2}     
\+{\fal}{\alpha}    
\+{\fep}{\gamma}    
\+{\fth}{\beta}     
\+{\fu}{f_u}        
\+{\fv}{f_v}        
\+{\G}{g}           
\+{\lG}{G}          
\+{\g}{\mx{h}}      
\+{\gam}{\gamma}    
\+{\gf}{\tilde\g}   
\+{\gp}{h}          
\+{\h}{h}           
\+{\hc}{\hat{h}}    
\+{\hp}{h_1}        
\+{\hr}{h_{\mathrm{r}}} 
\+{\INa}{I_{\mathrm{Na}}} 
\+{\Interval}{I}    
\+{\Irh}{I_{\mathrm{rh}}} 
\+{\i}{i}           
\+{\ia}{a}          
\+{\ib}{b}          
\+{\ic}{c}          
\+{\ii}{j}          
\+{\iter}{i}        
\+{\J}{\mx{F}}      
\+{\j}{j}           
\+{\jj}{j}          
\+{\k}{k}           
\+{\kk}{\kappa}     
\+{\Int}{\mathcal{I}} 
\+{\Ker}{\mx{K}}    
\+{\L}{\mathcal{L}} 
\+{\Length}{L}      
\+{\Lp}{\L^{+}}     
\+{\LV}{\mx{W}}     
\+{\l}{\ell}        
\+{\ltwo}{\ell^2}   
\+{\lv}{W}          
\+{\lw}{\rw}        
\+{\m}{m}           
\+{\mm}{m}          
\+{\mth}{a}         
\2{\mur}{\mu^{#1}_{#2}} 
\2{\mul}{\tilde\mu^{#1}_{#2}} 
\+{\mult}{\mu}      
\+{\N}{N}           
\+{\Num}{\mathcal{N}}         
\+{\n}{n}           
\+{\nn}{n}          
\+{\nup}{p}         
\+{\nuq}{q}         
\+{\nus}{\nu}       
\1{\nur}{\nu_{#1}}  
\2{\Nur}{\nu_{#1}^{#2}}  
\1{\Nup}{p^{#1}}    
\+{\Per}{P}         
\+{\pp}{p}          
\+{\p}{\rho}        
\+{\phir}{\phi}     
\+{\phil}{\tilde{\phi}}     
\+{\postf}{\omega}  
\+{\pref}{\alpha}   
\+{\precomp}{\eta}  

\+{\psir}{\psi}     
\+{\mur}{\mu}     

\+{\psil}{\tilde{\psi}}     
\2{\Q}{Q^{#1}_{#2}} 
\2{\R}{R_{#1,#2}}   
\+{\RV}{\mx{V}}     
\+{\r}{r}           
\+{\rv}{V}          
\+{\rw}{\lambda}    
\+{\Speed}{S}       
\+{\Scale}{\mx{S}}  
\+{\shift}{s}       
\+{\shiftwave}{\shift\wave} 
\+{\sig}{\sigma}    
\+{\solvr}{f_e}     
\+{\solvl}{\tilde{f}_e} 
\+{\solvln}{\tilde{v}_e} 
\+{\st}{_{\mathrm{s}}} 
\+{\suc}{\overline\buc} 
\+{\sRV}{\overline\RV}  
\+{\sLV}{\overline\LV}  
\+{\sprecomp}{\overline\precomp}  
\+{\Tp}{\mx{T}}     
\+{\t}{t}           
\+{\test}{^\#}      
\+{\tf}{\tau}       
\+{\tfs}{\tau'}     
\+{\ts}{\t'}        
\+{\tst}{\t\st}     
\+{\tint}{\delta\t} 
\+{\ttest}{\t\test} 
\+{\bu}{\mx{u}}     
\+{\buc}{\hat{\bu}} 
\+{\buct}{\hat{\bu}\test} 
\+{\bui}{\bu_0}     
\+{\Uwave}{\bu\wave}
\+{\Uper}{\bu_{\Per}} 
\+{\U}{U}           
\+{\u}{u}           
\+{\ub}{\u_-}       
\+{\uc}{\hat{\u}}   
\+{\uct}{\hat{\u}\test}   
\+{\uf}{\tilde\bu}  
\+{\ui}{u_0}        
\+{\Ust}{\bu\st}    
\+{\up}{\mx{v}}     
\+{\uq}{\mx{w}}     
\+{\Ur}{\bu_r}      
\+{\Ub}{\bu_-}      
\+{\ur}{\u_r}       
\+{\us}{\tilde{\bu}}
\+{\ust}{\bu\st}    
\+{\Val}{Z}         
\+{\v}{v}           
\+{\w}{w}           

\+{\val}{\zeta}     
\+{\W}{\bar{W}}     
\+{\wave}{_{\mathrm{w}}}
\+{\Xp}{\mx{X}}     
\+{\Xir}{\mx{\Xi}}  
\+{\Xil}{\tilde{\mx{\Xi}}} 
\+{\x}{x}           
\+{\xa}{x_*}        
\+{\xf}{\xi}          
\+{\xir}{\mx{q}}  
\+{\xil}{\tilde\mx{q}} 
\+{\xs}{\tilde{\x}} 
\+{\xsa}{\x_a}       
\+{\xst}{\x\st}     
\+{\xsts}{\tilde\xst}     
\+{\xm}{\Delta}
\+{\ya}{y_*}        
\+{\z}{z}           
\+{\zth}{\theta}    
\+{\etas}{\eta^*}
\+{\phis}{\phi^*}
\+{\psis}{\psi^*}
\+{\leftvec}{\mx{a}}    
\+{\rightvec}{\mx{b}}   
\1{\numer}{\check{#1}} 
\+{\un}{\numer{\u}} 
\+{\und}{\un}       
\+{\Und}{\numer{\mx{u}}} 
\+{\vn}{\numer{\v}} 
\+{\vnd}{\vn}       
\+{\Vnd}{\numer{\mx{v}}} 
\+{\w}{w}           
\+{\wn}{\numer{\u}} 
\+{\wnd}{\wn}       
\+{\Wnd}{\numer{\mx{u}}} 
\+{\TF}{\Fem}       

\+{\epsi}{\epsilon}
\+{\ut}{\tilde\u}  
\+{\vt}{\tilde\v}  
\+{\wt}{\tilde\w}  

\+{\uh}{\hat\u}  
\+{\vh}{\hat\v}  
\+{\wh}{\hat\w}  

\+{\U}{U}
\+{\V}{V}    
\+{\myxi}{\xi}
\+{\bU}{\mx{U}}     
\+{\bC}{\mx{C}}     

\+{\QW}{Q}
\+{\QA}{a}
\+{\QB}{b}
\+{\IRH}{I_{rh}}
\+{\CHRO}{\tau}
\+{\UEN}{U_E}
\+{\QK}{k}
\+{\QK}{k}
\+{\RC}{RC}
\+{\RONLY}{R}
\+{\CONLY}{C}
\+{\HILLLAM}{\lambda}
\+{\HILLKAP}{\kappa}
\+{\bM}{\mx{M}} 
\+{\bnabla}{\mx{\nabla}} 

\+{\ampes}{E\st}     
\+{\xft}{\x_f\leri{\t}}     
\+{\myxi}{\xi}     
\+{\T}{T}    
\+{\xfonly}{\x_f}     
\+{\Estar}{\E_*}     
\+{\Tfinal}{\T_{f}}     
\+{\dxi}{\Delta_\myxi}   
\+{\delT}{\Delta_\T}   
\+{\numn}{N}   
\+{\numm}{M}   
\+{\spina}{c}   

\+{\fetheta}{\theta}   

\+{\feD}{\mx{\fed}} 
\+{\fed}{D}

\+{\IBIR}{I_1}
\+{\IIKI}{I_2}
\+{\IUC}{I_3}
\+{\IDORT}{I_4}

\+{\mybe}{\beta}  
\+{\myal}{\alpha}  
\+{\myga}{\gamma}  
\+{\myde}{\delta}  
\+{\feG}{\mx{\feg}} 
\+{\feg}{G}
\+{\IBES}{I_5}
\+{\IALTI}{I_6}

\+{\hov}{\overline{h}}    
\+{\Eov}{\overline{E}}    
\+{\feK}{\mx{\fek}} 
\+{\fek}{K}
\+{\feL}{\mx{\fel}} 
\+{\fel}{L}
\+{\feMM}{\mx{\femm}} 
\+{\femm}{M}
\+{\IYEDI}{I_7}
\+{\ISEKIZ}{I_8}

\+{\hbar}{\bar{h}}    
\+{\Ebar}{\bar{E}} 
\+{\Ephi}{\phi} 
\+{\hpsi}{\psi} 

\+{\feg}{G}         
\+{\geg}{G}         
\+{\geG}{\mx{\geg}} 
\+{\fed}{D}         
\+{\feD}{\mx{\fed}} 
\+{\ampit}{I\st\test}  

\+{\mymu}{\mu}

\+{\Iext}{I_{ext}}       
\+{\seq}{{\overline{\mu}}}       
\+{\AIS}{\zeta_1}       
\+{\BIS}{\zeta_2}      
\+{\CIS}{\zeta_3}      

\+{\End}{\numer{\E}}       
\+{\EEnd}{\numer{\mx{\E}}} %
\+{\hnd}{\numer{\h}}       
\+{\Hnd}{\numer{\mx{\h}}} %

\+{\ses}{S}  
\+{\seS}{\mx{\ses}} 

\+{\nen}{N}  
\+{\neN}{\mx{\nen}} 

\+{\tol}{\mathrm{tolerance}}      

\+{\ET}{\mathcal{\E}^{t>1}}
\+{\EM}{\mathcal{\E}^m}
\+{\EB}{\mathcal{\E}^b}

\+{\xib}{\overline{\xi}}
\+{\xibnew}{\numer{\xi}}
\+{\Eover}{\overline{\E}}

\begin{document}

\let\WriteBookmarks\relax
\def\floatpagepagefraction{1}
\def\textpagefraction{.001}
\shorttitle{Strength-Duration Relationship}
\shortauthors{B Bezekci and VN Biktashev}

\title[mode=title]{Strength-Duration Relationship in an Excitable Medium}
\author[1,2]{B. Bezekci}
\ead{burhanbezekci@kilis.edu.tr}
\author[1]{V. N. Biktashev}
\cormark[1]
\fnmark[1]
\ead{v.n.biktashev@exeter.ac.uk}
\cortext[cor1]{Corresponding author}

\address[1]{College of Engineering, Mathematics and Physical Sciences, University of Exeter, Exeter EX4 4QF, UK}
\address[2]{Present address: Faculty of Arts and Sciences, Kilis 7 Aralik University, Kilis 79000, Turkey}


\begin{abstract}
We consider the strength-duration relationship in one-dimensional spatially  extended excitable media. In a previous study~\cite{idris2008analytical} set out to separate initial (or boundary) conditions leading to propagation wave  solutions from those leading to decay solutions, an analytical criterion based on an approximation of the (center-)stable manifold of a certain critical solution was presented. The theoretical prediction in the case of strength-extent curve was later on extended to cover a wider class of excitable systems including multicomponent reaction-diffusion systems, systems with non-self-adjoint linearized operators and in particular, systems with moving critical solutions (critical fronts and critical pulses)~\cite{bezekci2015semianalytical}. In the present work, we consider extension of the theory to the case of strength-duration curve. 
\end{abstract}

\begin{highlights}
\item Approximate semi-analytical description for strength-duration curves for initiation of propagating waves in excitable media can be obtained by linearisation around the critical solution
\item For the one-component systems, where the critical solution is a critical nucleus, the resulting curve coincides with the classical Lapicque-Blair's exponential formula
\item For multicomponent systems, where the critical solution is a critical front or critical pulse, the theory results in a transcendental equation involving functions to be determined numerically
\item Using quadratic instead of linear approximation can improve the accuracy of the solution, but leads to considerable compication of the answer
\end{highlights}


\maketitle


\section{Introduction}

\subsection{Motivation}

The threshold phenomenon \textit{``deals with the minimal, an event, or stimulus just 
strong enough to be perceived or to produce a response''} \cite{archaeology1985advances}  
and the presence of it \textit{``imposes the restriction on the types of mathematical 
model suitable to describe''} biological/chemical systems \cite{fitzhugh1955mathematical}.
Its extreme importance can be highlighted through examples. For instance, propagation of 
excitation in the heart involves action potential and threshold value controls if an applied 
stimulus is sufficient enough to generate an action potential. Understanding the mechanisms 
of initiation of propagating is extremely crucial as successful propagation enables continuous 
electrical and chemical communication between cells and failure may lead to serious medical 
conditions \cite{zipes2009cardiac}. Threshold phenomenon also plays a key role in understanding 
many age related diseases such as Alzheimer and Parkinson. Studies on neuronal changes in brain 
suggest that the threshold hypothesis helps to explain \textit{``some of the associations 
between clinical and pathological findings''} \cite{roth1986association,arendt1987alzheimer}. 

Originally, the term \textit{excitability} has come to be used to refer to the \textit{``property 
of living organisms to respond strongly to the action of a relatively weak external stimulus''} 
\cite{zykov2008excitable}. A well-known example of excitability is the ability of nerve cells 
to generate and propagate electrical activity. By definition, an excitable medium is a  spatially 
distributed system, each element of which possesses the property of excitability and it is usually 
defined as nonlinear reaction-diffusion system, where the reaction term defines how the constituents 
of the system are transformed into each other, and the diffusion part provides propagation of 
information \cite{barkley1991model,guo2010identification,zykov2008excitable}. There are a wide 
variety of areas where the term ``excitable medium'' has been used repeatably for decades in many 
fields including physical, chemical and biological systems and so on \cite{cross1993pattern,mccormick1991interrupted,kaplan1996subthreshold,zykov2008excitable,farkas2002social,seiden2015tongue}. 

\subsection{Problem statement}

We consider the problems of initiation of propagating
waves in one-dimensional reaction-diffusion systems,
\begin{align}                                     
\eqlabel{INTRDS}
  \df{\bu}{\t} = \bD\ddf{\bu}{\x} + \ff(\bu),
\end{align}
where $\bu:\Real\times\Real\to\Real^{\dim}$ is a
$\dim$-component reagents field, $\dim\ge1$, defined for $\x\in\Real$
and $\t\in\Real_+$, vector-function $\ff:\Real^\dim\to\Real^\dim$
describes the reaction rates and $\bD\in\Real^{\dim\times\dim}$ is the
matrix of diffusivity. Equation \eq{INTRDS} is assumed to describe an excitable 
medium as a system
\textit{``composed of elementary segments or cells, each of which
  possesses the following properties: 1. a well-defined rest state,
  2. a threshold for excitation, and 3. a diffusive-type coupling to
  its nearest neighbors. \dots\ Stimuli below the
  threshold are damped out and produce no persistent change in the
  system, \dots\ stimuli
  above the threshold induce the cell to change from its rest state to
  an excited state.''}\cite{fenton2002real} A closely related class
are bistable systems: whereas an excitable system proper returns to
the resting state after spending some time in the excitable state, a
bistable system remains in the excitable state for ever.

A definitive feature of an excitable or bistable medium is existence
of traveling wave solutions of \eq{INTRDS}. These can be described by
transforming the system of partial differential equations (PDEs) to a moving frame of reference,
\begin{align}
\myxi = \x-\c\t, \qquad  \bu\left(\x,\t\right)=\bU\left(\myxi,\t\right). 
\end{align}
We are interested in the solutions that are stationary in this frame
of reference, for a fixed $\c$, \ie\ 
\begin{align}
\bD \partial_{\myxi\myxi}\bU +\c \partial_{\myxi}\bU +\ff(\bU)=0.
  \eqlabel{comovfram}
\end{align}
If the velocity $\c = 0$, then the traveling wave is called the
standing wave. The traveling wave is a front if
$\bU\left(-\infty\right)$ and $\bU\left(\infty\right)$ exist and
different from each other (this is typical for bistable systems), and
it is a pulse if $\bU\left(\infty\right)=\bU\left(-\infty\right)=\Ur$
(this happens in excitable systems).

Travelling wave solutions of \eq{INTRDS} have been a topic of intense
study.  For applications, for instance modelling of biological media and
chemical processes, the question of particular importance is emergence
of such solutions as a result of a perturbation of the resting state,
localized in space and time. For a problem set on the half-infinite
interval $\x\in[0,\infty)$, this can be formalized by initial and
boundary conditions
\begin{align}
  \bu(\x,0)=\Ur+\ampu\,\Xp(\x),
  \quad
  \bD\bu_\x(0,\t)=-\ampi\,\Tp(\t), 
      \eqlabel{semi-cableaim}
\end{align}
where $\Xp$ and $\Tp$ describe the shapes of the initial and boundary
profiles, and $\ampu$ and $\ampi$ are the strengths of those profiles.
The cases of non-homogeneous initial condition and non-homogeneous
boundary condition are usually handled separately. In
electrophysiological terms, these can be described as follows: 
\begin{enumerate}
\item Stimulation by current: $\ampu=0, \ampi\neq 0$. This is the case
  when the current is injected at the boundary point $\x=0$ during
  some time interval. For a fixed boundary profile $\Tp(\t)$, there
  exist a corresponding threshold strength value $\ampic$ such that
  the solution tends to propagating wave (``ignition'') as
  $\t\to\infty$ whenever $\ampi>\ampic$, and the solution tends to
  resting state (``failure'') otherwise. For a one-parametric family
  of profiles, parametrized by the stimulus duration $\tst$, the
  corresponding curve $\ampic(\tst)$ is called a strength-duration
  curve (see \fig{decaysuc}).
\item Stimulation by voltage: $\ampu\neq 0, \ampi=0$. Here the
  perturbation is instantaneous at $\t=0$, but is spread in space.
  For a fixed initial profile $\Xp(\x)$, there exist a corresponding
  threshold strength value $\ampuc$ such that the solution tends to
  propagating wave as $\t\to\infty$ whenever $\ampu>\ampuc$, and to
  resting state otherwise. For a one-parametric family of profiles,
  parametrized by the stimulus extent $\xst$, we shall have the
  corresponding critical curve $\ampuc(\xst)$, called a
  strength-extent curve.
\end{enumerate}
In our previous paper~\cite{bezekci2015semianalytical} we have
analysed some analytical and semi-analytical approaches to description
of the strength-extent curves. In this paper, we focus on the
strength-duration curves. In all specific examples below we shall
consider a rectangular profile of duration $\tst$,
\begin{align}\eqlabel{tsrecprof}
   \Tp(\t)=\Heav(\tst-\t)\best, 
\end{align}
where the fixed vector $\best$ determines which reagents are being
injected, and $\Heav(\cdot)$ is the Heaviside step function. 

\dblfigure{\includegraphics{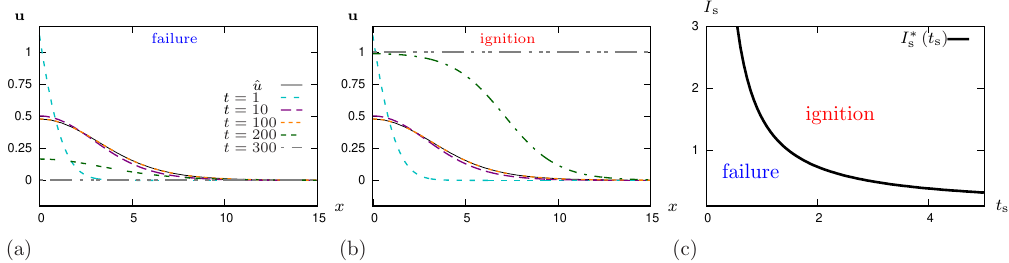}}{ %
  (color online) %
   Response to a below- and above-threshold initial perturbation in ZFK equation. 
   Parameter values: $\zth=0.05$, $\ampu=0$, $\tst=0.6$ for both, sub-threshold
   $\ampi=0.557123722019382$ (a) and super-threshold $\ampi=0.557123722019383$ (b) stimulus strengths.
}{decaysuc}

\subsection{A brief history of the mathematical approaches}

Mathematically, the problem of determining the conditions of
initiation of propagating waves in excitable or bistable media is
spatially-distributed, nonstationary, nonlinear and has generally no
helpful symmetries, so the accurate treatment is feasible only
numerically. However, the practical value of these conditions is so
high that analytical answers, even if very approximate, are in high
demand. Historically, there have been numerous attempts to obtain such
answers, based on various phenomenological and heuristic
approaches. Here we review some of these attempts, in chronological
order.

Phenomenological models describing experimental relationship between
the minimum stimulus amplitude required to excite an axon and the
duration for which the stimulus is applied first appeared well before
the physical mechanisms of biological excitability have been
discovered. The study of the charge-duration relation was first
carried out by Weiss~\cite{weiss1990possibilite} who experimentally
derived the following linear equation
\begin{align} \eqlabel{weissori}
	\QW=\QA+\QB \tst,
\end{align}
where $\QW$ is the threshold charge and $\QA$ and $\QB$ are fitted
parameters.  In his original formula, Weiss did not interpret the
constants $\QA$ and $\QB$ physically and hence they were later on
replaced by rheobasic current $\CHRO$ and chronaxie $\IRH$, so that
$\QA=\CHRO\IRH$ and $\QB=\IRH$~\cite{bostock1983strength} so
\eq{weissori} becomes
\begin{align} \eqlabel{weisscharge}
	\QW=\IRH\leri{\CHRO+\tst},
\end{align}
which is known as the Weiss excitation law for the charge.

An empirical equation developed by Lapicque
\cite{lapicque1907recherches,brunel2007quantitative}) reiterated
Weiss's equation in a different form, for the relation between the
stimulus strength and duration, \ie\ the strength-duration curve.
Lapicque observed that the strength of the current $\ampi$ required to
stimulate an action potential increased as the duration $\tst$ was
decreased. Lapicque proposed the following current law for excitation
\begin{align}\eqlabel{lapiccur}
	\ampi=\IRH\leri{1+\frac{\CHRO}{\tst}}
\end{align}
which is equivalent to \eq{weissori} as $\QW=\ampi\tst$. 

Note that the rheobase current, $\IRH$ may be defined as the minimal current amplitude 
of infinite duration for which threshold can be reached and that the chronaxie time of 
cell, $\CHRO$ refers to the value of the stimulus duration $\tst$ at twice the rheobase current.

An alternative expression for threshold stimulating current was based
on the idea that the nerve cell membrane could be represented by a
parallel resistance $\RONLY$ and capacitance $\CONLY$. The same
Lapicque paper, and also later
Blair~\cite{blair1932intensity,blair1932intensitytwo} discussed a
speculative model relying on an $\RONLY \CONLY $ network to formulate
the strength-duration curve. This resulted in the strength-duration
relationship of the form
\begin{align}\eqlabel{blair1932}
	\ampi=\frac{\IRH}{1-\exp\leri{-\tst/ \CHRO}}.
\end{align}
Lapicque-Blair's exponential strength-duration curve looks similar to that given by the hyperbolic
Weiss-Lapicque law \eq{weisscharge}, \eq{lapiccur}, and they nearly fit the
same data~\cite{blair1932intensitytwo}.

Lapicque-Blair's model combines fairly accurately fit experimental outcomes
with mathematical simplicity.  Thus, a number of researchers began to
focus on it, among which three important ones are
Rashevsky~\cite{rashevsky1933outline}, Monnier~\cite{monnier1934} and
Hill~\cite{hill1936excitation}.
Their results are equivalent up to the interpretation, but Hill's
article is the most cited.  Hill examined the relationship between the
stimulus, the excitability of the tissue, and its accommodation, where
the term ``accommodation''~\cite{nernst1908theorie} is used to
describe the membrane potential response to a sufficiently slow
increase in the stimulating current without exciting. A plausible
mathematical description for a speculative dynamic variable describing
the accommodation resulted in Hill's two time-constant model
\begin{align}\eqlabel{hillists}
	\ampi=\frac{\IRH\leri{1-\HILLKAP/\HILLLAM}}{\exp\leri{-\tst/\HILLLAM}-\exp\leri{-\tst/\HILLKAP}},
\end{align}
where $\HILLKAP$, $\HILLLAM$ are the time constant of excitation and
the time constant of accommodation, respectively. When
$\HILLLAM\to\infty$ and $\HILLKAP=\CHRO$, Hill's equation
\eq{hillists} reduces to Lapicque-Blair's equation \eq{blair1932}.

All above approaches were phenomenological and the parameters in the
strength-duration relationships were to be fitted to experimental data
rather than derived from ``first principles''.

Study of spatial aspects of the initiation problem dates back at least
to 1937, when Rushton \cite{rushton1937initiation} introduced the
concept of the ``liminal length'', to represent the idea that in order
to be successful, the stimulating current should excite a sufficiently
large portion of the excitable cable. He supported this idea by a
mathematical model of the nerve axon, which was of course linear
(passive), and the active character of the membrane and the existence
of the excitation threshold were taken into account in a speculative,
axiomatic manner.

The situation of course changed radically after Hodgkin and Huxley
have succeeded in producing a mathematical model describing the work
of a nerve membrane based on experimentally established physical
mechanisms. There were no more need for speculative modelling, but the
real equations were strongly nonlinear, apparenly making analytical
studies unfeasible and necessitating use of numerical methods.  We
note one such early study done by Noble and Stein~\cite{Noble:1966hb},
who used a simplification of the Hodgkin-Huxley membrane model to
explore numerically the influence of the membrane activation time and
accommodation on the strength-duration curve. They also deduced that
in the spatially-extended context, the strength-duration curve is
highly dependent on the geometry of the stimulus.

The analytical, or at least partly analytical, approaches started
looking less unfeasible after the papers by McKean and
Moll~\cite{mckean1985threshold} and Flores~\cite{flores1989stable}.
They worked with one simple class, scalar bistable models.  Considering
the corresponding PDE on half-line with
homogeneous boundary conditions as a dynamical system in a functional
space, they have identified a critical role of one special solution,
dubbed ``standing wave'' or, later, ``critical nucleus''. This is a
stationary, spatially non-uniform, unstable solution, with exactly one
unstable eigenvalue. Its special role is that its stable manifold
forms the boundary between basins of attraction of ``successful'' and
``unsuccessful'' outcomes of a stimulation attempt. 

An example of an approach seeking to take advantage of this
understanding is the work by Neu \etal~\cite{Neu1997}. They considered
a Galerkin projection of the infinite-dimensional dynamical system
described by the PDE onto a two-dimensional manifold of spatial
profiles, ``resembling'' the shape of a developing excitation wave on
a half-line (specifically, they used Gaussians). This resulted in a
second-order system of ordinary differential equations (ODEs), in
which the critical nucleus is represented by a saddle point, and the
boundary between the basins is the stable separatrix of this saddle
point. This however still left an open problem of describing
analytically this separatrix. One possibility was explored by
Idris~\cite[2.5.2]{idris2008initiation}, who approximated this
separatrix by the stable space of the saddle, which yielded an
analytical expression for the strength-extent curve. This expression,
however, produced a result that was only qualitatively correct.

The next step was using the linear approximation of the stable
manifold right in the functional space. This idea was implemented by
Idris and Biktashev~\cite{idris2008analytical} and rendered
surprisingly good approximations of both strength-extent and
strength-duration curves. 

Naturally, this approach is applicable only to systems where there
exists a critical nucleus. It excludes, for instance, cardiac
excitable models. Hence a question arises, what if anything is the
equivalent of the critical nucleus in such systems. An answer to this
question was proposed in another work by Idris and
Biktashev~\cite{idris2007critical}. It happens that in excitable
systems, the role of the critical nucleus is played not by stationary,
but by moving solutions with one unstable eigenvalue, ``critical
pulses'' and ``critical fronts''. These are unstable ``counterparts''
of the propagating wave solutions, existence of which was realised
long before.

Finally in this brief review, the approach  of
\cite{idris2008analytical} was extended to the moving critical
solutions by Bezekci \etal~\cite{bezekci2015semianalytical}, who also
explored the possibility of using quadratic rather than linear
approximation of the stable manifold. However, that paper only
considered the strength-extent curve, which in the dynamical systems
parlance is easier as it is about an autonomous systems. The question
of strength-duration curve involves studying non-autonomous systems,
and it is the subject of the present contribution.

\subsection{Aims}

The purpose of this paper is to quantify the strength-duration curves,
as an extension of the study~\cite{idris2008analytical}.  We
investigate how the quality of approximation produced by our method
depends on the parameters that define various test systems. Moreover,
we investigate the feasibility of improving the accuracy by using a
quadratic rather than a linear approximation of the critical manifold,
and related problems. Finally, we extend the method to the case where
there are no critical nucleus solutions.  This is observed in
multicomponent reaction-diffusion systems, where it has been
previously demonstrated that instead of a critical nucleus, one has
unstable propagating waves, such as critical
pulses~\cite{flores1991stability} or critical
fronts~\cite{idris2007critical}.

The structure of the paper is as follows. After this introductory
section, Section \secref{anathe} describes the proposed analytical
approach to the problem of the ignition of propagation waves in
one-dimensional bistable or excitable media, from one-component with a
critical nucleus to multi-component systems having moving critical
fronts and pulses, including linear and quadratic approximation of the
critical manifold. The strongly non-linear nature of the equations
makes it unlikely that the ingredients of these approximations can be
found analytically; thus, a short outline of the numerical techniques
used in the ``hybrid'' approach is given in the section
\secref{hybapp}. The applicability of the approach will be illistrated
for five different models from one-component examples
Zeldovich-Frank-Kamenetsky (ZFK) and McKean detailed in section
\secref{onecomp}, to multicomponent examples $\INa$-caricature,
FitzHugh-Nagumo (FHN) and Beeler-Reuter (BR) detailed in section
\secref{mutlicomp}. Finally, section \secref{disc} concludes the paper
with a short review of the results and some possible further research
directions.

\section{Analytical theory}
\seclabel{anathe}

We aim at classification of the solutions of
the system \eq{INTRDS} set on $\x\in[0,\infty)$, $\t\in[0,\infty)$,
supplied with the following initial and boundary conditions,
\begin{align}
  \bu(\x,0)=\Ur, \quad
  \bD\bu_\x(0,\t)=-\ampi\,\Tp(\t), \quad \x,\t>0,  \eqlabel{semi-cable}
\end{align}
in terms of their behaviour as $\t\to\infty$: whether it will approach
the propagating wave solution (``ignition'') or the resting state
(``failure''). 
We find it convenient to formalize the initiation problem as one posed
on the whole real line $\x\in\Real$, 
\begin{align}
  & \df{\bu}{\t} = \bD\ddf{\bu}{\x} + \ff(\bu) + \g(\x,\t), 
  \quad
  (\x,\t)\in\Real\times\Real_+, \nonumber\\
  & \bu(\x,0)=\Ur, 
  \qquad
  \g(\x,\t)\equiv\0\;\textrm{ for }\;\t>\tst,  \eqlabel{RDS-R}
\end{align}
where the boundary condition at $\x=0$ in \eq{semi-cable} is formally
represented by
\begin{align}                           \eqlabel{by-current-R}
  \g(\x,\t) = 2\ampi\,\Tp(\t)\,\dirac(\x) ,
\end{align}
where $\dirac(\cdot)$ is the Dirac delta function. 
 
The principal assumption of our approach is existence
of a \emph{critical solution}, which is defined as a self-similar
solution,
\begin{eqsplit}\eqlabel{buc}
   \bu(\x,\t) = \buc(\x-\c\t) , \\
   \0 = \bD\Ddf{\buc}{\xf} + \c\,\Df{\buc}{\xf} + \ff(\buc), \\
   \buc(\infty)=\Ur,\quad \buc(-\infty)=\Ub,
\end{eqsplit}
which is unstable with one unstable eigenvalue. 
Here $\Ub$ is the asymptotic state behind the critical
  solution: $\Ub=\Ur$ for a critical nucleus or critical pulse, 
  but $\Ub\ne\Ur$  for a critical front.

Similar to the stable wave solution, there is then a whole
one-parametric family of critical solutions,
\begin{align}
  \bu(\x,\t)=\buc(\x-\c\t-\shift), \qquad \shift\in\Real .
\end{align}
Due to this translation invariance, the critical solution always has
one zero eigenvalue. Hence its stable manifold has codimension two,
whereas its center-stable manifold has codimension one and as such it
can partition the phase space, i.e. it can serve as a boundary between
basins of different attractors. Our strategy is to approximate this
center-stable manifold. In the first instance, we consider a linear
approximation, and in selected cases, we also explore the feasibility
of the quadratic approximation.

\subsection{Linear approximation}

Let us rewrite the  system \eq{RDS-R} 
in a frame of reference moving with a constant speed $\c$, so that
$\bu(\x,\t)=\uf(\xf,\tf)$, $\xf=\x-\c\t-\shift$, $\tf=\t$,
\begin{align*}
  & \df{\uf}{\tf} = \bD\ddf{\uf}{\xf} + \c\df{\uf}{\xf} + \ff(\uf)+ \g(\xf+\c\tf+\shift,\tf) ,
  \nonumber\\
  & \uf(\xf,0)=\Ur.
\end{align*}

We linearize this equation on the critical solution, which is
stationary in the moving frame
\begin{align}
  \uf(\xf,\tf) = \buc(\xf) + \up(\xf,\tf) .
\end{align}
The linearization gives
\begin{align}
  & \df{\up}{\tf} = \bD\ddf{\up}{\xf} + \c\df{\up}{\xf} + \J(\xf)\up + \gf(\xf,\tf),
  \nonumber\\
  & \up(\xf,0) = \Ur- \buc(\xf),       \eqlabel{linearized}
\end{align}
where
\begin{align}
  \J(\xf) = \left.\df{\ff}{\bu}\right|_{\bu=\buc(\xf)}          \eqlabel{Jacob}
\end{align}
is the Jacobian matrix of the kinetic term, evaluated at the critical
solution and 
\begin{align}
  \gf(\xf,\tf) = \g(\xf+\c\tf+\shift,\tf)
\end{align}
is the forcing term as measured in the moving frame of reference. 
Equation \eq{linearized} is a linear non-homogeneous
equation, with time-independent linear operator, 
\begin{align}
  \partial_{\tf} \up = \L\up + \gf, 
  \quad
  \L \bydef \bD\ddf{}{\xf} + \c\df{}{\xf} + \J(\xf) .    \eqlabel{linop}
\end{align}
For simplicity of the argument, we assume that the
eigenfunctions of $\L$,
\begin{align}
  \L \RV_{\j}(\xf) = \rw_{\j}\RV_{\j}(\xf)        \eqlabel{RV}
\end{align}
are simple and form a basis in an appropriate functional
space, and the same is true for the adjoint $\Lp$~\footnote{%
  This assumption will, of course, have to be verified in each
  particular case. %
}. Another assumption, which simplifies formulas and is
true for all examples considered, is that all eigenvalues important
for the theory are real. We shall enumerate the eigenpairs in the
decreasing order of $\rw_\j$, so by assumption we always have
$\rw_1>\rw_2=0>\rw_3>\dots$.
Then the general solution of problem \eq{linearized} in that
space can be written as a generalized Fourier series
\begin{align}
  \up(\xf,\tf)=\sum\limits_{\j} \a_{\j}(\tf) \RV_{\j}(\xf) .   \eqlabel{Fourier}
\end{align}
The coefficients $\a_\j$ will then satisfy decoupled ODEs, 
\begin{align}
  \Df{\a_\j}{\tf} &= \rw_\j \a_\j + \gp_\j(\tf), \nonumber \\  
  \a_\j(0) &= \inner{\LV_\j(\xf)}{\up(\xf,0)},  \eqlabel{ajeqistsch3}
\end{align}
where
\begin{align}
  \gp_\j(\tf) = \inner{\LV_\j(\xf)}{\gf(\xf,\tf)},
\end{align}
the scalar product $\inner{\cdot}{\cdot}$ is defined as
\begin{align*}
  \inner{\leftvec}{\rightvec} = \intinf \leftvec\Tr \rightvec \,\d{\xf},
\end{align*}
and $\LV_\j$ are eigenfunctions of the adjoint operator, 
\begin{align}
  \Lp \LV_\j = \lw_\j \LV_\j, 
  \quad
  \Lp = \bD^\top\ddf{}{\xf} - \c\df{}{\xf} + \J\Tr(\xf),
\end{align}
normalized so that
\begin{align}
  \inner{\LV_{\j}}{\RV_{\k}} = \kron{\j}{\k}.
\end{align}
The solution of \eq{ajeqistsch3} is 
\begin{align*}
  \a_{\j}(\tf) = \e^{\rw_{\j}\tf} \left( \a_{\j}(0)+\int\limits_{0}^{\tf} \,
   \gp_{\j}(\tfs)\e^{-\rw_{\j}\tfs} \,\d{\tfs} \right).
\end{align*}
By assumption, $\rw_1>0$, and due to translational symmetry, $\rw_2=0$,
and the rest of the spectrum is assumed within the left half-plane.
Since the stimulation is supposed to be finite in time,
$\gp_{\j}(\tf)\equiv0$ for $\tf>\tst$. Therefore, the condition of
criticality is
\begin{align*}
  \a_1(\tst)=0
\end{align*}
which implies
\begin{align*}
  \a_1(0) + \int\limits_0^{\tst} \gp_1(\tfs)\,e^{-\rw_1\tfs} \,\d{\tfs}
  = 0,
\end{align*}
from which we seek to obtain the critical curve based on this linear
approximation.

\subsubsection*{General Setting}
Using the definitions of $\a_1(0)$ and $\gp_1(\tfs)$, we have, in terms
of the original model,
\begin{align}  
   & \int\limits_0^{\tst} e^{-\rw_1\tfs} 
    \inner{\LV_1(\xf)}{\g(\xf+\c\tfs+\shift,\tfs)}
    \,\d{\tfs} \nonumber \\
& = \inner{\LV_1(\xf)}{\buc(\xf)-\Ur}. \eqlabel{crit-generalists}
\end{align}
The forcing term is defined as
\begin{align}                                     \eqlabel{tsis-bc}
 \g(\x,\t) = 2\ampi\,\best\,\Heav(\tst-\t)\,\dirac(\x),
\end{align}
hence \eq{crit-generalists} gives
\begin{align}
  & 2\ampi \int\limits_0^{\tst} e^{-\rw_1\tfs} 
   \LV_1(-\c\tfs-\shift)^{ \!\top} \best 
    \,\d{\tfs} \nonumber \\
  &= \intinf \LV_1(\xf)^{ \!\top}\left(\buc(\xf)-\Ur\right)\,\d{\xf}. \eqlabel{istslinformch3}
\end{align}
This is a finite equation relating $\tst$ and $\ampi$ so, in
principle, gives the answer. However, it contains the parameter
$\shift$ which still has to be decided upon. This question occurs
already for the strength-extent curve, and we refer the reader
to~\cite{bezekci2015semianalytical} for a detailed discussion. The new
issue here is that for $\t\in[0,\tst]$ we are now dealing with a
time-dependent problem.

\subsubsection{The case of critical nucleus}

This is the case when $\c=0$, \ie\ the critical solution is
stationary, and moreover it is even in $\x$. Then there is a natural
choice of $\shift=0$ prescribed by symmetry.  Hence,
\eq{istslinformch3} gives the classical Lapicque-Blair
formula~\cite{lapicque1907recherches,blair1932intensity}
\begin{align}\eqlabel{LBH}
   \ampi = \dfrac{\Irh}{1 - e^{-\rw_1\tst}},
\end{align}
where the rheobase is
\begin{align}\eqlabel{rheobase}
   \Irh
 = \frac{
     \rw_1 \intoinf \LV_1(\xf)^{ \!\top}\leri{\buc(\xf)-\Ur}\,\d{\xf}
   }{
     \LV_1(0)^{ \!\top} \best 
   }.
\end{align}

\subsubsection{The case of moving critical solution}

In this case there is no $\x\to-\x$ symmetry and the choice of
$\shift$ is no longer trivial. %
We follow the ideas discussed in
  \cite{bezekci2015semianalytical}. According to those, we assume that
  the linear approximation works best if the initial value for $\v$
  is the smallest in some sense. Our heuristic is that it will be
  the smallest in the $\Ltwo$ norm if not only $\a_1=0$, but also
  $\a_2=0$. Moreover, small shifts of the critical solutions are
  equivalent to adding a small amount of $\RV_2$.  Hence an
  appropriate shift $\shift$ can achieve that $\a_2=0$. So we adopt
  $\a_2=0$ as the criterion of selecting $\shift$.
The only modification of this idea for the present case
is that we apply this condition not at $\t=0$, but at the moment from
which the system is autonomous, \ie\ at $\t=\tst$.
Employing both $\a_1\left(\tst\right)=0$ and $\a_2\left(\tst\right)=0$ results in following 
system of equations 
\begin{align}
  \begin{cases}
2\ampi\int\limits_0^{\tst}\e ^{-\rw_1 \tfs}\LV_1\left(-\c\tfs-\shift\right)\Tr \best\, \d {\tfs} & = \Num_1, \\
2\ampi\int\limits_0^{\tst}\e ^{-\rw_2 \tfs}\LV_2\left(-\c\tfs-\shift\right)\Tr \best\,  \d {\tfs}  & = \Num_2,
  \end{cases}                                     \eqlabel{crit-sys}
\end{align}
where the right hand sides $\Num_1$ and $\Num_2$ are constants, defined entirely
by the properties of the model,
\begin{align}
  \Num_l=\inner{\LV_l(\xf)}{\buc(\xf)-\Ur},  \qquad  l=1,2.        \eqlabel{Numnboth}
\end{align}
System \eq{crit-sys} is  a nonlinear system of two equations for two unknown parameters,
$\ampi$ and $\shift$. By eliminating the parameter $\ampi$, we find the compatibility 
condition as follows:
\begin{align}
\int\limits_0^{\tst} &  \left[ 
\Num_2 \e^{-\rw_1\tfs} \LV_1\left(-\c\tfs-\shift\right)\Tr \best\, \right.\nonumber \\
& \left.
 - \Num_1\e^{-\rw_2\tfs}\LV_2\left(-\c\tfs-\shift\right)\Tr \best\, \right]\d {\tfs}=0.
\end{align}
This can be further simplified by using the following change of variable,
\begin{align*}
\tfs=\frac{-\left(\val+\shift\right)}{\c}
\end{align*}
that leads to
\begin{align}\nonumber
& \seq(\shift) \bydef
\frac{\Num_1\e^{\rw_2 \shift/\c}}{\c}\int\limits_{-\shift}^{-\c \tst-\shift} 
\e^{\rw_2\val/\c}\LV_2\left(\val\right)\Tr \best\, \d  \val 
 \nonumber \\
& 
-\frac{\Num_2\e^{\rw_1 \shift/\c}}{\c}\int\limits_{-\shift}^{-\c \tst-\shift} \e^{\rw_1 \val/\c}
\LV_1\left(\val\right)\Tr \best\,\d \val=0.
\eqlabel{zeroofs}
\end{align}

The finite equation \eq{zeroofs} imposes a connection between the
shift $\shift$ and the stimulus duration $\tst$. This connection 
defines $\shift$ as an implicit function of $\tst$.
After finding the value of $\shift$,  one only needs to employ this value in one of the 
compatibility conditions in \eq{crit-sys}  in order to find the amplitude $\ampi$ since 
both produce the same result. This completes the construction of the linear approximation 
of the strength-duration curve $\ampi(\tst)$. 

\subsection{Quadratic approximation of the stable manifold}

In this subsection, we restrict consideration to the case of a critical
nucleus.

For simplicity, rather than using the matrix notation as in the linear
approximation, we shall now proceed with an explicit notation
for the components of the reaction-diffusion systems.
We use Greek letters for superscripts to enumerate
them, and adopt Einstein's summation convention
for those indices. In this way we start from the generic
reaction-diffusion system
\begin{align*}
  \df{\u^\alp}{\t} = \D^{\alp\bet} \ddf{\u^\bet}{\x} + \f^\alp(\u^\bet)
  + 2\ampi\,\est^\alp\,\Heav(\tst-\t)\delta(\x)
\end{align*}
then consider the deviation $\v^\alp$
of the solution $\u^\alp$  from
the critical nucleus $\uc^\alp$,
\begin{align*}
  \u^\alp(\x,\t) = \uc^\alp(\x) + \v^\alp(\x,\t),
\end{align*}
the equation defining the critical nucleus,
\begin{align*}
  \D^{\alp\bet} \ddf{\uc^\bet}{\x} + \f^\alp(\buc) = 0
\end{align*}
and the Taylor expansion of the equation for the deviation,
\begin{align*} 
  \dot\v^\alp
 = & \D^{\alp\bet} \v^\bet_{\x\x}
  + \f^\alp_{,\bet}(\buc) \v^\bet  
  + \f^\alp_{,\bet\gam}(\buc) \v^\bet\v^\gam  \\
  &
  + 2\ampi\,\est^\alp\,\Heav(\tst-\t)\delta(\x)+ \dots,
\end{align*}
where overdots denote differentiation with respect to
time, subscripts $\leri{\cdot}_{\x}$ denote differentiation with respect
to space and Greek subscripts after a comma designate a
partial differentiation by the corresponding reactive components.
The right and left eigenfunctions are defined
respectively by
\begin{align*}
  \D^{\alp\bet} \@_{\x\x}\rv^\bet_\j(\x) + \f^\alp_{,\bet}(\x) \rv^\bet_\j(\x) = \rw_\j\rv^\alp_\j(\x)
\end{align*}
and 
\begin{align*}
  \D^{\bet\alp} \@_{\x\x}\lv^\bet_\j(\x) + \f^\bet_{,\alp}(\x) \lv^\bet_\j(\x) = \rw_\j\lv^\alp_\j(\x)
\end{align*}
where $\j\in\{1, 2, 3,\cdots\}$, and the biorthogonality condition is
\begin{align*}
  \inner{\lv_\j}{\rv_\k}
  \triangleq
  \intinf \overline{\lv^\alp_\j(\x)}\,\rv^\alp_\k(\x)\,\d{\x}
  = \kron{\j}{\k}.
\end{align*}
We consider only even solutions, so in subsequent sums
only those $\j$ that correspond to even eigenfunctions are
assumed. We seek solutions in the form of generalized
Fourier series in the right eigenfunctions,
\begin{align*}
  \v^\alp(\x,\t) = \sum_\j \a_\j(\t) \rv^\alp_\j(\x)
\end{align*}
where the Fourier coefficients are defined by
\begin{align*}
  \a_\j(\t) = \inner{\lv_\j(\x)}{\v(\x,\t)}
  \triangleq \intinf \overline{\lv^\alp_\j(\x)}\,\v^\alp(\x,\t)\,\d{\x}.
\end{align*}
Time-differentiation of this gives
\begin{align}
 \dot\a_\j(\t) =& \inner{\lv^\alp_\j(\x)}{\dot\v^\alp(\x,\t)}\nonumber \\
 =&\rw_\j\a_\j + \sum_{\m,\n} \Q{\j}{\m,\n}\a_\m\a_\n
 + 2\ampi\,\Est_\j\,\Heav(\tst-\t)
 \eqlabel{diffofaj}
\end{align}
where
\begin{multline}                                     \eqlabel{Q}
  \Q{\j}{\m,\n} = \Q{\j}{\n,\m} \\
  \bydef  \frac{1}{2} \intinf \overline{\lv^\alp_\j(\x)} \, \f^\alp_{,\bet\gam}(\buc(\x)) 
  \rv^\bet_\m(\x)\rv^\gam_\n(\x)\,\d{\x},
\end{multline}
and 
\begin{align} \eqlabel{ejsanatheo}
  \Est_\j=\lv^\alp_\j(0)\best^\alp.
\end{align}
We assume that eigenvalues are real and ordered from
larger to smaller, $\rw_1>0$, $\rw_2=0$ is of course the eigenvalue corresponding to 
the translational symmetry and an odd eigenfunction $\rv_{2}=\buc'$ , and $\rw_\j<0$ 
for $\j\geq 3$. Our task is to determine the conditions on the initial values of the Fourier
coefficients
\begin{align}\eqlabel{Ajsanatheo}
  \ai_\j \triangleq \a_\j(0) = 
  \intinf \overline{\lv^\alp_\j(\x)} \v^\alp(\x,0) \,\d{\x}
\end{align}
that would ensure that
\begin{align*}
  \a_1(\infty)=0, 
\end{align*}
which means that the trajectory approaches the critical nucleus, so the initial condition is 
precisely at the threshold.

Let us rewrite the system \eq{diffofaj} as an equivalent system of integral equations,
\begin{align*}
  \a_\j(\t) =& \e^{\rw_\j\t}\left[
\ai_\j+ \frac{2\ampi\,\Est_\j}{\rw_\j}\left(1 -\e^{-\rw_\j\Min{\t}{\tst}} \right)\right. \\ & \left.
 +\int\limits_0^\t \e^{-\rw_\j\ts}
    \sum\limits_{\m,\n}
    \Q{\j}{\m,\n} \a_\m(\ts)  \a_\n(\ts)
    \,\d\ts \right]
\end{align*}
where we use the notation $\Min{a}{b}\bydef\min(a,b)$. 
Successive approximations to the solution can be 
obtained by direct iterations of this system,
\begin{align*}
  \a^{(\iter+1)}_\j(\t)
  =& \e^{\rw_\j\t} \left[
    \ai_\j 
    +
    2\ampi\,\Est_\j
    \frac1{\rw_\j}\left(
      1 - \e^{-\rw_\j\Min{\t}{\tst}}
    \right)
    \right. \\ & \left.
    +\int\limits_0^\t \e^{-\rw_\j\ts}
    \sum\limits_{\m,\n}
    \Q{\j}{\m,\n} \a^{(\iter)}_\m(\ts)  \a^{(\iter)}_\n(\ts)
    \,\d\ts
 \right].
\end{align*}
Taking $\a^{(0)}_\j = 0$ for all $\j$, we have
\begin{align*}
  \a^{(1)}_\j(\t)
  = \e^{\rw_\j\t} \left[
    \ai_\j 
    +
    \ampi\,
    \frac{2\Est_\j}{\rw_\j}\left(
      1 - \e^{-\rw_\j\Min{\t}{\tst}}
    \right)
  \right].
\end{align*}
The requirement $\a^{(1)}_1(\infty)=0$ recovers the linear
approximation. 
The next iteration produces  
%
\begin{align*}
  \a^{(2)}_\j(\t)
& = \e^{\rw_\j\t} \left\{
    \ai_\j 
    + 
    \ampi\,
    \frac{2\Est_\j}{\rw_\j}\left(
      1 - \e^{-\rw_\j\Min{\t}{\tst}}
    \right)
\rile
    +
    \int\limits_0^\t \,\d\ts \, \e^{-\rw_\j\ts}
    \sum\limits_{\m,\n} \Q{\j}{\m,\n} 
  \right. \\ & \left.
    \times \e^{\rw_\m\ts} \left[
      \ai_\m+\ampi\,\frac{2\Est_\m}{\rw_\m}\left(1 - \e^{-\rw_\m\Min{\ts}{\tst}}\right)
  \right]\right. \\ & \left.
   \times \e^{\rw_\n\ts} \left[
      \ai_\n+\ampi\,\frac{2\Est_\n}{\rw_\n}\left(1 - \e^{-\rw_\n\Min{\ts}{\tst}}\right)
  \right]
  \right\}.
\end{align*}
%
Note that $\e^{\left(\rw_\m+\rw_\n-\rw_1\right)\t}\to 0$ as $\t\to\infty$ 
because $\rw_{\m,\n}\le\rw_3<0$ for $\m,\n\ge3$, so upon exchanging
the order of intergration and summation, we have converging
improper integrals. The requirement 
$\a^{(2)}_1(\infty)=0$ leads to a quadratic equation for $\ampi$,
\begin{align} \eqlabel{istsquadch3}
  \AIS \ampi^2+\BIS\ampi+\CIS=0,
\end{align}
where 
\begin{align*}
 \AIS & =  4\sum\limits_{\m,\n}\Q{1}{\m,\n} \biggl[ \frac{\Est_\m\Est_\n }{\rw_\m\rw_\n} 
 \biggl\{ \frac{1-\e^{-\rw_1 \tst}}{\rw_1}-\frac{\e^{\left(\rw_\n-\rw_1\right)\tst}-1}{\rw_\n-\rw_1}
 \\ & 
 -\frac{\e^{-\rw_1\tst}-\e^{\left(\rw_\m-\rw_1\right)\tst}-\e^{\left(\rw_\n-\rw_1\right)\tst}+1}{\rw_\m+\rw_\n-\rw_1}\\ &
 -\frac{\e^{\left(\rw_\m-\rw_1\right)\tst}-1}{\rw_\m-\rw_1}
 \biggl\}
\biggl], \\
\BIS &= -\frac{2\Est_1\left(\e^{-\rw_1 \tst}-1\right)}{\rw_1}+ 2  \sum\limits_{\m,\n}\Q{1}{\m,\n} \biggl[ \\ &
\frac{\ai_\m \Est_\n}{\rw_\n} \biggl\{\frac{\e^{\left(\rw_\m-\rw_1\right)\tst}-1}{\rw_\m+\rw_\n-\rw_1}
-\frac{\e^{\left(\rw_\m-\rw_1\right)\tst}-1}{\rw_\m-\rw_1}
\biggl\}\\
&+\frac{\ai_\n\Est_\m}{\rw_\m}\biggl\{
\frac{\e^{\left(\rw_\n-\rw_1\right)\tst}-1}{\rw_\m+\rw_\n-\rw_1}-\frac{\e^{\left(\rw_\n-\rw_1\right)\tst}-1}{\rw_\n-\rw_1}
\biggl\}
\biggl],
\\
\CIS&=\ai_1-\sum\limits_{\m,\n}\Q{1}{\m,\n} \frac{\ai_\m\ai_\n}{\rw_\m+\rw_\n-\rw_1}.
\end{align*}

\subsection{A priori bound for critical nucleus case}

We conclude this section with an \apriori\ bound for the
strength-duration curve, obtained by Mornev~\cite{Mornev_1981}.  It is
applicable to scalar equations, $\dim=1$, $\bu=\Mx{\u}$,
$\ff=\Mx{\f}$, such that $\f(\u_\ii)=0$, $\ii=1,2,3$,
$\ur=\u_1<\u_2<\u_3=\ub$, $\f'(\u_{1,3})<0$, $\f'(\u_2)>0$. Then
  \begin{align}
    \ampic(\tst) \searrow \ampicl, \qquad \tst\to\infty ,
  \end{align}
  where
\begin{align} \eqlabel{ampicl}
  \ampicl=\max|\uc'(\x)|=|\uc'(\xa)|=\left(-2\int_{\u_1}^{\u_2}\f(\u)\,\d\u\right)^{1/2},
\end{align}
and $\xa$ is the coordinate of the inflexion point of the graph of
$\uc(\x)$, i.e. $\uc(\xa)=\u_2$.

\section{Hybrid approach}
\seclabel{hybapp}

With a few exceptions, the ingredients for the expressions used in the
linear and quadratic approximations of the strength-duration curve,
starting from the critical solution itself, are not available
analytically and have to be found numerically. In this section we
describe numerical methods we used to find these ingredients. We
divided this section into two subsections.  for models with
self-adjoint and non-self-adjoint linearization operator.

\subsection{Ingredients of the one-component systems}
\seclabel{ingonecomphyb}

This corresponds to the case of the critical nucleus and for the linear approximation, 
we need to have the knowledge of $\uc$, $\rw_{1}$, $\rv_{1}$ while for the quadratic 
approximation, ideally  the whole spectrum of  $\rw_{\l}$, $\rv_{\l}$, $\l=1,3,5,\cdots$ is needed.
Here we shortly describe the methods to obtain the mentioned ingredients in algorithmic 
forms. For a more detailed explanation, see~\cite{bezekci2015semianalytical}. 

In order to find the critical nucleus, we take advantage of the fact
that its center-stable manifold has codimension one, and divides the
phase space into two open sets, one leading to successful initiation
and the other to
decay~\cite{%
  flores1989stable,%
  flores1991stability,%
  mckean1970nagumo,%
  moll1990calculation,%
  Neu1997,%
  idris2008analytical%
}.
This means that the critical trajectories, corresponding to the
stimulus strength exactly equal to the threshold, tend towards the
critical nucleus as $\t\to \infty$, whereas the stimulus strength
slightly above or below the threshold produces the solution that gets
close to the critical nucleus and stays in its vicinity for a long
time, before deviating from it to propagate or to collapse. This can
be used to calculate an approximation of the critical nucleus, see
\alg{numcritnuc}. The calculations are done for
\eqtwo{INTRDS,semi-cable} for $\x\in[0,\Length]$, where $\Length$ is
chosen large enough for the results to be not significantly different
from those for $x\in[0,\infty)$.

\begin{algorithm}
\begin{algorithmic}
 \Algwidget{Input:}{Pre-found value of $\ampic$ for a chosen $\tst$}
 \Algwidget{Output:}{Critical nucleus $\uc$}
 \Algstep{Find $\u(\x,\t)$ by solving initial value problem \eqtwo{INTRDS,semi-cable}.}
 \Algstep{$\Speed(\t) \gets ||\dot\u||_{\Ltwo}^2=\int_0^{\Length} \u_{\t}^2(\x,\t)\,\d{\x}$}
 \Algstep{$\ttest \gets \argmin(\Speed(\t))$.}
 \Algstep{$\uc(\x) \gets \u(\x,\ttest)$}
\end{algorithmic}
\caption{Numerical critical nucleus by ``shooting''.}
 \alglabel{numcritnuc}
\end{algorithm}

We calculate the eigenpairs of the linearized operator $\L$ defined by
\eq{linop} (note that in the present case $\c=0$) using a variant
of the power iteration method. We use a random number generator to
assign linear independent initial guesses for $\RV_1$, $\RV_2$, \dots,
choose a time domain $\t\in\leri{0,\T}$, and then follow
\alg{numeigonecomp} until the desired convergence criterion is
fulfilled. 
The appropriate values of $\T$ of course depend on the
  spectrum of $\L$ which may not be known \apriori; our choice was
  entirely empirical.
Note that the algorithm is described without prejudice
to the choice of the norm used for the normalization; if
$\norm{\RV}=\norm{\RV}_{\Ltwo}=\inner{\RV}{\RV}^{1/2}$ then obvious simplifications
  are possible. Also, we use the convergence criterion based on the change in each
eigenvalue; it can also be done in terms of the change, say in
$\Ltwo$-norm, in each eigenfunction.

\begin{algorithm}
\begin{algorithmic}
  \Algwidget{Input:}{A linearly independent set $\left(\RV_1^0,\RV_2^0,\cdots,\RV_\n^0\right)$}
  \Algwidget{Output:}{$\left(\rw_1,\RV_1\right), \left(\rw_2,\RV_2\right), \cdots, \left(\rw_\n,\RV_\n\right)$}
  \Algstep{$\rw_1^0, \rw_2^0, \cdots, \rw_{\n}^0 \gets 0$}
  \Algstep{$\i\gets0$}
  \REPEAT
    \Algstep{$\i\gets\i+1$}
    \FOR{$\k=1,2,\ldots,\n$}
      \Algstep{Solve IVP: $\RV_{\k}^{\i}\gets\exp\leri{\L\T} \RV_{\k}^{\i-1}$}
      \Algstep{Orthogonalize:
        $\RV_{\k}^{\i}\gets\RV_{\k}^{\i}-\sum\limits_{\m=1}^{\k-1}\frac{
          \inner{\RV_{\k}^{\i}}{\RV_{\m}^{\i}}
        }{
          \inner{\RV_{\m}^{\i}}{\RV_{\m}^{\i}}
        }\RV_{\m}^{\i}
        $}
      \Algstep{Eigenvalue: $\rw_{\k}^\i\gets \dfrac{1}{\T}\ln\leri{\inner {\RV_{\k}^\i}{\RV_{\k}^{\i-1}}}$}
      \Algstep{Normalize: $\RV_{\k}^\i \gets \dfrac{\RV_{\k}^\i}{\left\Vert\RV_{\k}^\i \right\Vert}$}
    \ENDFOR
    \UNTIL{$|\rw_{\k}^\i-\rw_{\k}^{\i-1}|\le \tol$ $\forall\k$}
\end{algorithmic}
\caption{Numerical computation of $\n$ principal eigenpairs of self-adjoint
  operator $\L$ by ``marching''.}
 \alglabel{numeigonecomp}
\end{algorithm}

\subsection{Ingredients of the multi-component systems}


The non-stationary critical solutions, observed in multi-component
systems, can be found using an appropriate modification of~\alg{numcritnuc},
exploiting computations in a co-moving frame of reference, as
described in~\cite{bezekci2015semianalytical}. However, more accurate
results can be obtained by continuation of the boundary-value problem
\eq{buc}, an autonomous system for vector-function $\buc(\xf)$ and
scalar $\c$. 
For the critical pulses $\Ub=\Ur$, and our strategy is
  to use periodic solutions with very long periods as approximation to pulses.
We aim to calculate conduction velocity restitution
curve~\cite{simitev2011asymptotics}, that is, a one-parametric family
of solutions of the following periodic boundary-value problem:
\begin{eqsplit}                                   \eqlabel{Uper}
  & \0 = \bD\Ddf{\Uper}{\xf} + \cP\,\Df{\Uper}{\xf} + \ff(\Uper), \\
  & \Uper\left(\xf+\Per\right)\equiv \Uper\left(\xf\right),
\end{eqsplit}
where $\Per$ is the spatial period of the waves. When the problem is well posed, 
\eq{Uper} defines a curve in the $\left(\Per,\cP,\Uper(\xf)\right)$ space. In the 
limit $\Per\to\infty$, this curve splits into two branches, the upper
branch with a stable 
propagating pulse solution, $(\cwave,\Uwave(\xf))$ and the lower branch with an unstable 
critical pulse solution, $(\c,\buc(\xf))$, which is of interest to
us. We performed the continuation using AUTO~\cite{doedel1986auto}.
To obtain the periodic solutions, we consider an extension of
~\eq{Uper} by an extra parameter corresponding to ``stimulation
current'' added to the transmembrane voltage equation. Starting from an initial guess 
of $\cwave$, the continuation is done in accordance with~\alg{critpulse}.

\begin{algorithm}
\begin{algorithmic}
 \Algwidget{Input:}{An initial guess of $\cwave$.} 
 \Algwidget{Output:}{$\c$, $\buc(\xf)$.} 
 \begin{align*}                                   
   & \bD\Ddf{\Uper}{\xf} + \cP\,\Df{\Uper}{\xf} + \ff(\Uper)+\Iext\best=0, \\
   & \Uper\left(\xf+\Per\right)\equiv \Uper\left(\xf\right).
  \end{align*}
  \Algstep{Equilibrium, resting state: $\Iext\gets0$, $\c\gets\cwave$, $\buc(\xf)\gets \Ur$.}
  \Algstep{Continue the equilibrium by increasing $\Iext$, until a Hopf bifurcation is reached.}
  \Algstep{Continue the periodic orbit from the Hopf bifurcation in the $\left(\cP, \Per\right)$ plane, down 
      by $\cP$ until the fold is reached.}
  \Algstep{Continue the periodic orbit in the
      $\left(\Iext,\cP\right)$ plane, down by $\Iext$ until $\Iext=0$ is
      reached.}
  \Algstep{Continue the periodic orbit in the
      $\left(\cP,\Per\right)$ plane both ways.}
  \Algstep{For the branch with smaller $\c$, select a sufficiently
      large $\Per$, and 
      take $\c\gets\cP$, $\buc(\xf)\gets\Uper(\xf)$, in a suitably
      chosen interval of $\xf$. }
\end{algorithmic}
\caption{Numerical critical pulse by AUTO.}
 \alglabel{critpulse}
\end{algorithm}

As a final step, we calculate the left and right eigenfunctions by
means of the Gram-Schmidt orthogonalization process, modified with
account of the fact that $\L$ is now not self-adjoint, 
\alg{eigfunadj}.
%
%

\begin{algorithm}
\begin{algorithmic}
  \Algwidget{Input:}{A linearly independent set 
    $\left(\RV_1^0,\LV_1^0\right), \left(\RV_2^0, \LV_2^0\right),\cdots, \left(\RV_n^0, \LV_n^0\right)$}
  \Algwidget{Output:}{$\leri{\rw_1,\RV_1,\LV_1}$,
    $\leri{\rw_2,\RV_2,\LV_2}$, \dots, $\leri{\rw_n,\RV_n,\LV_n}$}
  \Algstep{$\rw_1^0,\rw_2^0,\cdots,\rw_n^0 \gets0$}
  \Algstep{$\i\gets0$}
  \REPEAT
    \Algstep{$\i\gets\i+1$}
    \FOR{$k=1,2,\ldots,n$}
      \Algstep{Solve IVP: $\RV_{\k}^{\i}\gets\exp\leri{\L\T} \RV_{\k}^{i-1}$}
      \Algstep{Solve IVP: $\LV_{\k}^{\i}\gets\exp\leri{\Lp\T} \LV_{\k}^{i-1}$}
      \Algstep{Biorthogonality: 
        $\RV_{\k}^{\i}\gets\RV_{\k}^{\i}-\sum\limits_{\m=1}^{\k-1}
        \frac{\inner {\LV_{\m}^{\i}}{\RV_{\k}^{\i}}}{\inner{\LV_{\m}^{\i}}{\RV_{\m}^{\i}}}
        \RV_{\m}^{\i}$
      }
      \Algstep{Biorthogonality:
        $\LV_{\k}^{\i}\gets\LV_{\k}^{\i}-\sum\limits_{\m=1}^{\k-1}
        \frac{\inner{\LV_{\k}^{\i}}{\RV_{\m}^{\i}}}{\inner{\LV_{\m}^{\i}}{\RV_{\m}^{\i}}} \LV_{\m}^{\i}$}
      \Algstep{Eigenvalue: $\rw_{\k}^{\i}\gets\frac{1}{\T}\ln\leri{\inner{\RV_{\k}^{\i}}{\RV_{\k}^{\i-1}}}$}
      \Algstep{Normalization: $ \RV_{\k}^{\i} \gets \frac{\RV_{\k}^{\i}}{\norm{\RV_{\k}^{\i}}}$}
      \Algstep{Normalization: $\LV_{\k}^{\i} \gets \frac{\LV_{\k}^{\i}}{\norm{\LV_{\k}^{\i}}}$}
    \ENDFOR
  \UNTIL{$|\rw_k^{\i}-\rw_k^{\i-1}|\le \tol$ $\forall \k$}
\end{algorithmic}
\caption{Numerical computation of $\n$ principal eigenpairs of
  non-self-adjoint operator $\L$ by ``marching''.}
 \alglabel{eigfunadj}
\end{algorithm}

\section{One component systems}
\seclabel{onecomp}

\subsection{Zeldovich-Frank-Kamenetsky equation}

Our first application example is the one-component reaction-diffusion equation,
first introduced by Zeldovich and Frank-Kamenetsky
(ZFK)~\cite{zel1938towards} to describe propagation of flames; it is also
known as ``Nagumo equation''~\cite{mckean1970nagumo} and ``Schl\"ogl
model''~\cite{schlogl1972chemical}:
\begin{align}
  & 
  \dim=1, \quad
  \bD=\Mx{1}, \quad 
  \bu=\Mx{\u},
  \nonumber\\&
  \ff(\bu)=\Mx{\f(\u)}, \quad
  \f(\u)=\u(\u-\zth)(1-\u) ,                      \eqlabel{ZFK}
\end{align}
where we assume that $\zth\in(0,1/2)$. 
The critical nucleus solution $\buc=\Mx{\uc}$ for this equation can be found
analytically~\cite{flores1989stable,idris2008analytical}
~\footnote{
  Actually, expressions given in both of these works contain typos.
}
\begin{align}                                     \eqlabel{ZFK-nucleus}
 \uc(\x) = \frac{
    3\zth\sqrt{2}
  }{
    (1+\zth)\sqrt{2}+\cosh(\x\sqrt{\zth}) \sqrt{2-5\zth+2\zth^2}
  } .
\end{align}
The other two components required for the definition of critical curves in the linear 
approximation are  $\rw_1$ and $\LV_1=\RV_1=\Mx{\rv_1}$ which are solutions of 
\begin{align}
  &
  \Ddf{\rv_1}{\x} + \left(-3\uc^2+2(\zth+1)\uc-\zth\right)\rv_1=\lw_1\rv_1, 
  \nonumber\\&
  \lw_1>0, 
  \quad
  \rv_1(\pm\infty)=0.                             \eqlabel{ZFK-ignition}
\end{align}
We have been unable to find solution of this eigenvalue problem
analytically. We note, however, that $\uc$ given by \eq{ZFK-nucleus}
is unimodal, therefore $\uc'$, which is the eigenfunction of $\L$
corresponding to $\rw=0$, has one root, hence by Sturm's oscillation
theorem, $\uc'=\rv_2$ and $\rw_2=0$, and there is indeed exactly one
simple eigenvalue $\rw_1>0$ and the corresponding
$\rv_1$ solving \eq{ZFK-ignition} has no roots.

\sglfigure{\includegraphics{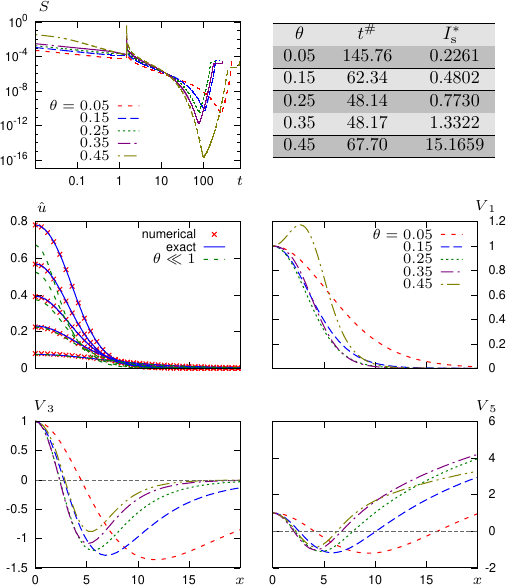}}{ %
  (color online) %
  ZFK ingredients for different threshold parameters. Top row: The
  illustration of the typical function $\Speed(\t)$ along with at
  near-threshold boundary conditions.  Middle row: Comparison between
  analytical and numerical critical nuclei and first
  eigenfunction. Bottom row: Second and third eigenfunctions obtained
  using Gram-Schmidt orthogonalization method. Parameters:
  $\zth=0.05,0.15,0.25,0 35,0.45$, $\dx=0.03$, $\dt=4\dx^2/9$,
  $\Length=30$, $\tst=1.5$, $\tol=10^{-6}$. %
}{zfkbetaieg}

\sglfigure{\includegraphics{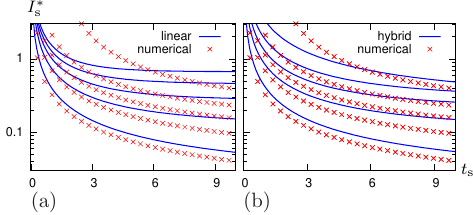}}{%
  (color online) %
  Strength-duration curves for the ZFK model, for $\zth=0.05$, $0.15$,
  $0.25$, $0.35$, $0.45$ (bottom to top), comparison of direct
  numerical simulations (lines with symbols) with theoretical
  predictions (dashed lines), %
  (a) for the explicit analytical answers in
  the $\zth\ll1$ limit, linear approximation; %
  (b) for the hybrid method, using the
  numerically found ignition eigenpairs, linear approximation.
   Parameters: $\dx=0.03$, $\dt=4\dx^2/9$, $\Length=100$, $\tol=10^{-5}$.
}{zfk-tsis}

\subsubsection{The small-threshold limit and the ``fully analytical''  result}

In this subsection we extend
the results of \cite{idris2008analytical} in the parameter space and
correct some typos found in the paper.
For $\zth\ll1$, the critical nucleus \eq{ZFK-nucleus} is $\O{\zth}$
uniformly in $\x$, and is approximately
\begin{align}
  \uc(\x) \approx \frac{3\zth}{1+\cosh(x\sqrt\zth)}
  =\frac32\,\zth\,\sech^2(x\sqrt\zth/2) .         \eqlabel{ZFK-nucleus-small}
\end{align}
In the same limit, the nonlinearity can be approximated by
$\f(\u)\approx\u(\u-\zth)$. With these approximations, problem
\eq{ZFK-ignition} has the solution
\begin{align}                                   \eqlabel{ZFK-ignition-small}
  \rw_1 \approx \frac54\zth, 
  \qquad
  \rv_1 \approx \sech^3(\x\sqrt\zth/2) ,
\end{align}
and \eq{LBH} then gives an explicit expression for
the strength-duration curve in the form
\begin{align}\eqlabel{LBHzfk}
   \ampi = \dfrac{\Irh}{1 - \e^{-\tst/ \CHRO }}
\end{align}
with the following rheobase and chronaxie form \cite{idris2008analytical}
\begin{align}\eqlabel{irhtauzfk}
    \Irh= \frac{45}{64}\pi \zth^{3/2}, \quad 
    \CHRO=\leri{\rw_1}^{-1}=\frac{4}{5\zth}.
\end{align}
\Fig{zfk-tsis}(a) illustrates this approximate strength-duration curve, compared to the 
direct numerical simulations. For chosen parameter values, the comparison is 
significantly better with smaller values of $\zth$, as expected.

Remark that the performance of the resulting approximation based on the analytical 
expression for the strength-duration curve \eq{LBHzfk} and \eq{irhtauzfk} can be 
further improved by obtaining the essential ingredients numerically. This is done 
considering \eq{LBHzfk}, in  which the rheobase is, instead, defined according to \eq{rheobase}, as
\begin{align}\eqlabel{rheobasech5}
   \Irh
 = \frac{
     \rw_1 \intoinf \rv_1(\x)\uc(\x)\,\d{\x}
   }{
     \rv_1(0) 
   }.
\end{align}

The plot of the hybrid numeric-asymptotic prediction is compared with  the direct numerical
simulations as shown in \fig{zfk-tsis}(b).
As depicted in the figure, reasonable agreement between the two data sets is observed when 
the threshold parameter is small.

It should be noted that the strength-duration curve approximation remains above 
the \apriori\  lower bound \eq{ampicl} 
\begin{align*}
   \ampicl=\leri{-2\int\limits_{0}^{\zth}\u(\u-\zth)(1-\u)\, \d \u }^{1/2}= 
   \frac{\zth^{3/2} \sqrt{\leri{2-\zth}}}{\sqrt{6}}, 
\end{align*}
for all $\tst$. 

\subsubsection{Hybrid approach}

Numerical computation of the essential ingredients needed for linear approximation
of the critical curves is carried out using \algs{numcritnuc} and 
\alref{numeigonecomp} described in \secref{ingonecomphyb}. \Fig{zfkbetaieg}
illustrates the processes of numerical computation of the critical 
nucleus and the first eigenmodes in the ZFK equation, for 
the threshold parameter varying from $0.05$ to $0.45$ with the increment $0.1$. 
The stimulation is done by fixing the duration time at the value $\tst=1.5$.
To obtain the minimum of $\Speed(\t)$ and $\ttest=\argmin(\Speed(\t))$, 
the bisection loop is terminated as soon as the absolute difference between 
upper and lower estimate for the threshold is sufficiently small, 
\ie~$|\overline\ampi-\underline\ampi|<10^{-5}$. For each case, the 
solution $\u(\x,\ttest)$ of the nonlinear problem $\u_\t = \u_{\x\x} + \f(\u)$
provides an estimate of the critical nucleus.

\subsubsection{Quadratic theory}

To estimate a few principal eigenmodes of the ZFK equation, we have
considered 
a finite interval 
$\x\in[0,\Length)$ as an approximation of  $\x\in[0,\infty)$. 
We find only a few approximate eigenvalues in 
the discrete spectrum, while the remaining ones are in the continuous spectrum. 
The eigenfunctions corresponding to the discrete eigenvalues are well 
localized whereas those corresponding to the continuous eigenvalues are evidently 
non-localized, and thus they cannot be taken into account in quadratic approximation. 
We observe that at increasing values of $\Length$, the eigenfunctions $\rv_1$ and $\rv_3$ 
corresponding to the discrete eigenvalues $\rw_1$ and $\rw_3$ are well localized 
towards the left end of the interval $\x\in[0,\Length]$, whereas those corresponding 
to the continuous eigenvalues are evidently non-localized, \ie~vary significantly 
throughout $\x\in[0,\Length]$.
 
Thus, for the quadratic approximation, we retain in \eq{istsquadch3} only the leading
term. Setting $\n=\m=3$ gives a closed expression for the critical curve in the 
strength-duration plane,
\begin{align}
  \AIS \ampic^2+\BIS\ampic+\CIS=0,
\end{align}
where
\begin{align*}
\AIS & =\frac{4\Q{1}{3,3}\Est_3^2}{\rw_3^2} 
 \left\{ \frac{1-\e^{-\rw_1 \tst}}{\rw_1}-2\frac{\e^{\left(\rw_3-\rw_1\right)\tst}-1}
 {\rw_3-\rw_1} \right. \\ & \left.
 -\frac{\e^{-\rw_1\tst}-2\e^{\left(\rw_3-\rw_1\right)\tst}+1}{2\rw_3-\rw_1}
 \right\}, \\
\BIS &=  
\frac{4\Q{1}{3,3} \ai_3 \Est_3\leri{1- \e^{\left(\rw_3-\rw_1\right)\tst}}}
{\left(2\rw_3-\rw_1\right)\left(\rw_3-\rw_1\right)}-\frac{2\Est_1\left(\e^{-\rw_1 \tst}-1\right)}{\rw_1},
\\
\CIS &=- \frac{\Q{1}{3,3}\ai_3^2}{2\rw_3-\rw_1}+\ai_1,
\end{align*}
the coefficients in which are defined by \eq{Q}, \eq{ejsanatheo}  and \eq{Ajsanatheo}.
\Fig{zfkquadtsis} shows the comparison between quadratic approximation of the critical 
curves and the numerical curves. Compared to the linear approximation, one can see some
significant improvement for $\zth=0.05$, $0.15$, $0.25$, $0.35$, while the discrepancy 
between analytical and numerical results continues for $\zth=0.45$.

\sglfigure{\includegraphics{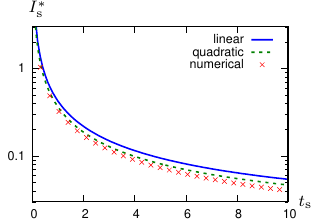}}{%
  (color online) %
  Quadratic approximation of the strength-extent curve for ZFK model for $\zth=0.05$ 
  (green line) compared with direct numerical simulations (lines with symbols) and linear 
  approximation (blue line). Parameters: $\Length=100$,
  $\dx=0.03$, $\dt=4\dx^2/9$. 
}{zfkquadtsis}

\subsection{McKean equation}
\seclabel{mckean}

\subsubsection{Model formulation}

Our second example is a piece-wise linear version of the ZFK equation,
considered by McKean in \cite{mckean1970nagumo} and then also in
\cite{rinzel1973traveling}:
\begin{align}                                     
  & 
  \dim=1, \quad
  \bD=\Mx{1}, \quad 
  \bu=\Mx{\u}, 
  \nonumber\\ &
  \ff(\bu)=\Mx{\f(\u)}, \quad
  \f(\u)=-\u+\Heav(\u-\mth) ,                     \eqlabel{McK}
\end{align}
where we assume that $\mth\in(0,1/2)$.
The critical nucleus solution in this equation is 
found in a closed form, 
\begin{align}                                     \eqlabel{McK-uc}
  \uc(\x) =
  \begin{cases}\displaystyle
    1-(1-\mth)\frac{\cosh(\x)}{\cosh(\xa)}, & \x\le\xa, \\
    \mth\,\exp(\xa-\x), & \x\ge\xa,
  \end{cases}
\end{align}
where
\begin{align}  \eqlabel{McK-xa}
  \xa = \frac12\ln\left(\frac{1}{1-2\mth}\right)
\end{align}
obtained by the fact that $\uc(\x)$ and its derivative are continuous at this point.
The eigenvalue problem  can be expressed as 
\begin{align}\eqlabel{eigprobmckeanch5}
\L\rv=\lw\rv
\end{align}
where the linearization operator contains the Dirac delta function:
\begin{align}                                     \eqlabel{McK-L}
  \L \triangleq \ddf{}{\xf} - 1 - \frac{1}{\mth}\dirac(\x-\xa).
\end{align}
The principal eigenvalue and the corresponding eigenfunction can be
written in the form
\begin{align}                                     
  & \lw_1=-1+\kk^2,
  \nonumber\\
  & \rv_1=\begin{cases}
    \dfrac{\cosh\left(\kk\x\right)}{\cosh\left(\kk\xa\right)}, 
    \quad \x\le\xa, \\
    \exp\left(
      \kk(\xa-\x)
      \right), \quad \x\ge\xa,
  \end{cases}
                                                  \eqlabel{McK-ignition}
\end{align}
where
\begin{align}                                     \eqlabel{McK-k}
  \kk
   =
   \frac{1}{2\mth} 
    + 
    \frac{1}{2\xa}\Lamb\left(
      \frac{\xa}{\mth}\,\e^{-\xa/\mth}
    \right)
\end{align}
and $\Lamb(\cdot)$ is the principal branch of the Lambert W-function as
defined \eg\ in \cite{corless1996lambertw}.

\subsubsection{Hybrid approach}

In this model, since the exact analytical solution for the critical
nucleus and the ignition eigenpair are known for an arbitrary
$\mth\in(0,1/2)$, the ``hybrid approach'' is not necessary.
However, for technical purposes, we address it here as well, to show  
the numerical computation of the essential ingredients based on 
\algs{numcritnuc} and \alref{numeigonecomp} works satisfactorily even for 
the models with discontinuous right hand sides.

Due to the discontinuous terms, the numerical computation of the 
ingredients of the McKean equation requires the finite-element 
treatment which was outlined in our previous paper \cite{bezekci2015semianalytical}. 
Hence, we skip the details here. \Fig{McK-lin-funs}  illustrates the processes 
involved in obtaining the critical nucleus and ignition mode.
The results of these ingredients are compared with their analytical counterparts 
and we see a good agreement between the two. 

\sglfigure{\includegraphics{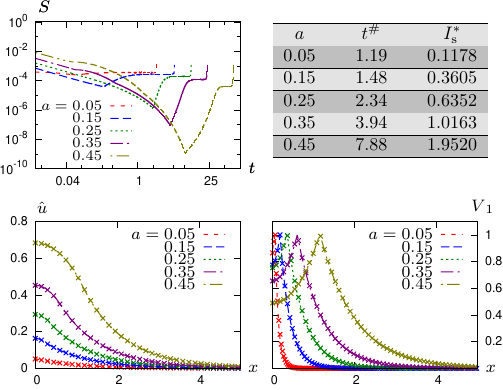}}{%
  (color online) %
  Illustration of the numerical computation of the critical nucleus
  and ignition mode by ``shooting'' and ``marching'' in McKean. Top
  panel: Typical function $\Speed(\t)$ at near threshold boundary
  conditions. Bottom panel: Critical nucleus solutions and %
  ignition modes of the McKean model \eq{McK} %
  for various values of the parameter $\mth$.  Parameters:
  $\mth=0.05,0.15,0.25,0.35,0.45$, $\dx=0.03$, $\dt=4\dx/9$,
  $\Length=10$, $\tst=1.2$, $\tol=10^{-5}$. %
}{McK-lin-funs}

\subsubsection{Linear theory}

Linear approximation of the strength-duration curve can be found using 
the analytically derived expression given by \eq{LBH}. However, it must 
be noted that in this case, the rheobase is found as
\begin{align}\eqlabel{rheobasech5mckean}
   \Irh
 & = \rw_1\Num,
\end{align}
where
{\small
\begin{align} \eqlabel{McK-num}
    \Num&=\frac{\sinh\left(\kk\xa\right)}{\kk}+\frac{\mth}{\kk+1}\cosh\left(\kk\xa\right) \\ \nonumber
    &-\frac{1-\mth}{2\cosh\left(\xa\right)}
    \left(\frac{\sinh\left(\left(\kk+1\right)\xa\right)}{\kk+1}+\frac{\sinh
    \left(\left(\kk-1\right)\xa\right)}{\kk-1}\right).
\end{align}
} 

\sglfigure{\includegraphics{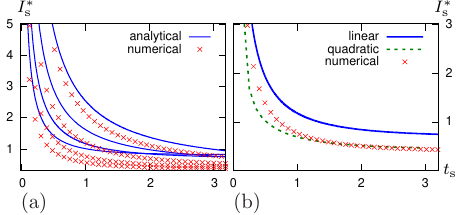}}{%
  (color online) %
  Strength-duration curves in McKean model: %
  direct numerical simulations (red circles) vs %
  (a) linear theory, for $\mth=0.35$ at the bottom , $\mth=0.4$, $\mth=0.45$
   to $\mth=0.48$ at the top, and %
  (b) linear and quadratic theories, for $\mth=0.4$. %
  Blue long-dashed lines: analytical dependencies given by
  \eqthree{LBH,rheobasech5mckean,McK-num}. 
  Green short-dashed lines: 
  the predictions given by quadratic theory. %
  Discretization: $\dx=0.03$, $\dt=4\dx^2/9$, $\Length=10$, $\tol=10^{-5}$. %
}{mck-tsis}

This linear prediction formalism compared with the direct numerical simulations 
is depicted in \fig{mck-tsis}(a). The \apriori\ bound for these 
chosen threshold parameters is outside of the duration domain. 
As shown in the figure, the linear approximation for parameter $\mth$ values close 
to $1/2$ better fits to the numerical simulation than that for small values. 
This may be related to the fact that the leading eigenvalue is inversely proportional to the 
threshold parameter $\mth$. Even for larger $\mth$, there are still some deviations  
between  the linear theory and numerical simulations, which can be reduced by 
considering second order approximation that will be outlined in the following subsection.  

\subsubsection{Quadratic theory}

Linear approximation of the strength-duration curve can be  improved by considering 
the second order approximation. For the quadratic approximation, the knowledge of 
the whole spectrum is ideally required. We know that the linearization spectrum 
of the critical nucleus has only one unstable eigenvalue $\rw_1>0$, and due to 
translational symmetry, $\rw_2=0$, and the rest of the spectrum lies entirely 
in the left half-plane. The first aim of this part of the subsection is to find 
these remaining stable eigenvalues and  corresponding eigenfunctions. To obtain 
these eigenpairs, we replace the infinite interval $[0, \infty)$ with a finite 
interval $[0,\Length]$ with a homogeneous Dirichlet boundary condition at $\x=\Length$, 
aiming to consider the limit $\Length\to \infty$.

The solution of eigenvalue problem \eq{eigprobmckeanch5} in this case  is
{\small
\begin{align*}
  \rv(\x;\rw)=\p\cos\left(\p\x\right) - \cos\left(\p\xa\right) \sin\left(\p(\x-\xa)\right) \Heav(\x-\xa)
\end{align*}
}
where $\p=\sqrt{-1-\rw}$ and the eigenvalues are expressed in terms of 
the following transcendental equation,
\begin{align}
h(\p)=\tan\left(\p \Length\right)-\tan\left(\p \xa\right)
-\frac{\mth\p}{\cos^2\left(\p \xa\right)}=0,
\end{align}
which is dependent of the domain size $\Length$ as opposed to the eigenfunction
expression. After finding analytical expressions for the eigenpairs, the next step 
is to calculate \eq{Q}, \eq{ejsanatheo}  and \eq{Ajsanatheo}, and then substitute 
them back in the coefficients of the quadratic equation for $\ampi$ \eq{istsquadch3} 
so that the second-order approximation of the strength-duration curve can be generated.
 
\Fig{mck-tsis}(b) shows graphs of the linear and quadratic approximation of the 
strength-duration curve along with its numerical result for $\mth=0.4$. The 
quadratic approximation was obtained for $\Length = 10$ and $287$
eigenvalues. 
The accuracy of the second-order approximation is much 
closer to the direct numerical simulation compared to the first-order approximation.

\section{Multi component systems}
\seclabel{mutlicomp}

\subsection{$\INa$-caricature model}
\seclabel{ina}

\subsubsection{Model formulation}

Our next example is the caricature model of an $\INa$-driven cardiac
excitation front suggested in~\cite{biktashev2002dissipation}. It is a
two-component reaction-diffusion system~\eq{INTRDS} with
$\bu=(\E,\h)^\top$, $\bD= \begin{pmatrix}
    1 & 0 \\
    0 & 0
  \end{pmatrix} $ and $\ff=(\fE,\fh)^\top$,
where
\begin{align}
  \fE(\E,\h) &= \Heav(\E-1)\h, \nonumber\\
  \fh(\E,\h) &= \frac{1}{\exty}\left(\Heav(-\E) - \h\right),      \eqlabel{b02kin}  
\end{align}
and $\Heav(\cdot)$ is the Heaviside step function. The component $\E$
of the solution corresponds to the nondimensionalized transmembrane
voltage, and the component $\h$ describes the inactivation gate of the
fast sodium current, which is known in electrophysiology as $\INa$ and
which is mainly responsible for the propagation of excitation in
cardiac muscle in the norm.

A special feature of this model is that there is a
  continuum of potential resting/pre-front states,
\[
  \Ur =  \lim\limits_{\xf\to\infty} \buc = \Mx{ -\pref , 1}^\top,
  \qquad \pref>0, 
\]
and a continuum of potential post-front states, 
\[
  \Ub =  \lim\limits_{\xf\to-\infty} \buc = \Mx{ \postf , 0}^\top, 
  \qquad \postf>1 ,
\]
so any front solution connects a point from one continuum to a point
from another continuum. 
The critical solution $\buc=(\Ec,\hc)^\top$ is described by
\begin{align}
  \Ec(\xf)&=
  \begin{cases}
    \postf -\dfrac{\exty^2\c^2}{1+\exty\c^2}\,\e^{\,\xf/(\exty \c)}, 
    & \xf \leq -\xm, \nonumber\\[2ex]
    -\pref + \pref \e^{-\c\xf}, & \xf \geq -\xm,   
  \end{cases}
\\[2ex]
  \hc(\xf)&=
  \begin{cases}
    \e^{\,\xf/(\exty \c)}, & \xf \leq 0,\\[2ex]
    1, & \xf \geq 0,   
  \end{cases}                                   \eqlabel{bv-exactsol-c5}
\end{align}
where the post-front voltage $\postf$ and front thickness $\xm$ are
given by
\begin{align}
  \postf = 1+\exty\c^2\,(1+\pref), 
  \quad
  \xm = \dfrac{1}{\c}\,\ln\left(\dfrac{1+\pref}{\pref}\right),      \eqlabel{bv-vars-pars-c5}
\end{align}
and the front speed $\c$ is defined by an implicit equation
\begin{align}
\exty \c^2 \ln\left(\frac{(1+\pref)(1+\exty\c^2)}{\exty}\right)
 + \ln\left(\frac{\pref+1}{\pref}\right)=0,                  \eqlabel{transalpha}
\end{align}
or equivalently
\begin{align}
  \exty = \G(\bet,\sig) \bydef \frac{1+\sig}{1-\bet}\,\bet^{-1/\sig} ,
                                                           \eqlabel{nlchareq}
\end{align}
where
\begin{align}
  \sig=\exty \c^2, \; \bet=\pref/(\pref+1).                  \eqlabel{nlsubs}
\end{align}
For the analytical expression of the first two left and right eigenfunctions, 
please see \cite{bezekci2015semianalytical,idris2008initiation}.  

\subsubsection{Hybrid approach}

Even though we know the ingredients of the linear theory for this
model analytically, we still found them numerically as well. The
hybrid approach is needed not only because it helps to validate the
analytical result but also because, in some cases, it is the only
option as the analytical derivation is not always possible.  Due to
the discontinuous right-hand sides, it is essential to use the
standard finite element method, at least when dealing with these
discontinuous terms.  The complete discretization formula for the
critical front of the model is presented in \Appn{inadisc}.

\sglfigure{\includegraphics{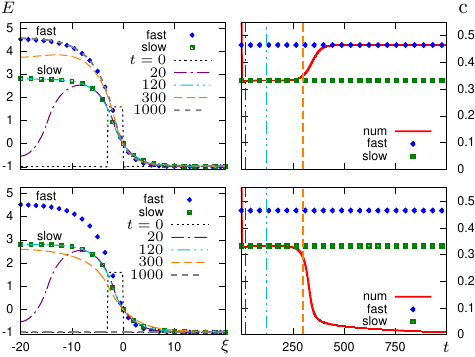}}{ %
  (color online) %
  Evolution of $\E$ component of the $\INa$-caricature model with
  chosen sub- and super-threshold initial condition in the comoving
  frame of reference. Parameters used: $\dx=0.05$, $\dt=4\dx^2/9$,
  $\pref=1$, $\exty=8.2$, $\xst=1.5$ for both, sub-threshold
  $\ampu=2.59403$ (a) and super-threshold $\ampu=2.59404$ (b) cases.
}{inaeval}

For two initial conditions, \fig{inaeval} shows the evolution of $\E$ 
component in the comoving frame of reference. For each case, the solution 
approaches the critical front, \ie~the solution at $\t=120$ in the figure 
and then gives rise to the stable propagating wave if the initial condition 
is above the threshold or decays back to the resting state otherwise. 
\Fig{inauc} gives the comparison of the numerical critical front obtained 
using operator splitting method and its analytical closed-form solution 
given by \eq{bv-exactsol-c5}. We can see that the shooting procedure 
provides a good approximation of the critical front for the selected parameters. 

Numerical experiments suggest that there are two values of the speed
$\c$, $\c_{slow}$ and $\c_{fast}$, 
satisfying $0<\c_{slow}<\c_{fast}<\infty$,  such that the faster 
fronts are higher and stable, and the slower fronts are lower and
unstable, hence a slower front
either dissipates or increases in the speed and magnitude to the fast branch solution 
depending on the initial condition being below- and above-threshold, 
respectively \cite{biktashev2002dissipation,hinch2004stability}. 
This can be seen in the right panel of \fig{inaeval} where the blue 
circle and green square symbols represent fast and slow front speeds 
for the selected pair of $\pref=1$, $\exty=8.2$, and the red line indicates
how the speed of the front changes in time. 
For initial conditions slightly above threshold, the front speed gets closer 
to the slow speed and stays in the vicinity of it for a long time before 
developing into the fast speed while the initial condition slightly below 
threshold results in the front speed to drop to zero eventually.

\sglfigure{\includegraphics{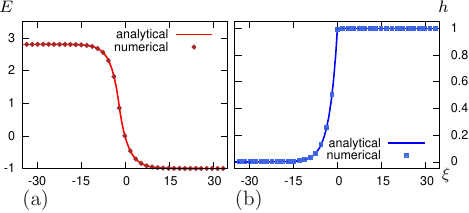}}{ %
  (color online) %
  (a) Comparison between analytical and numerical critical front of
  the $\INa$-caricature model.  For the numerical front, we used
  following discretization parameters:$\pref=1$, $\exty=8.2$,
  $\dx=0.05$, $\dt=4\dx/9$, $\Length=20$.  %
}{inauc}

The next step after finding the critical front of the $\INa$-caricature model  is
the determination of the right and left eigenfunctions along with the 
corresponding eigenvalues employing \alg{eigfunadj} detailed in \secref{ingonecomphyb}.
In \fig{inaeigfun}, the eigenfunctions obtained using the hybrid method fairly 
resemble exact analytical eigenfunctions. The largest difference between the 
numerical and analytical eigenfunctions is observed in the vicinity of the 
discontinuous values, $\myxi=0$ and $\myxi=-\xm$. This is totally expected as the 
numerical scheme used to calculate the eigenfunctions is the second-order accurate, 
except near discontinuities, where it reduces to first order accuracy and introduces 
spurious oscillations due to the Beam-Warming method~\cite{ewing2001summary}.

\sglfigure{\includegraphics{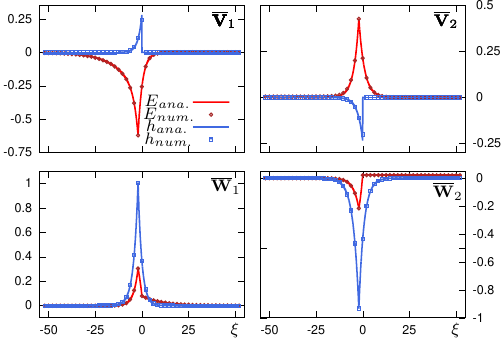}}{ %
  (color online) %
  Comparison between first two right and left eigenfunctions of the
  $\INa$-caricature model.  Parameters same as previous figure. %
}{inaeigfun}

\subsubsection{Linear theory}

The result of the calculation of the strength-duration curve for $\INa$-caricature 
model is visualized in \fig{ina-ists}, where the linear approximation is 
based on the formulas \eq{crit-sys} and \eq{zeroofs}.  For every chosen 
stimulus duration, $\tst$, we calculate the zeros of \eq{zeroofs} in 
order to find the value of the shift $\shift$. We then substitute this 
value of $\shift$ into one of  the equations in \eq{crit-sys} 
(both produce the same result) to get the corresponding value of $\ampi$.
In the simulations, we choose two different set of the model parameters,
$\exty=7.8$, $\pref=2/3$ and $\exty=8$, $\pref=9/11$ from which the resulting 
curves are respectively shown in \fig{ina-ists}(a,c) and \fig{ina-ists}(b,d).
The shape of $\seq(\shift)$ is rather similar for both cases and the main 
difference between the two is the closeness of the  $\shift$ values for two 
different duration of stimulus values, $\tst=3$ and $\tst=10$.
We observe that the theoretical critical curve for the first set of 
parameter values is well adapted to the direct numerical simulation threshold 
curve for smaller values of $\tst$, and then bends down dramatically as the value 
of $\tst$ increases. The theoretical prediction for the second set of parameters values, 
however, gets better with $\tst$. 

\sglfigure{\includegraphics{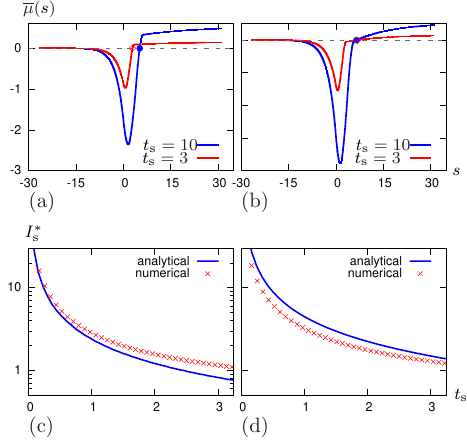}}{ %
  (color online) %
  Comparison of analytical and numerical strength duration curve for
  $\INa$-caricature model for the choice of parameters $\exty=7.8$,
  $\pref=2/3$ (a,c) and $\exty=8$, $\pref=9/11$ (b,d).  Panels (a) and
  (b) show functions $\seq(\shift)$ defined by \eq{zeroofs} for two
  selected values of $\tst$ and their roots. Panels (c) and (d) are
  strength-duration curves. %
}{ina-ists}

\subsection{FitzHugh-Nagumo system}
\seclabel{fhn}

\subsubsection{Model formulation}

The FitzHugh-Nagumo (FHN) system is a two-component reaction-diffusion system,
which could be considered as a ZFK equation extended by adding a
second, slow variable, describing inhibition of excitation. It is
probably the single historically most important model describing
excitable media. We consider it in the form
\begin{align}
  & 
  \dim=2, \quad
  \bD=\diag(1,0), \quad 
  \bu=\Mx{\u,\v}\Tr,
  \nonumber\\&
  \ff(\bu)=\Mx{\fu(\u,\v),\fv(\u,\v)}\Tr, \quad
  \nonumber\\&
  \fu(\u,\v)=\u(\u-\fth)(1-\u)-\v, \qquad
  \nonumber\\&
  \fv(\u,\v)=\fep(\fal\u-\v) . 
  \eqlabel{FHN}
\end{align}
for fixed values of the slow dynamics parameters,
$\fep=0.01$ and $\fal=0.37$, and two values of the excitation
threshold for the fast dynamics, $\fth=0.05$ and $\fth=0.13$.

\subsubsection{Hybrid approach}

System~\eq{FHN} has an unstable propagating pulse solution as opposed 
to its reduced form, the ZFK equation with nontrivial stationary solution. 
It is known (see \eg~\cite{flores1991stability}
and references therein) that in the limit $\fep\searrow0$, the cricial
pulse solution whose $\v$-component is small and $\u$-component is
close to the critical pulse of the corresponding ZFK equation. 
This makes it feasible to obtain explicit analytical solutions by
using perturbation techniques in the
double limit $\fep\searrow0$, $\fth\searrow0$. These asymptotics will be
described in a
separate publication, and here 
we describe only the hybrid approach. The critical pulse is obtained 
by applying \alg{critpulse} by means of AUTO. The corresponding CV restitution 
curves are illustrated in~\fig{fhncvrcch6}. For the critical pulses, we take 
the solutions at lower branches at $\Per>7.5\X3$ (see top row of \fig{fhn-ingredients}). 
Other essential ingredients of the theory are the first two left eigenfunctions and 
the first leading eigenvalue, which are computed using \alg{eigfunadj}. 
The eigenfunctions for the two selected cases look rather similar, as 
shown in \fig{fhn-ingredients}.

\sglfigure{\includegraphics{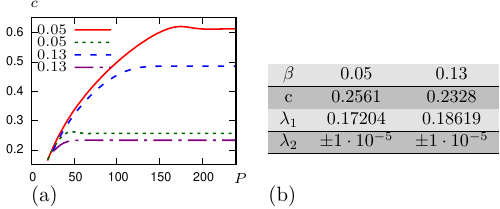}}{ %
  (color online) %
  CV restitution curves for the FHN model for two selected
  values of the model parameter. %
}{fhncvrcch6}

\sglfigure{\includegraphics{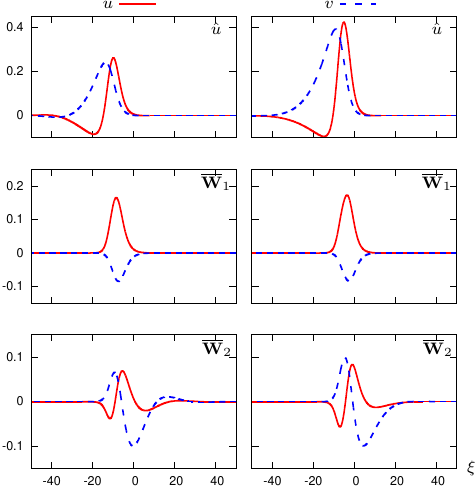}}{ %
  (color online) %
  FHN theory ingredients for (a) $\fal=0.05$ and (b) $\fal=0.13$. %
  Shown are components of scaled vector functions, indicated in top
  right corner of each panel, where %
  $\suc=\Scale\buc$, %
  $\sLV_\j=\Scale^{-1}\LV_\j$,  %
  and $\Scale=\diag(1,10)$. %
  The space coordinate is chosen so that $\xf=0$ at the maximum of
  $\uc$. %
  Correspondence of lines with components is according to the legends
  at the top. %
}{fhn-ingredients}

\subsubsection{Linear theory}

\Fig{fhn-ists} illustrates the calculation of the strength-duration curve for 
FHN model for $\fal=0.05$ and $\fal=0.13$ according to the formulas \eq{crit-sys} 
and \eq{zeroofs}. The equation \eq{zeroofs} has two roots, one of them is 
negative close to zero and the other is positive. We find that in both cases 
the smaller root denoted by blue circle and red square points in 
\fig{fhn-ists}(a,b) gives the corresponding value of $\shift$. The critical 
curves compared with those obtained from direct numerical simulation are 
sketched in \fig{fhn-ists}(c,d). From this plot, it can be seen that the 
theoretical prediction for both values of parameter is almost equally close to the 
numerical prediction.

\sglfigure{\includegraphics{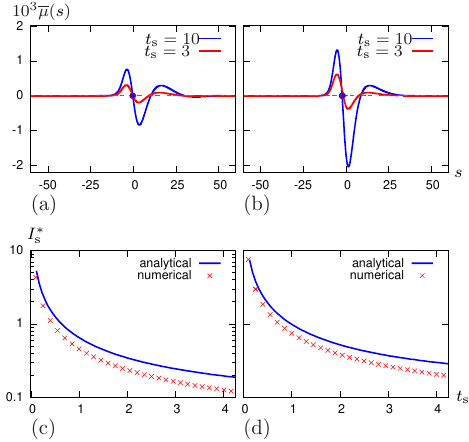}}{ %
  (color online) %
  Equation \eq{zeroofs} that defines the shift $\shift$ in terms of $\tst$ and
  comparison of analytical and numerical strength duration-curve for the FHN 
  model for $\fth=0.05$ and $\fth=0.13$. Other parameters: $\fep=0.01$, $
  \fal=0.37$, $\dx=0.03$, $\dt=4\dx^2/9$, $\Length=60$.
}{fhn-ists}

\subsection{The modified Beeler-Reuter model}
\seclabel{brp-model}

\subsubsection{Model formulation}

Here, we looked at a variant of the classical Beeler-Reuter (BR) model of 
mammalian ventricular cardiac myocytes \cite{beeler1977reconstruction}, 
modified to describe phenomenologically the dynamics of neonatal rat cells \cite{arutunyan2003behavior,pumir2005genesis,biktashev2008generation,biktashev2011evolution}:
\begin{align}
  & 
  \dim=7, \quad
  \bD=\diag(1,0,0,0,0,0,0), 
  \\
  &
  \bu=\Mx{\brV,\brh,\brj,\brx,\brd,\brf,\brc}\Tr,
  \\ 
  &
  \ff(\bu)=\Mx{
    -(\brik+\brix+\brin+\bris) \\
    \bralh (1-\brh)-\brbeh \brh \\
    \bralj (1-\brj)-\brbej \brj \\
    \bralx (1-\brx)-\brbex \brx \\
    \brald (1-\brd)-\brbed \brd \\
    \bralf (1-\brf)-\brbef \brf \\
    -10^{-7} \bris+0.07 (10^{-7}-\brc)
  } .
  \eqlabel{BRP}
\end{align}
For the detailed description of the components of $\ff(\bu)$, please see 
the appendix of \cite{bezekci2015semianalytical}.

\subsubsection{Hybrid approach}

As in the FitzHugh-Nagumo system, the critical solution is a moving pulse, 
and thus, we obtain the CV restitution curves and the critical pulse in a 
similar way. The CV restitution curves for the modified Beeler-Reuter model 
are shown in \fig{brp-cvrc}. 
Apart from the critical pulse, the solution at lower branches, the knowledge of 
the first two left eigenfunctions and the first leading eigenvalue are also 
required. These ingredients have been found by the marching 
method given by \alg{eigfunadj}. The essential ingredients of the theory for 
two different data sets are sketched in \fig{brp-ingredients}.

\sglfigure{\includegraphics{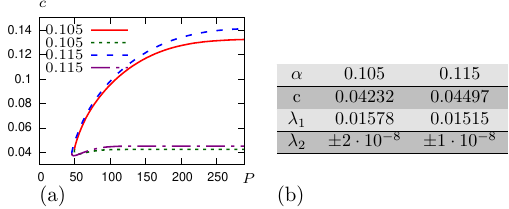}}{ %
  (color online) %
  CV restitution curves for the BR model for two selected
  values of the model parameter. Stable (upper) and unstable (lower) branches are
    shown by different line types.
}{brp-cvrc}

\sglfigure{\includegraphics{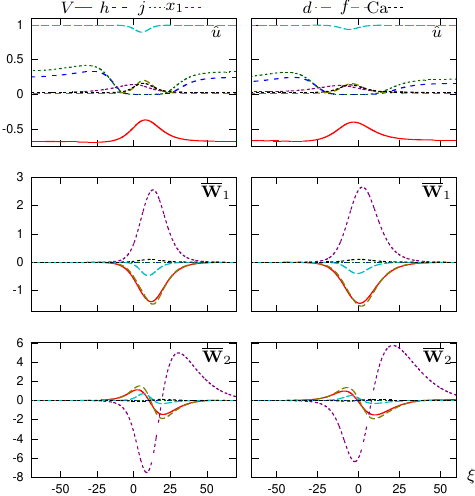}}{ %
  (color online) %
  BR theory ingredients for (a) $\bralp=0.105$ and (b) $\bralp=0.115$. %
  Shown are components of scaled vector functions, indicated in top
  right corner of each panel, where \dots %
  $\suc=\Scale\buc$, %
  $\sLV_\j=10^4\Scale^{-1}\LV_\j$, %
  and
  $\Scale=\diag(10^{-2},1,1,1,1,1,10^5)$. %
  The space coordinate is chosen so that $\xf=0$ at the maximum of
  $\brVc$. %
  Correspondence of lines with components is according to the legends
  at the top. %
}{brp-ingredients}

\subsubsection{Linear theory}

\Fig{br-ists} exhibits the strength-duration threshold curve analysis 
for BR model for two different excitability parameters, $\bralp=0.105$
and $\bralp=0.115$. 
The resulting theoretical critical curves are derived according to the 
formulas \eq{crit-sys} and \eq{zeroofs}. Firstly, the values of $\shift$ 
are determined by the transcendental equation \eq{zeroofs} and compared 
to FHN system, it is easier to detect the zeros of this equation, two of 
which are shown in the top panel of the figure for $\tst=3$ and $\tst=10$. 
Then, the remaining part is to insert the found value of $\shift$ back into 
theoretical threshold curve generated by \eq{crit-sys}. The bottom panel of 
the figure shows these threshold curves being compared with numerical critical 
curves. As can be seen from the figure, the analytical estimate for $\bralp=0.115$ 
provides a somewhat better approximation than that for $\bralp=0.105$.

\sglfigure{\includegraphics{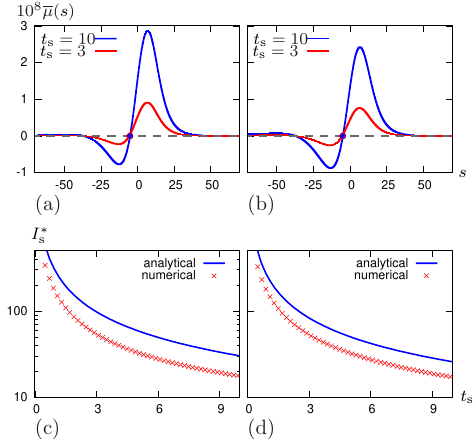}}{ %
  (color online) %
  Comparison of analytical and numerical strength duration curve for BR 
  for $\bralp=0.105$ (a) and $\bralp=0.115$ (b). Other parameters:
   $\dx=0.03$, $\dt=4\dx/9$, $\Length=30$, $\tol=10^{-3}$.
}{br-ists}

\section{Discussion}
\seclabel{disc}

The main aim of this paper was to extend the method
proposed in~\cite{idris2008analytical} for analytical description of 
the threshold curves that separate the basins of attraction of propagating
 wave solutions and of decaying solutions of certain reaction-diffusion models
  of spatially-extended excitable media. Specific aims are:
\begin{itemize}
	\item Extending the proposed theory to analysis of a wider class of excitable systems, 
	including multicomponent reaction-diffusion systems, systems with non-self-adjoint 
	linearized operators and in particular, systems with moving critical solutions 
	(critical fronts and critical pulses).
	\item  Building an extension of this method from a linear to a quadratic approximation 
	of the (center-)stable manifold of the critical solution to demonstrate the discrepancy 
	between the analytical based on linear approximation and  numerical threshold curves 
	encountered when considering this quadratic approximation.
\end{itemize}

The essential ingredients of the theory are the critical solution itself, and the 
eigenfunctions of the corresponding linearized operator. For the linear approximation 
in the critical nucleus case, we need the leading left (adjoint) eigenfunction; in the 
moving critical solution case, we need two leading left eigenfunctions; and for the 
quadratic approximations we require as many eigenvalues and left and right eigenfunctions 
as possible to achieve better accuracy. Of course, closed analytical formulas for these 
ingredients can only be obtained in exceptional cases, and in a more typical situation 
a ``hybrid'' approach is required, where these ingredients are obtained numerically. 
We thus have provided insight into how the numerical computation of
these essential ingredients can be done. 
 
The theory have been demonstrated on five different test problems ranging from one-component 
reaction-diffusion systems where the critical solution is the critical nucleus to the 
multicomponent test problems with either critical front or critical pulse solution. In 
all models, the analytical threshold curves are compared with the numerical simulations 
obtained  using the bisection algorithm. We have applied both linear and quadratic 
approximations for one-component test problems, ZFK and McKean models. The quadratic 
approximation agreed much better with numerical threshold curves compared to the linear 
approximation's results, as would be expected. 

The accuracy and efficiency of the hybrid computation of the essential ingredients 
is dependent on the numerical scheme and mesh resolution. It is obvious that some 
of the numerical schemes discussed in this paper do not outperform other long-running 
and mathematically more complicated numerical schemes. In particular, the numerical 
study of $\INa$-caricature model has introduced the spurious oscillations and first-order 
accurate result near the discontinuities due to Beam-Warming method, which can be 
tackled using some advanced shape-preserving advection schemes 
(see, for example, \cite{rucong1994two}). Hence, such approaches that demand high 
computational cost can be carried out as an interesting direction for future 
research if higher accuracy is required. 

As the results of the theory pointed out, our method provides more accurate results 
for some parameter values than the others, especially in linear analysis. Even 
though the quadratic approximation offered for one-component test problems with 
the critical nucleus solutions provides more accurate estimates, still it is not 
fully understood why the choice of parameter values significantly matters. Hence, 
this remains quite important line of research. On the other hand, the 
proposed theory based on the moving critical solutions involves only the linear 
approximation. As an additional consideration, quadratic approximation for the 
cases of moving critical solutions can also be carried out. 

The theory established in this paper is limited to one spatial dimension. 
Therefore, it could be of interest for further research to adapt the theory to two 
and three dimensions.

Throughout the theory, we have made the assumption that the spectrum is real. 
This is, however, not necessarily the case for the non-self-adjoint problems, 
which remains an interesting direction for future research.

Another extension of the work worth considering would be to investigate the theory 
on some up-to-date realistic cardiac excitation models~\cite{clayton2011models}, 
simplified cardiac models with unusual properties~\cite{duckett2000modeling,hinch2002analytical}, 
and other excitable media such as combustible
media~\cite{initiandproghugbridson,scott1992simplenew,luke2011soil},
pipe flow~\cite{barkley2011modeling,barkley2015rise} \etc
In the context of this Special Issue, of particular interest
  is the link to the problem of threshold of front propagation in
  neural fields, discussed in the recent
  paper~\cite{Faye-Kilpatrick-2018}, where the Lapicque-Blair
  strength-duration relationships also naturally occurs under certain
  simplifying assumptions.

\paragraph{Funding: }
  BB gratefully acknowledges financial support from the Ministry of National Education of the Republic of Turkey. 
  VNB was supported in part the EPSRC Grants No. 
  GR/S75314/01, 
  EP/I029664, 
  EP/N014391/1 
  (UK), and
  National Science Foundation Grant No. NSF PHY-1748958, %
  NIH Grant No. R25GM067110, %
  and the Gordon and Betty Moore Foundation Grant No. 2919.01 %
  (USA).

\paragraph{Data statement:} The research data supporting this publication are provided within this paper.

\appendix
\section{Discretization formula for strength-duration curve}
\seclabel{disformula}

With the aim of comparing the explicit approximations for 
the strength-duration curve, we describe how the threshold
curve is obtained using direct numerical simulations in 
this section. The numerical strength-duration threshold 
curve is computed by solving the nonlinear system 
\eq{INTRDS} for the initial and boundary conditions 
given by \eq{semi-cableaim}-\eq{tsrecprof} using standard 
finite difference or finite element discretization.
More specifically, for ZFK, FHN and BR models we use finite 
differences, and for McKean and $\INa$-caricature
models we implement finite element method instead.

\subsection{Finite Difference Discretization Formula}

The discretization formula for the generic form of initial-boundary 
value problem
\begin{align} \eqlabel{genreacapp}
& \bu_t=\bD\bu_{xx}+\ff\leri{\bu}, \qquad
\bu(\x,0)=\Ur, \nonumber  \\ 
& 
\bD\bu_\x(0,\t)=-\ampi\,\Heav(\tst-\t)\best, \qquad \x,\t>0, 
\end{align} 
We employ explicit first order Euler forward difference 
approximation in time and explicit second order centered 
finite difference approximation in space and plugging these 
into \eq{genreacapp}, we obtain
\begin{align}
 \hat{\bu}_i^{\j+1}=\hat{\bu}_i^\j+\frac{\bD\dt}{\dx^2}
 \left(\hat{\bu}_{i-1}^\j-2\hat{\bu}_i^\j+\hat{\bu}_{i+1}^\j\right)
 +\dt \ff\left(\hat{\bu}_i^\j\right),
 \eqlabel{discfindif}
\end{align}
in conjunction with its initial and boundary conditions
\begin{align}
& \hat{\bu}_i^0=\Ur, \nonumber \\
& \hat{\bu}_0^0=\hat{\bu}_2^0+2\dx\ampi\,\Heav(\tst-\t)\best , 
\quad \hat{\bu}_{N+1}^0=\hat{\bu}_{N-1}^0, 
\eqlabel{dicsacctoboun}\\
& \hat{\bu}_0^{\j+1}=\hat{\bu}_2^{\j+1}, \quad \hat{\bu}_{N+1}^{\j+1}
=\hat{\bu}_{N-1}^{\j+1}. \nonumber
\end{align}

\subsection{Finite Element Discretization Formula} 

Two of our test problems namely, McKean and $\INa$-caricature models,
contain discontinuous kinetic terms.

\subsubsection*{Discretization Formula for McKean equation}

We use even extension of the problem 
and start with one component McKean model 
\begin{align}
  & \df{\u}{\t} = \ddf{\u}{\x} -\u+\Heav(\u-\mth) + 2\ampi\,
  \Heav(\tst-\t)\dirac(\x),\nonumber \\ & \u(\x,0)=0,
  \quad
  (\x,\t)\in\Real\times\Real_+.
  \eqlabel{evenexreacdif}
\end{align}
In the  Galerkin finite element method, this can be written 
in the vector form, for $\wnd_\j$, 
\begin{align}  \eqlabel{fulldisfinele}
  \feA\Df{\wnd}{\t}+\leri{\feA+\feB}\wnd=\feF+2\ampi\,\Heav(\tst-\t)\feD,
\end{align}
where  $\feA$ is the the mass matrix, $\feB$ is the  
stiffness matrix and $\feF$ is the load vector (see \cite{bezekci2015semianalytical} for a 
crude introduction to finite 
element method and the derivation of the matrices). 
The vector $\feD$ has only one nonzero entry by definition of 
the Dirac delta function.

Finally, we employ the generalized trapezoidal rule 
(also known as $\theta$ scheme) \cite{bochev2004stability},
in which the residual is evaluated at $\j+\theta$, with this 
notation implying
\begin{align*}
\wnd^{\n+\theta}= \theta \wnd^{\n+1} +\leri{1-\theta}\wnd^{\n}
\end{align*}
where $0\le \theta\le 1$ is a real parameter. 
Based on the above, the fully discrete problem 
\eq{fulldisfinele} becomes 
\begin{align}  
\left[\feA+\dt\theta  \leri{\feA+\feB}\right]\wnd^{\n+1}= &
\left[\feA-\dt\leri{1-\theta} \leri{\feA+\feB}\right]\wnd^{\n} 
\nonumber \\ & +
\feF+2\ampi\,\Heav(\tst-\t)\feD.
\eqlabel{linearsystem2}
\end{align}
For $\theta=0$ ,  the linear system \eq{linearsystem2}
is the explicit Euler method that has a stability condition 
to be satisfied and its truncation error is 
$\mathcal O\left(\dt\right)+\mathcal O \left(\dx^2\right)$,
 $\theta =1/2$ gives the second-order unconditionally stable
Crank-Nicolson method with truncation error 
$\mathcal O \left(\dt^2\right)+\mathcal O\left(\dx^2\right)$, 
and $\theta=1$ gives the first-order accurate implicit Euler rule 
that is also unconditionally stable and its truncation error 
is   $\mathcal O\left(\dt\right)+\mathcal O \left(\dx^2\right)$
\cite{bochev2004stability}.

\subsubsection*{Discretization Formula for the $\INa$-caricature model}
Even extended version of the model is in the following form 
\begin{align}
  & \df{\E}{\t} = \ddf{\E}{\x}+\Heav(\E-1)\h + 2\ampi\,
  \Heav(\tst-\t)\dirac(\x),\nonumber \\ 
  & \df{\h}{\t} =\frac{1}{\exty}\left(\Heav(-\E) - \h\right), \eqlabel{evenexapbikmodel} \\
  & \E(\x,0)=-\pref, \quad \h(\x,0)=1, \quad (\x,\t)\in\Real\times\Real_+. \nonumber
\end{align}
The fully discretized finite element discretization formula for 
the model is 
\begin{align}
    \left[\feA+\dt\theta \feB\right]\End^{\n+1}= &
\left[\feA-\dt\leri{1-\theta}\feB\right]\End^{\n} \nonumber \\ & +
\dt\seS\hnd^{\n}+2\dt\ampi\,\Heav(\tst-\t)\feD \nonumber \\ 
    \left[\exty+\dt\theta\right]\feA\hnd^{\n+1}= &
\left[\exty-\dt\leri{1-\theta}\right]\feA\hnd^{\n} + \dt \feg,
\end{align}
where
\begin{align*}
&  \seS=\left[s_{\i,\j}\right]=\int_{-\Length}^\Length 
  \Heav\left(\End-1\right)\Fem_\i(\myxi)\Fem_\j(\myxi)\,\d \myxi, \\
 &  \feg=\left[g_{\i}\right]=\int_{-L}^L \Fem_\i(\myxi)\Heav\left(-\End\right)\d \myxi.
\end{align*}
The matrix $\seS$ is a tridiagonal matrix with following diagonal elements:
\begin{align*}
s_{\i,\i}&=\int_{-\Length}^\Length \Heav\left(\End-1\right)\Fem_\i^2\left(\myxi\right)\d \myxi \\
& =\frac{1}{\dxi^2}\left(\int_ {\myxi_{\i-1}}^{\myxi_\i} 
\Heav\left(\End-1\right)\left(\myxi-\myxi_{\i-1}\right)^2\d \myxi \right. \\
    & + \left.
\int_{\myxi_\i}^{\myxi_{\i+1}}\Heav\left(\End-1\right) \left(\myxi_{\i+1}-\myxi\right)^2\d \myxi  
  \right)\\
&=\IUC^{\i}+\IDORT^{\i}, \quad \quad  \text{for} \quad \i=2, 3,\cdots,\numn-1  \\
\end{align*}
where 
\begin{align}\nonumber
  \IUC^{\i}=\frac{1}{3\dxi^2}
  \begin{cases}
    \dxi^3, & \ETT{t_1}{\i-1},  \\[2ex]
    \left(\xibnew_{\i-1}-\myxi_{\i-1}\right)^3, & \ETT{m_1}{\i-1}, \\[2ex]
    \dxi^3-\left(\xibnew_{\i-1}-\myxi_{\i-1}\right)^3, & \ETT{b_1}{\i-1}, \\[2ex]
    0, & \textrm{otherwise,}
  \end{cases}
\end{align}
\begin{align}\nonumber
  \IDORT^{\i} =\frac{1}{3\dxi^2}
  \begin{cases}
     \dxi^3, & \ETT{t_1}{\i},  \\[2ex]
    \dxi^3 -\left(\myxi_{\i+1}-\xibnew_{\i+1}\right)^3, & \ETT{m_1}{\i}, \\[2ex]
    \left(\myxi_{\i+1}-\xibnew_{\i+1}\right)^3, &\ETT{b_1}{\i}, \\[2ex]
    0, & \textrm{otherwise.}
  \end{cases}
\end{align}  
\begin{align}\nonumber
&s_{\i,\i+1} =\int_{-\Length}^\Length \Heav\left(\End-1\right)\Fem_\i(\myxi)\Fem_{\i+1}(\myxi)\d \myxi \\\nonumber
& =\frac{1}{\dxi} \int_{\myxi_\i}^{\myxi_{\i+1}}\Heav\left(\End-1\right) \left(\myxi_{\i+1}-\myxi\right)\left(\myxi-\myxi_\i\right) \d \myxi\\\nonumber
&=\frac{1}{6\dxi^2}
  \begin{cases}
    \dxi^3, & \ETT{t_1}{\i}, \\[2ex]
    3\myxi_{\i+1}\xibnew_{\i+1}^2+3\myxi_{\i+1}\myxi_\i^2 \\
    \mbox{} -2\xibnew_{\i+1}^3-\myxi_\i^3 , & \ETT{m_1}{\i}, \\
    \mbox{}-6\myxi_{\i+1}\myxi_\i\xibnew_{\i+1}+3\myxi_\i\xibnew_{\i+1}^2\\[2ex]
    \myxi_{\i+1}^3-3\myxi_{\i+1}\xibnew_{\i+1}^2 \\
   \mbox{}+2\xibnew_{\i+1}^3-3\myxi_\i\xibnew_{\i+1}^2,&\ETT{b_1}{\i},\\
  \mbox{}-3\myxi_\i\xibnew_{\i+1}^2+6\myxi_\i\myxi_{\i+1}\xibnew_{\i+1}\\[2ex]
    0, & \textrm{otherwise.}
  \end{cases}
\end{align}
\begin{align}\nonumber
& s_{\i-1,\i}=\int_{-\Length}^{\Length} \Heav\left(\End-1\right)\Fem_{\i}(\myxi)\Fem_{\i-1}(\myxi)\d \myxi \\\nonumber
&= \frac{1}{\dxi^2}\int_{\myxi-1}^{\myxi_\i}\Heav\left(\End-1\right)\left(\myxi-\myxi_{\i-1}\right)\left(\myxi_\i-\myxi\right)\d \myxi\\\nonumber
&=\frac{1}{6\dxi^2}
  \begin{cases}
    \dxi^3, & \ETT{t_1}{\i-1}, \\[2ex]
    3\myxi_\i\xibnew_{\i-1}^2+3\myxi_{\i-1}\xibnew_{\i-1}^2 \\
    \mbox{}+3\myxi_{\i}\myxi_{\i-1}^2-\myxi_{\i-1}^3 , & \ETT{m_1}{\i-1}, \\
    \mbox{}-6\myxi_{\i-1}\myxi_\i\xibnew_{\i-1}-2\xibnew_{\i-1}^3\\[2ex]
    \myxi_\i^3+6\myxi_\i\myxi_{\i-1}\xibnew_{\i-1}
     \\
    \mbox{}
    +2\xibnew_{\i-1}^3-3\myxi_{\i-1}\myxi_\i^2
     ,&\ETT{b_1}{\i-1}, \\
    \mbox{}-3\myxi_\i\xibnew_{\i-1}^2-3\myxi_{\i-1}\xibnew_{\i-1}^2\\[2ex]
    0, & \textrm{otherwise.}
  \end{cases}
\end{align}
We need to implement no flux boundary condition 
\begin{align}\nonumber
s_{1,1}
=\frac{1}{3\dxi^2}
  \begin{cases}
    2\dxi^3, & \ETT{t_1}{1}, \\[2ex]
    2\dxi^3-\left(\myxi_2-\xibnew_2\right)^3  \\
    \mbox{} -\left(\xibnew_2-\myxi_2+2\dxi\right)^3, & \ETT{m_1}{1}, \\[2ex]
    \left(\xibnew_2-\myxi_2+2\dxi\right)^3 \\
    \mbox{}+\left(\myxi_2-\xibnew_2\right)^3 ,&\ETT{b_1}{1}, \\[2ex]
    0, & \textrm{otherwise.}
  \end{cases}
  \end{align}
\begin{align}\nonumber
s_{\numn,\numn}
=\frac{1}{3\dxi ^2}
  \begin{cases}
    2\dxi^3, & \ETT{t_1}{\numn-1}, \\[2ex]
    \left(\myxi_{\numn-1}+2\dxi -\xibnew_{\numn-1}\right)^3 \\
    \mbox{}+\left(\xibnew_{\numn-1}-\myxi_{\numn-1}\right)^3 , &\ETT{m_1}{\numn-1}, \\[2ex]
    -\left(\myxi_{\numn-1} +2\dxi-\xibnew_{\numn-1}\right)^3
     \\
    \mbox{}2\dxi^3-\left(\xibnew_{\numn-1}-\myxi_{\numn-1}\right)^3, &\ETT{b_1}{\numn-1}, \\[2ex]
    0, & \textrm{otherwise.}
  \end{cases}
\end{align}
In these formulas, we use a shorthand notation,
\begin{align*}
& \begin{pmatrix} \ETT{t_{1}}{o} \\ \ETT{m_{1}}{o}\\\ETT{b_{1}}{o}  \end{pmatrix}=\begin{cases}
    \End_{o}>1,\; \End_{o+1}>1, \\[.2ex]
    \End_{o}>1,\; \End_{o+1}<1, \\[.2ex]
    \End_{o}<1,\; \End_{o+1}>1. \\[.2ex]
  \end{cases} 
\end{align*}

Having in mind that the tent functions we have chosen are piecewise linear, the points $\xibnew_{\i-1}$, $\xibnew_{\i+1}$, $\xibnew_{2}$ and $\xibnew_{\numn-1}$ are then obtained from the linear interpolation method
\begin{align*}
  \xibnew_{p+1}=\frac{\left[ \End\left(\myxi_{p+1}\right)-1\right] \myxi_{p}+\left[1- \End\left(\myxi_{p}\right)\right] \myxi_{p+1}}{\End\left(\myxi_{p+1}\right)-\End\left(\myxi_{p}\right)},
\end{align*}
for $p=1, \i-2,\i,\numn-2$. The vector $\feg$  has entries for $\i=2,3,\cdots,\numn-1$
\begin{align*}
  g_{\i}=&\frac{1}{\dxi}\left( \int_{\myxi_{\i-1}}^{\myxi_\i} \Heav\left(-\End\right) \left(\myxi-\myxi_{\i-1}\right)\d \myxi 
\right. \\ & \left.+\int_{\myxi_\i}^{\myxi_{\i+1}}\Heav\left(-\End\right)\left(\myxi_{\i+1}-\myxi\right)\d \myxi= \IBES^{\i}+\IALTI^{\i}
  \right),
\end{align*}
where 
\begin{align}\nonumber
 \IBES^{\i}=\frac{1}{2\dxi}
  \begin{cases}
    \dxi^2, & \ETT{t_0}{\i-1},  \\[2ex]
   \left(\xib_{\i-1}-\myxi_{\i-1}\right)^2, & \ETT{m_0}{\i-1}, \\[2ex]
   \dxi^2-\left(\xib_{\i-1}-\myxi_{\i-1}\right)^2, &\ETT{b_0}{\i-1}, \\[2ex]
    0, & \textrm{otherwise.}
  \end{cases}
\end{align}
and 
\begin{align}\nonumber
  \IALTI^{\i} =\frac{1}{2\dxi}
  \begin{cases}
     \dxi^2, & \ETT{t_0}{\i},  \\[2ex]
    \dxi^2 -\left(\myxi_{\i+1}-\xib_{\i+1}\right)^2, & \ETT{m_0}{\i}, \\[2ex]
    \left(\myxi_{\i+1}-\xib_{\i+1}\right)^2, &\ETT{b_0}{\i}, \\[2ex]
    0, & \textrm{otherwise.}
  \end{cases}
\end{align}  
and on the boundaries
\begin{align}\nonumber
  g_1=\frac{1}{2\dxi}
  \begin{cases}
    2\dx^2, & \ETT{t_0}{1}, \\[2ex]
    2\dxi^2 -\left(\xib_2-\myxi_2+2\dxi\right)^2 \\
    \quad\mbox{}-\left(\myxi_2-\xib_2\right)^2, & \ETT{m_0}{1}, \\[2ex]
    \left(\xib_2-\myxi_2+2\dxi\right)^2\\
    \quad\mbox{}+\left(\myxi_2-\xib_2\right)^2,& \ETT{b_0}{1}, \\[2ex]
    0, & \textrm{otherwise.}
  \end{cases}
\end{align}
\begin{align}\nonumber
  g_\numn=\frac{1}{2\dxi}
  \begin{cases}
    2\dx^2, &\ETT{t_0}{\numn-1}, \\[2ex]
    \left(\myxi_{\numn-1}+2\dxi-\xib_{\numn-1}\right)^2 \\
    \mbox{}+\left(\xib_{\numn-1}-\myxi_{\numn-1}\right)^2, & \ETT{m_0}{\numn-1}, \\[2ex]
    2\dxi^2-\left(\xib_{\numn-1}-\myxi_{\numn-1}\right)^2 
    \\
    -\left(\myxi_{\numn-1}+2\dxi-\xib_{\numn-1}\right)^2 ,& \ETT{b_0}{\numn-1}, \\[2ex]
    0, & \textrm{otherwise.}
  \end{cases}
\end{align}
The shorthand notations used above are, 
\begin{align*}
& \begin{pmatrix} \ETT{t_{0}}{o} \\ \ETT{m_{0}}{o}\\\ETT{b_{0}}{o}  \end{pmatrix}=\begin{cases}
    \End_{o}<0,\; \End_{o+1}<0, \\[.2ex]
    \End_{o}<0,\; \End_{o+1}>0, \\[.2ex]
    \End_{o}>0,\; \End_{o+1}<0, \\[.2ex]
  \end{cases} \\ 
\end{align*}
and $\xib$ are found via interpolation method
\begin{align}\nonumber
    \xib_{p+1}=\frac{ \End\left(\myxi_{p+1}\right)\myxi_{p}-\End\left(\myxi_{p}\right)\myxi_{p+1}}{\End\left(\myxi_{p+1}\right)-\End\left(\myxi_{p}\right)},
\end{align}
for $p=1,\i-2,\i,\numn-2$.

\subsection{Threshold curve} 

To obtain the threshold curve in the stimulus strength-duration plane, 
we solve a sequence  of the ``stimulation by current'' initial-value 
problem \eq{RDS-R} and \eq{by-current-R}. The choice of the numerical 
scheme changes according to the specific model as defined above. The 
computation is done by fixing stimulation time duration $\tst$ and 
varying the strength of the current $\ampi$. For any initial upper 
estimate $\overline\ampi$ (superthreshold), known to be sufficient 
for ignition, and lower estimate $\underline\ampi$, known to fail to 
ignite, the following bisection algorithm gives the threshold value $\ampic$:\\ 
\begin{algorithm}
\begin{algorithmic}
  \Algwidget{Input:}{$\tst$, $\overline\ampi$, $\underline\ampi$}
  \Algwidget{Output:}{$\ampic$}
  \WHILE{$\left(|\overline\ampi-\underline\ampi|\,\ge\,\tol \right)$} 
    \Algstep{$\ampit \gets \left(\overline\ampi+\underline\ampi\right)/2$}
    \Algstep{Solve \eq{RDS-R}-\eq{by-current-R} \emph{with} $\ampi=\ampit$}
    \IF{ignition} 
      \Algstep{$\overline\ampi \gets \ampit$}
    \ELSE
      \Algstep{$\underline\ampi \gets \ampit$}
    \ENDIF
  \ENDWHILE
  \Algstep{$\ampic \gets \ampit$}
\end{algorithmic}
\caption{Bisection loop for finding the strength of the current $\ampic$ 
for a fixed parameter $\tst$.}
\end{algorithm}

This procedure is repeated as many times for different $\tst$ as 
necessary to obtain the strength-duration curve.

\section{Numerical methods for simplified cardiac excitation model}
\seclabel{inadisc}

This appendix contains the discretization formula for the $\INa$-caricature model.
Specifically, we aim to provide the numerical procedure for finding the 
critical front and first two leading eigenvalues and the corresponding 
left and right eigenfunctions accordingly. To begin with, we introduce 
co-moving frame of reference by setting  $\myxi=\x-\xft$ and 
$\T=\t$ such that $\E\leri{\myxi,\T}$ 
is the voltage profile in the standard position and  $\xft$ is the 
movement of this profile. Problem \eq{evenexapbikmodel} then becomes
\begin{align}
 & \frac{\partial \E}{\partial \T}=\frac{\partial ^2 \E}{\partial \myxi^2}
+\xfonly'\leri{\t}\frac{\partial \E}{\partial \myxi} 
+ \Heav\leri{\E-1}\h, \nonumber \\ 
 & \frac{\partial \h}{\partial T}=\xfonly'\leri{t}
\frac{\partial \h}{\partial \myxi} +\frac{1}{\exty}
\left(\Heav\left(-\E\right)-\h\right).  \eqlabel{ehcomapp}
\end{align}
Here, we do the numerical computation through the initial condition 
\begin{align}
    \E\leri{\myxi,0}=\ampes\Heav\leri{-\myxi}\Heav\leri{\myxi+2\xst}-\pref.
\end{align}
We also need to impose a pinning condition in order to find the 
value of $\xfonly'$ which varies at each time step. 
This can be achieved by considering the shape of $\E$ component 
of the critical front solution. A common way to define such 
condition is to choose a constant $\Estar$ represented once 
in the front profile for every time step,  
\begin{align}
\E\leri{\xft,\T}=\Estar.
\end{align}
For simplicity, we take $\Estar=0$.

\subsection{Discretization formula for the critical front of the
  model}

Numerical analysis of the critical front for $\INa$-caricature model is based 
on a combination of finite element and finite difference methods by 
using the operator splitting method (see \eg~\cite{foulkes2010riding}). 
Finite element method is used to handle the right hand sides of the 
equations with discontinuity terms involving the Heaviside step function.
For the standard finite difference discretization, we use Beam-Warming 
scheme for the first spatial derivatives of both $\E$ and $\h$. We 
set the domain of $\myxi$ and $\T$ coordinates to be 
$-\Length\leq \myxi\leq \Length$, $0\leq \T\leq \Tfinal $ so that the 
grid points $\myxi_\i$, $\T_\j$ are
\begin{equation*}
   \myxi_\i=-\Length+\i\dxi,\quad  \T_\j=\j\delT,
\end{equation*} 
where $\dxi>0$ and $\delT>0$ are fixed space and time steps, 
and $\i=0,1,\cdots,\numn$, $\j=0,1,\cdots,\numm$ for $\numn,\numm>0$.

For convenience, we use the following shorthand notations: 
$\E_{\i}^{\j}$, $\h_{\i}^{\j}$ as the numerical approximation of  
$\E\leri{\myxi_\i,\T_\j}$ and $\h\leri{\myxi_\i,\T_\j}$, 
$\E_{\i_*}^{\j}=0$ as our pinning condition which will be 
further explained later, and finally $\spina^{\j}=\xfonly'\leri{\T_\j}$ 
as the speed at $\j$-th time step. Using these representations, 
we solve \eq{ehcomapp}  numerically using operator splitting 
method approach in the following seven steps:\\
\steps{Step 1(Finite element method):} As a first step, we solve $\E$ 
equation without the advection term as it is multiplied by the speed 
which is not determined yet
\begin{align}
\frac{\partial \E}{\partial \T}=\frac{\partial ^2 \E}{\partial \myxi^2}
+ \Heav\leri{\E-1}\h. \nonumber
\end{align}
We have to employ the finite element method due to the 
discontinuous Heaviside step function which gives
\begin{align}
\leri{\feA+\delT \feB \fetheta} \E^{\j+\frac{1}{2}}_\i &=  
\leri{\feA-\delT \feB\leri{1-\fetheta}} \E^{\j}_\i \nonumber \\ & + 
\delT \seS \h^{\j}_\i. \eqlabel{efinele}
\end{align}
\steps{Step 2(Thomas algorithm):} Inserting the elements of 
the matrices $\feA$, $\feB$, $\feD$ and the discretized 
solution $\E^{\j}_\i$ and $\h^{\j}_\i$ (both known) into  \eq{efinele} 
yields a tridiagonal system of $\numn$ equations in a following form
\begin{align*}
\left[ \begin{array}{ccccc}
\mybe_1 & \myga_1 & 0 & \dots & 0 \\
\myal_2 & \mybe_2 & \myga_2 & \dots & 0 \\
0  & \myal_3 & \mybe_3 &\ddots &  \vdots \\
\vdots & \vdots & \ddots & \ddots  & \myga_{\numn-1}  \\
0  & 0 & \dots & \myal_\numn & \mybe_\numn
\end{array} \right] \begin{pmatrix}
    \E_{1}^{\j+\frac{1}{2}} \\
    \vdots\\
    \vdots\\
    \vdots\\
    \E_{\numn}^{\j+\frac{1}{2}}
  \end{pmatrix} =  \begin{pmatrix}
    \myde_1 \\
    \vdots\\
    \vdots\\
    \vdots\\
    \myde_\numn
  \end{pmatrix}.
\end{align*}
This can be solved using a standard Gaussian elimination 
method such as Thomas algorithm and using such algorithm 
is sometimes crucial as it leads to a reduced computational cost.
The back substitution procedure (see \eg~\cite{weickert1998efficient}
for more detailed explanation) generates the solution as 
\begin{align}
& \myga_i'
= \begin{cases}
    \frac{\myga_\i}{\mybe_\i}, & \i=1, \\[.2ex]
    \frac{\myga_\i}{\mybe_\i-\myal_\i \myga_{\i-1}'} , & \i=2,3,\cdots,\numn-1,
  \end{cases} \nonumber \\
  & \myde_i'
= \begin{cases}
    \frac{\myde_\i}{\mybe_\i}, & \i=1, \\[.2ex]
    \frac{\myde_\i-\myal_\i \myde_{\i-1}'}{\mybe_\i-\myal_\i -\myga_{\i-1}'} , & \i=2,3,\cdots,\numn,
  \end{cases} \nonumber \\
  & \E_\numn^{\j+\frac{1}{2}}=\myde_\numn',  \label{thomasalE}  \\
  &  \E_{\i}^{\j+\frac{1}{2}}=\myde_i'-\myga_i'\E_{\i+1}^{\j+\frac{1}{2}}, \quad \quad \i=\numn-1,\numn-2,\cdots,1. \nonumber
\end{align}
\steps{Step 3(Finding the value of the speed):} We have divided $\E$ 
equation in \eq{ehcomapp} into two parts and it remains to find 
the solution of the advection step in the splitted scheme. 
Before we update the solution, it is necessary to find the 
value of the speed according to the pinning condition 
$\E_{\i_*}^{\j+\frac{1}{2}}=0$, where the index $\i_*$ corresponds to an 
integer value, indicating the spatial position at which the 
solution is equal to zero initially, \ie~$\myxi_{\i_*}=0$. As the 
Beam-Warming method is second order accurate, we find the speed 
value using the Beam-Warming scheme by means of the discretized 
solution found in the previous step as,
\begin{align}
  \spina^{\j} 
  &= - 
    \left(
      \left[\frac{\delT}{2\dxi} \left(3\E_{\i_*}^{\j+\frac{1}{2}}-4\E_{\i_*-1}^{\j+\frac{1}{2}}+\E_{\i_*-2}^{\j+\frac{1}{2}}\right) \right]^2
    \right.
  \nonumber \\ \lefteqn{\hspace*{-3em}
    \left.\mbox{}
      -\frac{2\delT^2}{\dxi^2} 
                 \left(\E_{\i_*}^{\j+\frac{1}{2}}-2\E_{\i_*-1}^{\j+\frac{1}{2}}+\E_{\i_*-2}^{\j+\frac{1}{2}}\right)
                 \left(\E_{\i_*}^{\j+\frac{1}{2}}-\Estar\right)
    \right)^{1/2} }
  \nonumber \\ & \mbox{}
     \times \left({\frac{\delT^2}{\dxi^2}\left(\E_{\i_*}^{\j+\frac{1}{2}}-2\E_{\i_*-1}^{\j+\frac{1}{2}}+\E_{\i_*-2}^{\j+\frac{1}{2}}\right)}  \right)^{-1}
  \nonumber \\ & \mbox{}
    - \frac{%
       \dxi  \left(3\E_{\i_*}^{\j+\frac{1}{2}}-4\E_{\i_*-1}^{\j+\frac{1}{2}}+\E_{\i_*-2}^{\j+\frac{1}{2}}\right)
      }{
        2\delT  \left(\E_{\i_*}^{\j+\frac{1}{2}}-2\E_{\i_*-1}^{\j+\frac{1}{2}}+\E_{\i_*-2}^{\j+\frac{1}{2}}\right)
      }. 
\end{align}
%
%
This formula is actually derived from the following standard Beam-Warming discretization, which is quadratic in $\spina$. \\ 
\steps{Step 4(Beam-Warming scheme):} The next step is to use Beam-Warming 
scheme to update the solution of advection term of $\E$ component at step $\j+1$,
\begin{align}
\E_{\i}^{\j+1}&=\E_{\i}^{\j+\frac{1}{2}}+\frac{\spina^{\j}\delT}{2\dxi}\left(3\E_{\i}^{\j+\frac{1}{2}}-4\E_{\i-1}^{\j+\frac{1}{2}}+\E_{\i-2}^{\j+\frac{1}{2}}\right)\nonumber\\
&+\left(\frac{\spina^{\j} \delT}{ \sqrt{2} \dxi}\right)^2\left(\E_{\i}^{\j+\frac{1}{2}}-2\E_{\i-1}^{\j+\frac{1}{2}}+\E_{\i-2}^{\j+\frac{1}{2}}\right).
\end{align}
\steps{Step 5(Finite element method):} In a similar manner, 
the equation of $\h$ component can be divided into two part and 
once again, we use finite element method to solve $\h$ equation 
with the advection term removed
\begin{align} \nonumber
    \frac{\partial \h}{\partial \T}=\frac{1}{\exty}
    \left(\Heav\left(-\E\right)-\h\right),
\end{align}
which gives
\begin{align}
\feA\left[1+\frac{\delT\theta}{\exty}\right]\h^{\j+\frac{1}{2}}_\i=
\feA\left[1-\frac{\delT\left(1-\theta\right)}{\exty}\right]
 \h^{\j}_\i+\frac{\delT \feg}{\exty}.
\eqlabel{eh1apddffd} 
\end{align}
\steps{Step 6 (Thomas algorithm)}:  As \eq{eh1apddffd} is also a 
tridiagonal matrix, we can employ the Thomas algorithm here 
as well, similarly to the case of $\E$ equation. \\
\steps{Step 7( Beam-Warming scheme):} Once again, we use second 
order accurate Beam-Warming scheme with truncation error  
$\mathcal{O}\left(\delT^2, \dxi^2\right)$ for the advection 
term of the $\h$ component
\begin{align}
 & \h_{\i}^{\j+1}=  \h_{\i}^{\j+\frac{1}{2}}+\frac{\spina^{\j}\delT}{2\dxi}
 \left(3\h_{\i}^{\j+\frac{1}{2}}-4\h_{\i-1}^{\j+\frac{1}{2}}+
 \h_{\i-2}^{\j+\frac{1}{2}}\right)\nonumber \\
 & +\left(\frac{\spina^{\j} \delT}{ \sqrt{2} \dxi}\right)^2
 \left(\h_{\i}^{\j+\frac{1}{2}}-2\h_{\i-1}^{\j+\frac{1}{2}}+\h_{\i-2}^{\j+\frac{1}{2}}\right).
\end{align}
The numerical computation of the critical front is 
achieved by solving \eq{ehcomapp} according to above 
seven-step procedure. As the $\INa$-caricature model is a two-component 
system, to find the numerical estimation of the critical front, 
we calculate $\Speed(\T)$ as
\begin{align}
  \Speed(\T)=\int\limits_{-\Length}^{\Length} 
  \leri{ \E_{\T}^2(\myxi,\T)+ \h_{\T}^2(\myxi,\T)} \,\d{\myxi}.  
\end{align}

\subsection{Discretization Formula for the Linearized Problem}

We linearize the system \eq{ehcomapp} about the critical 
front $\leri{\Ec,\hc}$ using 
\begin{align}
  \E\left(\myxi,\T\right)=\Ec\left(\myxi\right)+\epsilon 
  \Eov\left(\myxi,\T\right), \nonumber \\
  \h\left(\myxi,\T\right)=\hc\left(\myxi\right)+
  \epsilon \hov\left(\myxi,\T\right), 
\end{align}
where $\epsilon\ll 1,\, |\Eov\left(\myxi,\T\right)|\ll 1
 ,\,| \hov\left(\myxi,\T\right)|\ll 1$. Hence, we have 
 the following system of equations:
\begin{align}
 \frac{\partial \Eov}{\partial \T}=&\frac{\partial ^2 
\Eov}{\partial \myxi^2}+\spina\frac{\partial \Eov}
{\partial \myxi}-\frac{1}{\Ec'\left(-\xm\right)}
\dirac\left(\myxi+\xm \right)\hc\Eov\nonumber \\ & +\Heav
\left(-\xm-\myxi\right)\hov, \nonumber \\
 \frac{\partial \hov}{\partial \T}=&\spina
\frac{\partial \hov}{\partial \myxi}+
\left(\frac{1}{\Ec'(0)}\dirac\left(\myxi\right)\Eov-\hov\right)/\exty.
\eqlabel{folsys}
\end{align}
We solve this linearized equation with the operator 
splitting technique, by splitting the system into four equations. 
We use either the finite element or finite difference 
methods to obtain the solution of each of these four 
equations as follows:\\  
\steps{Step 1 (Finite element method):} First of all, we solve
\begin{align*}
  \frac{\partial \Eov}{\partial \T}=\frac{\partial ^2 
\Eov}{\partial \myxi^2}-\frac{1}{\Ec'
\left(-\xm\right)}\dirac\left(\myxi+\xm \right)
\hc\Eov+\Heav\left(-\xm-\myxi\right)\hov,
\end{align*}
using the finite element method this yields
\begin{align}\eqlabel{kleqsys}
 & \left[\feA+\delT \theta\leri{\feB+\frac{\feL}
 {\Ec'\left(-\xm\right)}} \right]\Eov_{\i}^{\j+\frac{1}{2}} \\
 & =\left[\feA-\delT \left(1-\theta\right)  
 \leri{\feB+\frac{\feL}{\Ec'\left(-\xm\right)}}\right]
 \Eov_{\i}^{\j}+\delT \feK \hov_{\i}^{\j}, \nonumber
\end{align}
where the matrices $\feK$ and $\feL$ are 
\begin{align*}
 \feK  = & \left[k_{\i,\j}\right] =\int_{-\Length} ^\Length
  \Heav\left(-\xm-\myxi\right) \Fem_\i(\myxi)\Fem_\j(\myxi)\,\d \myxi, \\
 \feL = & \left[l_{\i,\j}\right]=\int_{-\Length} ^\Length 
 \dirac \left(\myxi+\xm\right) \hc(\myxi) \Fem_\i(x)
 \Fem_\j(\myxi)\,\d \myxi \\ & \quad\quad =  \hc\left(-\xm\right) 
 \Fem_\i\left(-\xm\right) \Fem_\j\left(-\xm\right). 
\end{align*} 
The matrix $\feL$ has exactly $4$ non-zero elements and these are 
\begin{align}\nonumber
  l_{\i,\j}=\frac{-\hc\left(-\xm\right)}{\dxi ^2}
  \begin{cases} 
    -\left(\myxi_{l+1} + \xm \right)^2, & \i=\j=l, \\[0.2ex]
    \left(\myxi_{l+1}+\xm  \right)\left(\xm+ \myxi_l\right), & \i=l, \j=l+1, \\[.2ex]
\left(\xm+ \myxi_l\right)\left(\myxi_{l+1}+\xm \right),&\i=l+1, \j=l, \\[.2ex]
-\left(\xm+ \myxi_{l}\right)^2,&\i=\j=l+1, \\[.2ex]
    0, & \textrm{otherwise},
  \end{cases}
\end{align}
where $\myxi_{l}\leq\xm\leq\myxi_{l+1}$. Actually, 
this can be further simplified by discretizating the 
spatial domain in the way that $\xm$ is situated exactly 
on the grid that makes $\feL$ with only one non-zero element. 
On the other hand, the diagonal entries of the matrix $\feK$ 
are found as 
\begin{align}
  k_{\i,\i}=& \int_{-\Length}^\Length \Heav\left(-\xm -\myxi\right)
   \Fem_{\i}^2\left(\myxi\right)\,\d \myxi =\IYEDI^{\i}+\ISEKIZ^{\i}, \nonumber
\end{align}
where
\begin{align}\nonumber
 \IYEDI^{\i}=\frac{1}{3\dxi^2}
  \begin{cases}
    \dxi^3, & \ETT{t_{-\xm}}{\i-1},  \\[.2ex]
   -\left(\xm+ \myxi_{\i-1}\right)^3, & \ETT{m_{-\xm}}{\i-1}, \\[.2ex]
   \dxi^3+\left(\xm+ \myxi_{\i-1}\right)^3, &\ETT{b_{-\xm}}{\i-1}, \\[.2ex]
    0, & \textrm{otherwise,}
  \end{cases}
\end{align}
and
\begin{align}\nonumber
\ISEKIZ^{\i}= \frac{1}{3\dxi^2}
  \begin{cases}
    \dxi^3, & \ETT{t_{-\xm}}{\i},  \\[.2ex]
   \dxi^3 -\left(\myxi_{\i+1}+\xm \right)^3, & \ETT{m_{-\xm}}{\i}, \\[.2ex]
   \left(\myxi_{\i+1}+\xm\right)^3, &\ETT{b_{-\xm}}{\i}, \\[.2ex]
    0, & \textrm{otherwise.}
  \end{cases}
\end{align}
The supradiagonal elements of $\feK$ are evaluated as
\begin{align}\nonumber
&k_{\i,\i+1} =\int_{-\Length}^\Length \Heav\left(-\xm -\myxi\right) \Fem_\i(\myxi)\Fem_{\i+1}(\myxi)\d \myxi \\ \nonumber
& =\frac{1}{\dxi^2} \int_{\myxi_\i}^{\myxi_{\i+1}}\Heav\left(-\xm -\myxi\right) \left(\myxi_{\i+1}-\myxi\right)\left(\myxi-\myxi_\i\right) \d \myxi\\\nonumber
&=\frac{1}{6\dxi^2}
  \begin{cases}
    \dxi^3, & \ETT{t_{-\xm}}{\i}, \\[2ex]
    3\myxi_{\i+1}\xm^2+6\myxi_{\i+1}\myxi_\i\xm \\
    \mbox{}+2\xm^3+3\myxi_\i\xm^2,& \ETT{m_{-\xm}}{\i},\\
    \mbox{}+3\myxi_{\i+1}\myxi_{\i}^2-\xm^3 , \\[2ex]
    \myxi_{\i+1}^3-3\myxi_{\i}\myxi_{\i+1}^2 \\
       \mbox{}-2\xm^3-3\myxi_\i\xm^2 ,&\ETT{b_{-\xm}}{\i}, \\
    \mbox{}-3\myxi_{\i+1}\xm ^2-6\myxi_\i\myxi_{\i+1}\xm\\[2ex]
    0, & \textrm{otherwise.}
  \end{cases}
  \end{align}
The subdiagonal elements of $\feK$ are also found as
\begin{align}\nonumber
& k_{\i-1,\i}=\int_{-\Length}^{\Length} \Heav\left(-\xm -\myxi\right)\Fem_{\i}(\myxi)\Fem_{\i-1}(\myxi)\d \myxi \\\nonumber
&= \frac{1}{\dxi^2}\int_{\myxi_{{\i-1}}}^{\myxi_\i}\Heav\left(-\xm -\myxi\right)\left(\myxi-\myxi_{\i-1}\right)\left(\myxi_\i-\myxi\right)\d \myxi\\\nonumber
&=\frac{1}{6\dxi^2}
  \begin{cases}
    \dxi^3, & \ETT{t_{-\xm}}{\i-1}, \\[2ex]
    3\myxi_\i\xm ^2+2\xm ^3 \\
     \mbox{}+3\myxi_{\i}\myxi_{\i-1}^2-\myxi_{\i-1}^3 , & \ETT{m_{-\xm}}{\i-1}, \\
    \mbox{}+6\myxi_{\i-1}\myxi_\i\xm +3\myxi_{\i-1}\xm^2\\[2ex]
    \myxi_{\i}^3-3\myxi_{\i-1}\myxi_{\i}^2 \\
    \mbox{}-6\myxi_\i\myxi_{\i-1}\xm -2\xm^3 ,&\ETT{b_{-\xm}}{\i-1}, \\
    \mbox{}-3\myxi_\i\xm ^2-3\myxi_{\i-1}\xm^2\\[2ex]
    0, & \textrm{otherwise}
  \end{cases}
  \end{align}
and on the boundaries, we have
\begin{align}\nonumber
k_{1,1}
=\frac{1}{3\dxi^2}
  \begin{cases}
    2\dxi^3, & \ETT{t_{-\xm}}{1}, \\[2ex]
    2\dxi^3 -\left(2\dxi -\xm-\myxi_2\right)^3 \\
    \quad\mbox{}-\left(\myxi_2+\xm \right)^3 , & \ETT{m_{-\xm}}{1}, \\[2ex]
     \left(2\dxi-\xm-\myxi_2 \right)^3 \\
    \quad\mbox{}+\dxi^3 ,&\ETT{b_{-\xm}}{1}, \\[2ex]
    0, & \textrm{otherwise,}
  \end{cases}
  \end{align}
 \begin{align}\nonumber
k_{\numn,\numn}
=\frac{1}{3\dxi^2} 
  \begin{cases}
    2\dxi^3, & \ETT{t_{-\xm}}{\numn-1}, \\[2ex]
    \left(\myxi_{\numn-1}+2\dxi+\xm \right)^3 \\
    \quad\mbox{}-\left(\xm+ \myxi_{\numn-1}\right)^3 , & \ETT{m_{-\xm}}{\numn-1}, \\[2ex]
    2\dxi^3+\left(\xm-\myxi_{\numn-1}\right)^3
     \\
    \mbox{} -\left(\myxi_{\numn-1}+2\dxi+\xm \right)^3 ,&\ETT{b_{-\xm}}{\numn-1}, \\[2ex]
    0, & \textrm{otherwise.}
  \end{cases}
\end{align} 
The shorthand notations used above are, 
\begin{align*}
& \begin{pmatrix} \ETT{t_{-\xm}}{o} \\ \ETT{m_{-\xm}}{o}\\\ETT{b_{-\xm}}{o}  \end{pmatrix}=\begin{cases}
    \End_{o}<-\xm,\; \End_{o+1}<-\xm, \\[.2ex]
    \End_{o}<-\xm,\; \End_{o+1}>-\xm, \\[.2ex]
    \End_{o}>-\xm,\; \End_{o+1}<-\xm. \\[.2ex]
  \end{cases} \\ 
\end{align*}

\steps{Step 2 (Beam Warming scheme):} The second step is to solve 
the pure advection equation using the Beam-Warming scheme giving 
\begin{align} 
 \Eov_{\i}^{\j+1}&=\Eov_{\i}^{\j+\frac{1}{2}}+\frac{\spina\delT}{2\dxi}
 \left(4\Eov_{\i+1}^{\j+\frac{1}{2}}-3\Eov_{\i}^{\j+\frac{1}{2}}-
 \Eov_{\i+2}^{\j+\frac{1}{2}}\right)\nonumber\\
&+\left(\frac{\spina\delT}{ \sqrt{2} \dxi}\right)^2
\left(\Eov_{\i}^{\j+\frac{1}{2}}-2\Eov_{\i+1}^{\j+\frac{1}{2}}+\Eov_{\i+2}^{\j+\frac{1}{2}}\right),
\end{align}
that finishes the numerical scheme of $\Eov$ component of 
the linearized problem. \\
\steps{Step 3 (Finite element method):}  Again the finite element method 
is conveniently employed to numerically solve the first equation of 
$\hov$ component,
\begin{align*}
  \frac{\partial  \hov}{\partial \T}=\left(\frac{1}
  {\Ec'(0)}\dirac\left(\myxi\right)\Eov-\hov\right)/\exty,
\end{align*}
that results in
\begin{align}\eqlabel{syswithmapc}
 \left[ \feA+\frac{\delT \theta\feA}{\exty}\right]
 \hov_{\i}^{\j+\frac{1}{2}}= & \left[\feA-\frac{\delT 
 \left(1-\theta\right)\feA}{\exty}\right]\hov_{\i}^{\j} \nonumber \\
 & +\frac{\delT\feMM }{\exty \Ec'(0)}\Eov_{\i}^{\j}, 
\end{align}
where
\begin{align*}
\feMM=&\left[m_{\i,\j}\right]= \int_{-\Length}^\Length 
\dirac(\myxi)\Fem_\i(\myxi)\Fem_\j(\myxi)\,\d \myxi = \Fem_\i(0)\Fem_\j(0)
\\ =& \frac{1 }{\dxi^2}
  \begin{cases}
    \myxi_{r+1}^2, & \i=\j=r, \\[.2ex]
    -\myxi_{r+1}\myxi_r, & \i=r, \j=r+1, \\[.2ex]
-\myxi_r\myxi_{r+1},&\i=r+1, \j=r, \\[.2ex]
\myxi_{r}^2,&\i=\j=r+1, \\[.2ex]
    0, & \textrm{otherwise,}
  \end{cases} 
\end{align*}
such that $\myxi_r\leq 0\leq \myxi_{r+1}$. 
\steps{Step 4 (Beam Warming scheme):} Similar to $\Eov$ equation, 
we employ the Beam-Warming scheme for pure advection equation 
of  $\hov$ component,
\begin{align*}
\hov_{\i}^{\j+1}=&\hov_{\i}^{\j+\frac{1}{2}}+\frac{\spina\delT}
{2\dxi}\left(4\hov_{\i+1}^{\j+\frac{1}{2}}-3\hov_{\i}^{\j+\frac{1}{2}}-
\hov_{\i+2}^{\j+\frac{1}{2}}\right)\nonumber \\
&+\left(\frac{\spina \delT}{ \sqrt{2} \dxi}\right)^2
\left(\hov_{\i}^{\j+\frac{1}{2}}-2\hov_{\i+1}^{\j+\frac{1}{2}}+\hov_{\i+2}^{\j+\frac{1}{2}}\right),
\end{align*}
that completes the numerical solution of the linearized equation. 

\subsection{Discretization Formula for the Adjoint Linearized Problem}
\seclabel{apbikadjeq}

The adjoint linearized problem for the $\INa$-caricature model is
\begin{align}
 \frac{\partial \Ebar}{\partial \T}=&\frac{\partial ^2 
\Ebar}{\partial \myxi^2}-\spina\frac{\partial \Ebar}
{\partial \myxi}-\frac{1}{\Ec'\left(-\xm\right)}\dirac
\left(\myxi+\xm\right)\hc \Ebar&\nonumber \\ &+\frac{\dirac
\left(\myxi\right)}{\exty \Ec'(0)}\hbar, \nonumber \\
 \frac{\partial  \hbar}{\partial \T}=&-\spina
\frac{\partial \hbar}{\partial \myxi}+\Heav
\left(-\xm -\myxi\right)\Ebar-\frac{\hbar}{\exty}  .
\eqlabel{adeqina}
\end{align}
This problem can also be solved in $4$-steps as follows: \\
\steps{Step 1 (Finite element method):} We begin with the 
equation of $\Ebar$ component and solve first the following,
\begin{align*}
  \frac{\partial \Ebar}{\partial \T}=\frac{\partial ^2 
  \Ebar}{\partial \myxi^2}-\frac{1}{\Ec'\left(-\xm\right)}
  \dirac\left(\myxi+\xm\right)\hc \Ebar+
  \frac{\dirac\left(\myxi\right)}{\exty \Ec'(0)}\hbar
\end{align*}
with the solution based on the finite element method,
\begin{align} \eqlabel{advEadjapcapp}
& \left[\feA+\delT \theta\left(\feB+ \frac{\feL}
{\Ec'\left(-\xm\right)}\right) \right]\Ebar_{\i}^{\j+\frac{1}{2}}
=\frac{\delT \feMM}{\exty \Ec'(0)} \hbar_{\i}^{\j} \nonumber\\ 
& + \left[\feA-\delT \left(1-\theta\right) \left(\feB+\frac{\feL}
{\Ec'\left(-\xm\right)}\right) \right]\Ebar_{\i}^{\j}. 
\end{align}
\steps{Step 2 (Beam Warming scheme):}  As the advection term has 
negative sign in the front, the Beam-Warming numerical scheme 
is in this case,
\begin{align*}
\Ebar_{\i}^{\j+1}=& \Ebar_{\i}^{\j+\frac{1}{2}}-\frac{\spina\delT}{2\dxi}
\left(3\Ebar_{\i}^{\j+\frac{1}{2}}-4\Ebar_{\i-1}^{\j+\frac{1}{2}}+\Ebar_{\i-2}^{\j+\frac{1}{2}}\right)\nonumber\\
& + \left(\frac{\spina \delT}{ \sqrt{2} \dxi}\right)^2
\left(\Ebar_{\i}^{\j+\frac{1}{2}}-2\Ebar_{\i-1}^{\j+\frac{1}{2}}+\Ebar_{\i-2}^{\j+\frac{1}{2}}\right).
\end{align*}
\steps{Step 3 (Finite element method):} Using the finite element 
method to solve,
\begin{align*}
  \frac{\partial  \hbar}{\partial \T}=\Heav\left(-\myxi-\xm\right)
  \Ebar-\frac{\hbar}{\exty},
\end{align*}
let us obtain,
\begin{align}\eqlabel{advhadjapcapp}
 \left[\feA+\frac{\feA \delT \theta}{\exty}\right]
 \hbar_{\i}^{\j+\frac{1}{2}}=&\left[\feA-\frac{\delT \feA\left
 (1-\theta\right)}{\exty}\right]\hbar_{\i}^{\j}\nonumber \\  +& \delT \feMM  \Ebar_{\i}^{\j}. 
\end{align}
\steps{Step 4 (Beam Warming scheme):} We conclude the numerical 
solution of the adjoint problem by solving the pure advection 
equation for $\hbar$ component, 
\begin{align*}
\hbar_{\i}^{\j+1}=&\hbar_{\i}^{\j+\frac{1}{2}}-\frac{\spina\delT}
{2\dxi}\left(3\hbar_{\i}^{\j+\frac{1}{2}}-4\hbar_{\i-1}^{\j+\frac{1}{2}}+
\hbar_{\i-2}^{\j+\frac{1}{2}}\right) \\
& +\left(\frac{\spina \delT }{ \sqrt{2} \dxi}\right)^2
\left(\hbar_{\i}^{\j+\frac{1}{2}}-2\hbar_{\i-1}^{\j+\frac{1}{2}}+\hbar_{\i-2}^{\j+\frac{1}{2}}\right).
\end{align*}
We note that the Thomas algorithm can also be applied 
to the diagonal systems \eq{kleqsys}, \eq{syswithmapc}, 
\eq{advEadjapcapp} and \eq{advhadjapcapp} in a similar manner 
to the case of critical front steps. Hence, we skip the 
similar derivations here.

For this model, the linear approximation of the critical 
curves requires the knowledge of the critical front as well 
as first two leading eigenvalues and corresponding left 
and right eigenfunctions. The numerical calculating of 
the eigenpairs are determined by the  method discussed 
in \Secn{hybapp}, and these two last subsections are 
dedicated to how to derive the first two eigenmodes as 
a numerical solution of the linearized and adjoint linearized 
problems by means of the operator splitting method. 
Alternatively, the implicitly restarted Arnoldi method 
\cite{radke1996matlabnew} can be used to estimate these 
essential ingredients, in which case we use the matrix 
representations of the  discretized versions of the 
equations \eq{folsys} and \eq{adeqina}.  

\bibliographystyle{cas-model2-names}

\begin{thebibliography}{65}
\expandafter\ifx\csname natexlab\endcsname\relax\def\natexlab#1{#1}\fi
\providecommand{\url}[1]{\texttt{#1}}
\providecommand{\href}[2]{#2}
\providecommand{\path}[1]{#1}
\providecommand{\DOIprefix}{doi:}
\providecommand{\ArXivprefix}{arXiv:}
\providecommand{\URLprefix}{URL: }
\providecommand{\Pubmedprefix}{pmid:}
\providecommand{\doi}[1]{\href{http://dx.doi.org/#1}{\path{#1}}}
\providecommand{\Pubmed}[1]{\href{pmid:#1}{\path{#1}}}
\providecommand{\bibinfo}[2]{#2}
\ifx\xfnm\relax \def\xfnm[#1]{\unskip,\space#1}\fi
\bibitem[{Arendt and Bigl(1987)}]{arendt1987alzheimer}
\bibinfo{author}{Arendt, T.}, \bibinfo{author}{Bigl, V.}, \bibinfo{year}{1987}.
\newblock \bibinfo{title}{Alzheimer's disease as a presumptive threshold
  phenomenon}.
\newblock \bibinfo{journal}{Neurobiology of Aging} \bibinfo{volume}{8},
  \bibinfo{pages}{552--554}.
\bibitem[{Arutunyan et~al.(2003)Arutunyan, Pumir, Krinsky, Swift and
  Sarvazyan}]{arutunyan2003behavior}
\bibinfo{author}{Arutunyan, A.}, \bibinfo{author}{Pumir, A.},
  \bibinfo{author}{Krinsky, V.}, \bibinfo{author}{Swift, L.},
  \bibinfo{author}{Sarvazyan, N.}, \bibinfo{year}{2003}.
\newblock \bibinfo{title}{Behavior of ectopic surface: effects of
  $\beta$-adrenergic stimulation and uncoupling}.
\newblock \bibinfo{journal}{American Journal of Physiology-Heart and
  Circulatory Physiology} \bibinfo{volume}{285}, \bibinfo{pages}{H2531--H2542}.
\bibitem[{Barkley(1991)}]{barkley1991model}
\bibinfo{author}{Barkley, D.}, \bibinfo{year}{1991}.
\newblock \bibinfo{title}{A model for fast computer simulation of waves in
  excitable media}.
\newblock \bibinfo{journal}{Physica D: Nonlinear Phenomena}
  \bibinfo{volume}{49}, \bibinfo{pages}{61--70}.
\bibitem[{Barkley(2011)}]{barkley2011modeling}
\bibinfo{author}{Barkley, D.}, \bibinfo{year}{2011}.
\newblock \bibinfo{title}{Modeling the transition to turbulence in shear
  flows}.
\newblock \bibinfo{journal}{Journal of Physics: Conference Series}
  \bibinfo{volume}{318}, \bibinfo{pages}{032001}.
\bibitem[{Barkley et~al.(2015)Barkley, Song, Mukund, Lemoult, Avila and
  Hof}]{barkley2015rise}
\bibinfo{author}{Barkley, D.}, \bibinfo{author}{Song, B.},
  \bibinfo{author}{Mukund, V.}, \bibinfo{author}{Lemoult, G.},
  \bibinfo{author}{Avila, M.}, \bibinfo{author}{Hof, B.}, \bibinfo{year}{2015}.
\newblock \bibinfo{title}{The rise of fully turbulent flow}.
\newblock \bibinfo{journal}{Nature} \bibinfo{volume}{526},
  \bibinfo{pages}{550--553}.
\bibitem[{Beeler and Reuter(1977)}]{beeler1977reconstruction}
\bibinfo{author}{Beeler, G.W.}, \bibinfo{author}{Reuter, H.},
  \bibinfo{year}{1977}.
\newblock \bibinfo{title}{Reconstruction of the action potential of ventricular
  myocardial fibres}.
\newblock \bibinfo{journal}{The Journal of Physiology} \bibinfo{volume}{268},
  \bibinfo{pages}{177}.
\bibitem[{Bezekci et~al.(2015)Bezekci, Idris, Simitev and
  Biktashev}]{bezekci2015semianalytical}
\bibinfo{author}{Bezekci, B.}, \bibinfo{author}{Idris, I.},
  \bibinfo{author}{Simitev, R.D.}, \bibinfo{author}{Biktashev, V.N.},
  \bibinfo{year}{2015}.
\newblock \bibinfo{title}{Semianalytical approach to criteria for ignition of
  excitation waves}.
\newblock \bibinfo{journal}{Physical Review E} \bibinfo{volume}{92},
  \bibinfo{pages}{042917}.
\bibitem[{Biktashev(2002)}]{biktashev2002dissipation}
\bibinfo{author}{Biktashev, V.N.}, \bibinfo{year}{2002}.
\newblock \bibinfo{title}{Dissipation of the excitation wave fronts}.
\newblock \bibinfo{journal}{Physical Review Letters} \bibinfo{volume}{89},
  \bibinfo{pages}{168102}.
\bibitem[{Biktashev et~al.(2008)Biktashev, Arutunyan and
  Sarvazyan}]{biktashev2008generation}
\bibinfo{author}{Biktashev, V.N.}, \bibinfo{author}{Arutunyan, A.},
  \bibinfo{author}{Sarvazyan, N.A.}, \bibinfo{year}{2008}.
\newblock \bibinfo{title}{Generation and escape of local waves from the
  boundary of uncoupled cardiac tissue}.
\newblock \bibinfo{journal}{Biophysical Journal} \bibinfo{volume}{94},
  \bibinfo{pages}{3726--3738}.
\bibitem[{Biktashev et~al.(2011)Biktashev, Biktasheva and
  Sarvazyan}]{biktashev2011evolution}
\bibinfo{author}{Biktashev, V.N.}, \bibinfo{author}{Biktasheva, I.V.},
  \bibinfo{author}{Sarvazyan, N.A.}, \bibinfo{year}{2011}.
\newblock \bibinfo{title}{Evolution of spiral and scroll waves of excitation in
  a mathematical model of ischaemic border zone}.
\newblock \bibinfo{journal}{PloS One} \bibinfo{volume}{6},
  \bibinfo{pages}{e24388}.
\bibitem[{Blair(1932a)}]{blair1932intensity}
\bibinfo{author}{Blair, H.A.}, \bibinfo{year}{1932}a.
\newblock \bibinfo{title}{On the intensity-time relations for stimulation by
  electric currents. i}.
\newblock \bibinfo{journal}{The Journal of General Physiology}
  \bibinfo{volume}{15}, \bibinfo{pages}{709--729}.
\bibitem[{Blair(1932b)}]{blair1932intensitytwo}
\bibinfo{author}{Blair, H.A.}, \bibinfo{year}{1932}b.
\newblock \bibinfo{title}{On the intensity-time relations for stimulation by
  electric currents. ii}.
\newblock \bibinfo{journal}{The Journal of General Physiology}
  \bibinfo{volume}{15}, \bibinfo{pages}{731--755}.
\bibitem[{Bochev et~al.(2004)Bochev, Gunzburger and
  Shadid}]{bochev2004stability}
\bibinfo{author}{Bochev, P.B.}, \bibinfo{author}{Gunzburger, M.D.},
  \bibinfo{author}{Shadid, J.N.}, \bibinfo{year}{2004}.
\newblock \bibinfo{title}{Stability of the {SUPG} finite element method for
  transient advection--diffusion problems}.
\newblock \bibinfo{journal}{Computer Methods in Applied Mechanics and
  Engineering} \bibinfo{volume}{193}, \bibinfo{pages}{2301--2323}.
\bibitem[{Bostock(1983)}]{bostock1983strength}
\bibinfo{author}{Bostock, H.}, \bibinfo{year}{1983}.
\newblock \bibinfo{title}{The strength-duration relationship for excitation of
  myelinated nerve: computed dependence on membrane parameters.}
\newblock \bibinfo{journal}{The Journal of Physiology} \bibinfo{volume}{341},
  \bibinfo{pages}{59}.
\bibitem[{Brunel and van Rossum(2007)}]{brunel2007quantitative}
\bibinfo{author}{Brunel, N.}, \bibinfo{author}{van Rossum, M.C.W.},
  \bibinfo{year}{2007}.
\newblock \bibinfo{title}{Quantitative investigations of electrical nerve
  excitation treated as polarization}.
\newblock \bibinfo{journal}{Biological Cybernetics} \bibinfo{volume}{97},
  \bibinfo{pages}{341--349}.
\bibitem[{Clayton et~al.(2011)Clayton, Bernus, Cherry, Dierckx, Fenton,
  Mirabella, Panfilov, Sachse, Seemann and Zhang}]{clayton2011models}
\bibinfo{author}{Clayton, R.H.}, \bibinfo{author}{Bernus, O.},
  \bibinfo{author}{Cherry, E.M.}, \bibinfo{author}{Dierckx, H.},
  \bibinfo{author}{Fenton, F.H.}, \bibinfo{author}{Mirabella, L.},
  \bibinfo{author}{Panfilov, A.V.}, \bibinfo{author}{Sachse, F.B.},
  \bibinfo{author}{Seemann, G.}, \bibinfo{author}{Zhang, H.},
  \bibinfo{year}{2011}.
\newblock \bibinfo{title}{Models of cardiac tissue electrophysiology: progress,
  challenges and open questions}.
\newblock \bibinfo{journal}{Progress in Biophysics and Molecular Biology}
  \bibinfo{volume}{104}, \bibinfo{pages}{22--48}.
\bibitem[{Corless et~al.(1996)Corless, Gonnet, Hare, Jeffrey and
  Knuth}]{corless1996lambertw}
\bibinfo{author}{Corless, R.M.}, \bibinfo{author}{Gonnet, G.H.},
  \bibinfo{author}{Hare, D.E.}, \bibinfo{author}{Jeffrey, D.J.},
  \bibinfo{author}{Knuth, D.E.}, \bibinfo{year}{1996}.
\newblock \bibinfo{title}{On the {L}ambert {W} function}.
\newblock \bibinfo{journal}{Advances in Computational Mathematics}
  \bibinfo{volume}{5}, \bibinfo{pages}{329--359}.
\bibitem[{Cross and Hohenberg(1993)}]{cross1993pattern}
\bibinfo{author}{Cross, M.C.}, \bibinfo{author}{Hohenberg, P.C.},
  \bibinfo{year}{1993}.
\newblock \bibinfo{title}{Pattern formation outside of equilibrium}.
\newblock \bibinfo{journal}{Reviews of Modern Physics} \bibinfo{volume}{65},
  \bibinfo{pages}{851}.
\bibitem[{Doedel and Kernevez(1986)}]{doedel1986auto}
\bibinfo{author}{Doedel, E.}, \bibinfo{author}{Kernevez, J.P.},
  \bibinfo{year}{1986}.
\newblock \bibinfo{title}{AUTO, software for continuation and bifurcation
  problems in ordinary differential equations}.
\newblock \bibinfo{publisher}{California Institute of Technology}.
\bibitem[{Duckett and Barkley(2000)}]{duckett2000modeling}
\bibinfo{author}{Duckett, G.}, \bibinfo{author}{Barkley, D.},
  \bibinfo{year}{2000}.
\newblock \bibinfo{title}{Modeling the dynamics of cardiac action potentials}.
\newblock \bibinfo{journal}{Physical Review Letters} \bibinfo{volume}{85},
  \bibinfo{pages}{884--887}.
\bibitem[{Ewing and Wang(2001)}]{ewing2001summary}
\bibinfo{author}{Ewing, R.E.}, \bibinfo{author}{Wang, H.},
  \bibinfo{year}{2001}.
\newblock \bibinfo{title}{A summary of numerical methods for time-dependent
  advection-dominated partial differential equations}.
\newblock \bibinfo{journal}{Journal of Computational and Applied Mathematics}
  \bibinfo{volume}{128}, \bibinfo{pages}{423--445}.
\bibitem[{Farkas et~al.(2002)Farkas, Helbing and Vicsek}]{farkas2002social}
\bibinfo{author}{Farkas, I.}, \bibinfo{author}{Helbing, D.},
  \bibinfo{author}{Vicsek, T.}, \bibinfo{year}{2002}.
\newblock \bibinfo{title}{Social behaviour: Mexican waves in an excitable
  medium}.
\newblock \bibinfo{journal}{Nature} \bibinfo{volume}{419},
  \bibinfo{pages}{131--132}.
\bibitem[{Faye and Kilpatrick(2018)}]{Faye-Kilpatrick-2018}
\bibinfo{author}{Faye, G.}, \bibinfo{author}{Kilpatrick, Z.P.},
  \bibinfo{year}{2018}.
\newblock \bibinfo{title}{Threshold of front propagation in neural fields: an
  interface dynamics approach}.
\newblock \bibinfo{journal}{SIAM J. Appl. Math.} \bibinfo{volume}{78},
  \bibinfo{pages}{2575--2596}.
\newblock \DOIprefix\doi{10.1137/18M1165797}.
\bibitem[{Fenton et~al.(2002)Fenton, Cherry, Hastings and
  Evans}]{fenton2002real}
\bibinfo{author}{Fenton, F.H.}, \bibinfo{author}{Cherry, E.M.},
  \bibinfo{author}{Hastings, H.M.}, \bibinfo{author}{Evans, S.J.},
  \bibinfo{year}{2002}.
\newblock \bibinfo{title}{Real-time computer simulations of excitable media:
  {JAVA} as a scientific language and as a wrapper for {C} and {FORTRAN}
  programs}.
\newblock \bibinfo{journal}{Biosystems} \bibinfo{volume}{64},
  \bibinfo{pages}{73--96}.
\bibitem[{FitzHugh(1955)}]{fitzhugh1955mathematical}
\bibinfo{author}{FitzHugh, R.}, \bibinfo{year}{1955}.
\newblock \bibinfo{title}{Mathematical models of threshold phenomena in the
  nerve membrane}.
\newblock \bibinfo{journal}{The Bulletin of Mathematical Biophysics}
  \bibinfo{volume}{17}, \bibinfo{pages}{257--278}.
\bibitem[{Flores(1989)}]{flores1989stable}
\bibinfo{author}{Flores, G.}, \bibinfo{year}{1989}.
\newblock \bibinfo{title}{The stable manifold of the standing wave of the
  {N}agumo equation}.
\newblock \bibinfo{journal}{Journal of Differential Equations}
  \bibinfo{volume}{80}, \bibinfo{pages}{306--314}.
\bibitem[{Flores(1991)}]{flores1991stability}
\bibinfo{author}{Flores, G.}, \bibinfo{year}{1991}.
\newblock \bibinfo{title}{Stability analysis for the slow travelling pulse of
  the {F}itzhugh-{N}agumo system}.
\newblock \bibinfo{journal}{SIAM Journal on Mathematical Analysis}
  \bibinfo{volume}{22}, \bibinfo{pages}{392--399}.
\bibitem[{Foulkes and Biktashev(2010)}]{foulkes2010riding}
\bibinfo{author}{Foulkes, A.J.}, \bibinfo{author}{Biktashev, V.N.},
  \bibinfo{year}{2010}.
\newblock \bibinfo{title}{Riding a spiral wave: numerical simulation of spiral
  waves in a comoving frame of reference}.
\newblock \bibinfo{journal}{Physical Review E} \bibinfo{volume}{81},
  \bibinfo{pages}{046702}.
\bibitem[{Guo et~al.(2010)Guo, Zhao, Billings, Coca, Ristic and
  DeMatos}]{guo2010identification}
\bibinfo{author}{Guo, Y.}, \bibinfo{author}{Zhao, Y.},
  \bibinfo{author}{Billings, S.A.}, \bibinfo{author}{Coca, D.},
  \bibinfo{author}{Ristic, R.I.}, \bibinfo{author}{DeMatos, L.},
  \bibinfo{year}{2010}.
\newblock \bibinfo{title}{Identification of excitable media using a scalar
  coupled mapped lattice model}.
\newblock \bibinfo{journal}{International Journal of Bifurcation and Chaos}
  \bibinfo{volume}{20}, \bibinfo{pages}{2137--2150}.
\bibitem[{Hill(1936)}]{hill1936excitation}
\bibinfo{author}{Hill, A.V.}, \bibinfo{year}{1936}.
\newblock \bibinfo{title}{Excitation and accommodation in nerve}.
\newblock \bibinfo{journal}{Proceedings of the Royal Society of London. Series
  B, Biological Sciences} \bibinfo{volume}{119}, \bibinfo{pages}{305--355}.
\bibitem[{Hinch(2002)}]{hinch2002analytical}
\bibinfo{author}{Hinch, R.}, \bibinfo{year}{2002}.
\newblock \bibinfo{title}{An analytical study of the physiology and pathology
  of the propagation of cardiac action potentials}.
\newblock \bibinfo{journal}{Progress in Biophysics and Molecular Biology}
  \bibinfo{volume}{78}, \bibinfo{pages}{45--81}.
\bibitem[{Hinch(2004)}]{hinch2004stability}
\bibinfo{author}{Hinch, R.}, \bibinfo{year}{2004}.
\newblock \bibinfo{title}{Stability of cardiac waves}.
\newblock \bibinfo{journal}{Bulletin of Mathematical Biology}
  \bibinfo{volume}{66}, \bibinfo{pages}{1887--1908}.
\bibitem[{Hughes et~al.(2013)Hughes, Brindley and
  McIntosh}]{initiandproghugbridson}
\bibinfo{author}{Hughes, K.J.}, \bibinfo{author}{Brindley, J.},
  \bibinfo{author}{McIntosh, A.C.}, \bibinfo{year}{2013}.
\newblock \bibinfo{title}{Initiation and propagation of combustion waves with
  competitive reactions and water evaporation}.
\newblock \bibinfo{journal}{Proc. R. Soc. A} \bibinfo{volume}{469},
  \bibinfo{pages}{20130506}.
\bibitem[{Idris(2008)}]{idris2008initiation}
\bibinfo{author}{Idris, I.}, \bibinfo{year}{2008}.
\newblock \bibinfo{title}{Initiation Of Excitation Waves}.
\newblock Ph.D. thesis. University of Liverpool.
\bibitem[{Idris and Biktashev(2007)}]{idris2007critical}
\bibinfo{author}{Idris, I.}, \bibinfo{author}{Biktashev, V.N.},
  \bibinfo{year}{2007}.
\newblock \bibinfo{title}{Critical fronts in initiation of excitation waves}.
\newblock \bibinfo{journal}{Physical Review E} \bibinfo{volume}{76},
  \bibinfo{pages}{021906}.
\bibitem[{Idris and Biktashev(2008)}]{idris2008analytical}
\bibinfo{author}{Idris, I.}, \bibinfo{author}{Biktashev, V.N.},
  \bibinfo{year}{2008}.
\newblock \bibinfo{title}{Analytical approach to initiation of propagating
  fronts}.
\newblock \bibinfo{journal}{Physical Review Letters} \bibinfo{volume}{101},
  \bibinfo{pages}{244101}.
\bibitem[{Kaplan et~al.(1996)Kaplan, Clay, Manning, Glass, Guevara and
  Shrier}]{kaplan1996subthreshold}
\bibinfo{author}{Kaplan, D.T.}, \bibinfo{author}{Clay, J.R.},
  \bibinfo{author}{Manning, T.}, \bibinfo{author}{Glass, L.},
  \bibinfo{author}{Guevara, M.R.}, \bibinfo{author}{Shrier, A.},
  \bibinfo{year}{1996}.
\newblock \bibinfo{title}{Subthreshold dynamics in periodically stimulated
  squid giant axons}.
\newblock \bibinfo{journal}{Physical Review Letters} \bibinfo{volume}{76},
  \bibinfo{pages}{4074}.
\bibitem[{Lapicque(1907)}]{lapicque1907recherches}
\bibinfo{author}{Lapicque, L.}, \bibinfo{year}{1907}.
\newblock \bibinfo{title}{Recherches quantitatives sur l'excitation
  {\'e}lectrique des nerfs trait{\'e}e comme une polarisation}.
\newblock \bibinfo{journal}{J. Physiol. Pathol. Gen} \bibinfo{volume}{9},
  \bibinfo{pages}{620--635}.
\bibitem[{Luke and Cox(2011)}]{luke2011soil}
\bibinfo{author}{Luke, C.M.}, \bibinfo{author}{Cox, P.M.},
  \bibinfo{year}{2011}.
\newblock \bibinfo{title}{Soil carbon and climate change: from the {J}enkinson
  effect to the compost-bomb instability}.
\newblock \bibinfo{journal}{European Journal of Soil Science}
  \bibinfo{volume}{62}, \bibinfo{pages}{5--12}.
\bibitem[{McCormick et~al.(1991)McCormick, Noszticzius and
  Swinney}]{mccormick1991interrupted}
\bibinfo{author}{McCormick, W.D.}, \bibinfo{author}{Noszticzius, Z.},
  \bibinfo{author}{Swinney, H.L.}, \bibinfo{year}{1991}.
\newblock \bibinfo{title}{Interrupted separatrix excitability in a chemical
  system}.
\newblock \bibinfo{journal}{The Journal of Chemical Physics}
  \bibinfo{volume}{94}, \bibinfo{pages}{2159--2167}.
\bibitem[{McKean(1970)}]{mckean1970nagumo}
\bibinfo{author}{McKean, H.P.}, \bibinfo{year}{1970}.
\newblock \bibinfo{title}{Nagumo's equation}.
\newblock \bibinfo{journal}{Advances in Mathematics} \bibinfo{volume}{4},
  \bibinfo{pages}{209--223}.
\bibitem[{McKean and Moll(1985)}]{mckean1985threshold}
\bibinfo{author}{McKean, H.P.}, \bibinfo{author}{Moll, V.},
  \bibinfo{year}{1985}.
\newblock \bibinfo{title}{A threshold for a caricature of the nerve equation}.
\newblock \bibinfo{journal}{Bulletin of the American Mathematical Society}
  \bibinfo{volume}{12}, \bibinfo{pages}{255--259}.
\bibitem[{Moll and Rosencrans(1990)}]{moll1990calculation}
\bibinfo{author}{Moll, V.}, \bibinfo{author}{Rosencrans, S.I.},
  \bibinfo{year}{1990}.
\newblock \bibinfo{title}{Calculation of the threshold surface for nerve
  equations}.
\newblock \bibinfo{journal}{SIAM Journal on Applied Mathematics}
  \bibinfo{volume}{50}, \bibinfo{pages}{1419--1441}.
\bibitem[{Monnier and Lapicque(1934)}]{monnier1934}
\bibinfo{author}{Monnier, A.M.}, \bibinfo{author}{Lapicque, L.{\'E}.},
  \bibinfo{year}{1934}.
\newblock \bibinfo{title}{L'excitation {\'e}lectrique des tissus: essai
  d'interpr{\'e}tation physique}.
\bibitem[{Mornev(1981)}]{Mornev_1981}
\bibinfo{author}{Mornev, O.A.}, \bibinfo{year}{1981}.
\newblock \bibinfo{title}{On the conditions of excitation of one-dimensional
  autowave media}, in: \bibinfo{booktitle}{Autowave processes in systems with
  diffusion}, \bibinfo{publisher}{Institute of Applied Physics of the USSR
  Academy of Sciences}, \bibinfo{address}{Gorky}. pp. \bibinfo{pages}{92--98}.
\bibitem[{Nernst(1908)}]{nernst1908theorie}
\bibinfo{author}{Nernst, W.}, \bibinfo{year}{1908}.
\newblock \bibinfo{title}{Zur {T}heorie des elektrischen {R}eizes}.
\newblock \bibinfo{journal}{Pfl{\"u}gers Archiv European Journal of Physiology}
  \bibinfo{volume}{122}, \bibinfo{pages}{275--314}.
\bibitem[{Neu et~al.(1997)Neu, Preissig and Krassowska}]{Neu1997}
\bibinfo{author}{Neu, J.C.}, \bibinfo{author}{Preissig, R.S.},
  \bibinfo{author}{Krassowska, W.}, \bibinfo{year}{1997}.
\newblock \bibinfo{title}{Initiation of propagation in a one-dimensional
  excitable medium}.
\newblock \bibinfo{journal}{Physica D: Nonlinear Phenomena}
  \bibinfo{volume}{102}, \bibinfo{pages}{285--299}.
\bibitem[{Noble and Stein(1966)}]{Noble:1966hb}
\bibinfo{author}{Noble, D.}, \bibinfo{author}{Stein, R.B.},
  \bibinfo{year}{1966}.
\newblock \bibinfo{title}{The threshold conditions for initiation of action
  potentials by excitable cells}.
\newblock \bibinfo{journal}{J Physiol} \bibinfo{volume}{187},
  \bibinfo{pages}{129--162}.
\bibitem[{Pumir et~al.(2005)Pumir, Arutunyan, Krinsky and
  Sarvazyan}]{pumir2005genesis}
\bibinfo{author}{Pumir, A.}, \bibinfo{author}{Arutunyan, A.},
  \bibinfo{author}{Krinsky, V.}, \bibinfo{author}{Sarvazyan, N.},
  \bibinfo{year}{2005}.
\newblock \bibinfo{title}{Genesis of ectopic waves: role of coupling,
  automaticity, and heterogeneity}.
\newblock \bibinfo{journal}{Biophysical Journal} \bibinfo{volume}{89},
  \bibinfo{pages}{2332--2349}.
\bibitem[{Radke(1996)}]{radke1996matlabnew}
\bibinfo{author}{Radke, R.J.}, \bibinfo{year}{1996}.
\newblock \bibinfo{title}{A {M}atlab implementation of the implicitly restarted
  {A}rnoldi method for solving large-scale eigenvalue problems}.
\newblock Master's thesis. Rice University.
\bibitem[{Rashevsky(1933)}]{rashevsky1933outline}
\bibinfo{author}{Rashevsky, N.}, \bibinfo{year}{1933}.
\newblock \bibinfo{title}{Outline of a physico-mathematical theory of
  excitation and inhibition}.
\newblock \bibinfo{journal}{Protoplasma} \bibinfo{volume}{20},
  \bibinfo{pages}{42--56}.
\bibitem[{Rinzel and Keller(1973)}]{rinzel1973traveling}
\bibinfo{author}{Rinzel, J.}, \bibinfo{author}{Keller, J.B.},
  \bibinfo{year}{1973}.
\newblock \bibinfo{title}{Traveling wave solutions of a nerve conduction
  equation}.
\newblock \bibinfo{journal}{Biophysical Journal} \bibinfo{volume}{13},
  \bibinfo{pages}{1313}.
\bibitem[{Roth(1986)}]{roth1986association}
\bibinfo{author}{Roth, M.}, \bibinfo{year}{1986}.
\newblock \bibinfo{title}{The association of clinical and neurological findings
  and its bearing on the classification and aetiology of {A}lzheimer's
  disease}.
\newblock \bibinfo{journal}{British Medical Bulletin} \bibinfo{volume}{42},
  \bibinfo{pages}{42--50}.
\bibitem[{Rucong(1994)}]{rucong1994two}
\bibinfo{author}{Rucong, Y.}, \bibinfo{year}{1994}.
\newblock \bibinfo{title}{A two-step shape-preserving advection scheme}.
\newblock \bibinfo{journal}{Advances in Atmospheric Sciences}
  \bibinfo{volume}{11}, \bibinfo{pages}{479--490}.
\bibitem[{Rushton(1937)}]{rushton1937initiation}
\bibinfo{author}{Rushton, W.A.H.}, \bibinfo{year}{1937}.
\newblock \bibinfo{title}{Initiation of the propagated disturbance}.
\newblock \bibinfo{journal}{Proceedings of the Royal Society of London. Series
  B, Biological Sciences} \bibinfo{volume}{124}, \bibinfo{pages}{210--243}.
\bibitem[{Schiffer(1985)}]{archaeology1985advances}
\bibinfo{author}{Schiffer, M.B.}, \bibinfo{year}{1985}.
\newblock \bibinfo{title}{Advances in archaeological method and theory}.
\newblock \bibinfo{publisher}{New York: Academic Press}.
\bibitem[{Schl{\"o}gl(1972)}]{schlogl1972chemical}
\bibinfo{author}{Schl{\"o}gl, F.}, \bibinfo{year}{1972}.
\newblock \bibinfo{title}{Chemical reaction models for non-equilibrium phase
  transitions}.
\newblock \bibinfo{journal}{Zeitschrift f{\"u}r Physik} \bibinfo{volume}{253},
  \bibinfo{pages}{147--161}.
\bibitem[{Scott and Showalter(1992)}]{scott1992simplenew}
\bibinfo{author}{Scott, S.K.}, \bibinfo{author}{Showalter, K.},
  \bibinfo{year}{1992}.
\newblock \bibinfo{title}{Simple and complex propagating reaction-diffusion
  fronts}.
\newblock \bibinfo{journal}{The Journal of Physical Chemistry}
  \bibinfo{volume}{96}, \bibinfo{pages}{8702--8711}.
\bibitem[{Seiden and Curland(2015)}]{seiden2015tongue}
\bibinfo{author}{Seiden, G.}, \bibinfo{author}{Curland, S.},
  \bibinfo{year}{2015}.
\newblock \bibinfo{title}{The tongue as an excitable medium}.
\newblock \bibinfo{journal}{New Journal of Physics} \bibinfo{volume}{17},
  \bibinfo{pages}{033049}.
\bibitem[{Simitev and Biktashev(2011)}]{simitev2011asymptotics}
\bibinfo{author}{Simitev, R.D.}, \bibinfo{author}{Biktashev, V.N.},
  \bibinfo{year}{2011}.
\newblock \bibinfo{title}{Asymptotics of conduction velocity restitution in
  models of electrical excitation in the heart}.
\newblock \bibinfo{journal}{Bulletin of Mathematical Biology}
  \bibinfo{volume}{73}, \bibinfo{pages}{72--115}.
\bibitem[{Weickert et~al.(1998)Weickert, Romeny and
  Viergever}]{weickert1998efficient}
\bibinfo{author}{Weickert, J.}, \bibinfo{author}{Romeny, B.M.T.H.},
  \bibinfo{author}{Viergever, M.A.}, \bibinfo{year}{1998}.
\newblock \bibinfo{title}{Efficient and reliable schemes for nonlinear
  diffusion filtering}.
\newblock \bibinfo{journal}{IEEE Transactions on Image Processing}
  \bibinfo{volume}{7}, \bibinfo{pages}{398--410}.
\bibitem[{Weiss(1990)}]{weiss1990possibilite}
\bibinfo{author}{Weiss, G.}, \bibinfo{year}{1990}.
\newblock \bibinfo{title}{Sur la possibilit\'e de rendre comparables entre eux
  les appareils servant \`a l'excitation \'electrique.}
\newblock \bibinfo{journal}{Archives Italiennes de Biologie}
  \bibinfo{volume}{35}, \bibinfo{pages}{413--445}.
\bibitem[{Zel'dovich and Frank-Kamenetsky(1938)}]{zel1938towards}
\bibinfo{author}{Zel'dovich, Y.B.}, \bibinfo{author}{Frank-Kamenetsky, D.A.},
  \bibinfo{year}{1938}.
\newblock \bibinfo{title}{Towards the theory of uniformly propagating flames},
  in: \bibinfo{booktitle}{Doklady AN SSSR}, pp. \bibinfo{pages}{693--697}.
\bibitem[{Zipes and Jalife(2000)}]{zipes2009cardiac}
\bibinfo{author}{Zipes, D.}, \bibinfo{author}{Jalife, J.},
  \bibinfo{year}{2000}.
\newblock \bibinfo{title}{Cardiac electrophysiology: from cell to bedside}.
\newblock \bibinfo{publisher}{WB Saunders CO}.
\newblock \URLprefix \url{https://books.google.co.uk/books?id=Tp4TAQAAMAAJ}.
\bibitem[{Zykov(2008)}]{zykov2008excitable}
\bibinfo{author}{Zykov, V.S.}, \bibinfo{year}{2008}.
\newblock \bibinfo{title}{{E}xcitable media}.
\newblock \bibinfo{journal}{Scholarpedia} \bibinfo{volume}{3},
  \bibinfo{pages}{1834}.

\end{thebibliography}

\end{document}